\documentclass[pdflatex,sn-mathphys-ay]{sn-jnl}

\usepackage{graphicx}%
\usepackage{multirow}%
\usepackage{amsmath,amssymb,amsfonts}%
\usepackage{amsthm}%
\usepackage{mathrsfs}%
\usepackage[title]{appendix}%
\usepackage{xcolor}%
\usepackage{textcomp}%
\usepackage{manyfoot}%
\usepackage{booktabs}%
\usepackage{longtable}%
\usepackage{array}%
\usepackage{multicol}%
\usepackage{algorithm}%
\usepackage{algorithmicx}%
\usepackage{algpseudocode}%
\usepackage{listings}%
\usepackage[T1]{fontenc}
\usepackage{caption}
\usepackage{bbding}
\usepackage{enumerate}
\usepackage{threeparttablex}
\usepackage{pdflscape}

\usepackage{zref-clever}
\usepackage{tikz}
\zcsetup{cap=true, nameinlink=true}
\zcRefTypeSetup{figure}{
    abbrev=true,
}
\zcRefTypeSetup{equation}{
    abbrev=true
}

\providecommand{\burl}[1]{\url{#1}}

\input{Utils/acronyms}

\theoremstyle{thmstyleone}%
\theoremstyle{thmstyletwo}%

\theoremstyle{thmstylethree}%

\newcommand{\ourTitle}{LLMs in Cloud: A Discussion on Architecture, Orchestration, and Inference}
\newcommand{\ourShortTitle}{LLMs in Cloud}

\begin{document}

    \title[\ourShortTitle]{\ourTitle}


    \author[1]{\fnm{Jannatun} \sur{Noor}}\email{jannatun@cse.uiu.ac.bd}

    \author*[1,2]{\fnm{Athar Noor Mohammad} \sur{Rafee}}\email{noor.mohammad.rafee@g.bracu.ac.bd}

    \author[1]{\fnm{Md. Sadman} \sur{Haque}}\email{mhaque2620015@mscse.uiu.ac.bd}

    \author[2]{\fnm{Sreezon Das} \sur{Gupta}}\email{sreezon23@gmail.com}
    \equalcont{These authors contributed equally to this work.}

    \author[2]{\fnm{Atiar} \sur{Osman}}\email{atiar.osman@g.bracu.ac.bd}
    \equalcont{These authors contributed equally to this work.}

    \author[1,2]{\fnm{Md. Abu Obaida} \sur{Zishan}}\email{md.abu.obaida.zishan@g.bracu.ac.bd}

    \affil*[1]{\orgdiv{Computing for Sustainability and Social Good (C2SG) Research Group, Department of Computer Science and Engineering}, \orgname{United International University}, \orgaddress{\city{Dhaka}, \postcode{Dhaka 1212},  \country{Bangladesh}}}

    \affil[2]{\orgdiv{Department of Computer Science and Engineering, School of Data \& Computational Sciences}, \orgname{BRAC University}, \orgaddress{\city{Dhaka}, \postcode{Dhaka 1212}, \country{Bangladesh}}}

    \abstract{
    The widespread deployment of foundational Large Language Models (LLM) across cloud datacenters marks a fundamental paradigm shift from stateless deep learning toward distributed, memory-intensive generative workloads. However, serving massive foundation models under stringent Service Level Objectives (SLO) exposes critical physical and system bottlenecks. These challenges include the persistent High Bandwidth Memory (HBM) wall, operational asymmetry between compute-bound prefilling and memory-bandwidth-bound decoding, explosive Key-Value (KV) cache expansion, prohibitive serverless cold starts, and hardware reliability vulnerabilities. Existing literature reviews remain layer-siloed, decoupling low-level accelerator virtualization from cluster orchestration and runtime inference, while largely overlooking defining advancements from 2024 to 2026. To address these limitations, this study presents a comprehensive, cross-layer systematic review guided by the Preferred Reporting Items for Systematic Reviews and Meta-Analyses (PRISMA) 2020 framework. We synthesize a holistic taxonomy unifying hardware virtualization, cluster-level scheduling, and runtime inference optimizations, including iteration-level continuous batching, prefill-decoding disaggregation, and hierarchical KV cache offloading. Furthermore, we evaluate resilient cloud-native service architectures, examining scale-to-zero serverless paradigms, fault-tolerant checkpointing against silent data corruption, and hardware-enforced Trusted Execution Environments (TEEs) for multi-tenant isolation. Finally, we formalize the multi-dimensional Pareto trade-offs governing throughput, latency, monetary cost, and confidentiality, outlining a forward-looking research roadmap spanning Compute Express Link (CXL) memory pooling, in-memory computing, and autonomous multi-agent control planes to guide future cloud LLM infrastructure.
    }



    \maketitle

    \acresetall

    \printacronyms

    \section{Introduction}
    \label{sec:introduction}
    The rapid evolution of foundational
\acp{LLM} \citep{mohan-cloud24} has triggered
an unprecedented transformation
in modern enterprise computing,
transitioning generative intelligence
from experimental research laboratories
into mainstream production infrastructure
across public and private cloud datacenters
\citep{private-and-public-cloud,p27}.
This transition represents a fundamental paradigm shift
away from traditional deep learning microservices.
Conventional deep learning deployments
typically feature stateless,
sub-gigabyte convolutional or recurrent
networks executing uniform batch inference
on homogeneous compute architectures.
In contrast, contemporary generative foundation
models encompass tens to hundreds of billions of parameters,
requiring execution across distributed clusters of
heterogeneous hardware accelerators,
including \acp{GPU}, \acp{TPU},
and domain-specific \acp{NPU} \citep{p7,p27,p36}.

Serving these massive generative
models in cloud datacenters
introduces distinctive systems characteristics
governed by the transformer architecture.
Foremost among these is the stark bifurcation
between two computationally asymmetric execution phases:
the prompt prefill phase and the
autoregressive decoding phase \citep{p37,p38}.
The initial prefill phase processes the entire
sequence of input prompt tokens in parallel,
saturating compute units and rendering execution compute-bound.
In sharp contrast, the subsequent autoregressive
decoding phase generates response tokens sequentially,
one token per iteration.
Because each generated token
depends on all prior context,
every decoding forward pass requires
loading the entire model parameter set
and streaming dynamic attention history
across memory buses,
rendering decoding fundamentally
memory-bandwidth-bound \citep{p2,p38}.
When multiplexed across thousands of
concurrent tenants with wildly divergent
request lengths, arrival rates,
and generation targets,
this phase asymmetry severely
destabilizes cluster utilization,
introduces queue head-of-line blocking,
and complicates scheduling.

The operational economics of generative
\ac{AI} services in the cloud are dictated by
stringent, multi-dimensional \acp{SLO}
designed to guarantee satisfactory
user experience \citep{p4,p5}.
Interactive applications,
ranging from conversational
agents to code-completion
engines and complex reasoning workflows,
mandate strict limits on both \ac{TTFT}
and \ac{TBT}, alternatively designated as \ac{ITL} \citep{p2,p8}.
While \ac{TTFT} reflects the
responsiveness of prompt ingestion,
\ac{TBT} dictates the perceived
fluidity of output token generation.
Satisfying both metrics simultaneously
under dynamic multi-tenant workloads exposes
deep-rooted systems bottlenecks across
the entire cloud serving stack.

At the physical and runtime layer,
the memory wall in \ac{HBM} remains
a primary performance limiter,
as memory capacity and bus bandwidth
scale far more slowly than arithmetic
processing capability \citep{p7,p27}.
This memory wall is compounded by the
explosive expansion of the \ac{KV} cache,
which stores intermediate attention keys
and values for preceding tokens
to avoid redundant matrix computations.
Because the \ac{KV} cache expands linearly
with sequence length and batch size,
multi-turn conversations and long-context
documents consume tens of gigabytes
of scarce accelerator memory per session,
transforming previously stateless cloud
serving engines into intensely stateful systems
\citep{p21,p41,stateful-llm-serving}.
Concurrently, allocating static, monolithic
\acp{GPU} to individual services
yields severe resource underutilization,
frequently stranding compute capacity while exhausting memory,
or vice versa \citep{p12,p15,p35}.
In cloud-native serverless \ac{FaaS} environments,
scaling instances to zero introduces prohibitive
cold-start latencies of tens of seconds to minutes,
driven by the need to transfer multi-gigabyte
model weights and execution states
across host \ac{PCIe} and network interfaces
\citep{p3,p55,p59,p64}.
Furthermore, the sheer physical scale of multi-node
accelerator fabrics makes hardware faults,
soft errors, and \ac{SDC} an unavoidable operational
reality rather than a rare anomaly \citep{p4}.
Simultaneously, multi-tenant memory sharing
and opportunistic \ac{KV} cache sharing
expose critical security and privacy vulnerabilities,
including cross-tenant prompt leakage
and side-channel snooping \citep{p76}.
Navigating this nexus of latency, throughput, memory capacity,
hardware fault resilience, and multi-tenant security has
become one of the central
challenges in computer systems research.

\subsection{Research Gaps and Motivation}\label{subsec:research-gaps-and-motivation}

The unprecedented convergence of
distributed systems challenges, accelerator constraints,
and stateful generative workloads has
spurred an extensive body of literature.
However, existing surveys and systematization
exhibit significant shortcomings,
leaving practitioners and systems researchers without a holistic,
end-to-end framework for cloud-based \ac{LLM} serving.
Specifically, the existing survey landscape reveals four
critical research gaps that motivate the present study:

\begin{enumerate}[(I)]
    \item \textbf{Layer-siloed coverage.}
    Existing review articles predominantly concentrate
    on a single isolated tier of the computing stack.
    For instance, prior surveys focus narrowly on
    low-level hardware accelerators and
    microarchitectural optimizations
    \citep[e.g.,][]{p27,p106}, algorithmic transformer optimizations
    and local model-serving runtime engines
    \citep[e.g.,][]{p32,p48}, generic cloud autoscaling protocols and
    \ac{AIOps} \citep[e.g.,][]{p29,p108},
    or safety, debiasing, and ethical
    guardrails \citep[e.g.,][]{p103,p104,p111}.
    While valuable within their specialized scopes,
    these works fail to construct cross-layer
    connections between physical accelerator virtualization,
    distributed cluster orchestration,
    runtime serving engines, and cloud infrastructure resilience.

    \item \textbf{Disconnect between cloud orchestration and \ac{LLM} workload characteristics.}
    General-purpose cloud autoscaling and
    cluster orchestration surveys \citep[e.g.,][]{p29}
    are formulated around stateless
    microservices or classical batch jobs.
    They omit the profound operational
    implications of stateful \ac{KV} cache migration,
    phase-asymmetric chunked prefilling,
    and distributed model sharding.
    Conversely, recent operational and
    \ac{AIOps} surveys \citep[e.g.,][]{p28,p49,p108}
    predominantly evaluate \acp{LLM} as diagnostic
    tools used to automate cloud management,
    rather than investigating \ac{LLM} inference itself as a primary,
    heterogeneous workload that must be scheduled
    and autoscaled across cloud fabrics.

    \item \textbf{Decoupling of reliability and security from systems performance.}
    Surveys examining \ac{LLM} trustworthiness, robustness,
    and governance \citep[e.g.,][]{p103,p104,p111}
    typically analyze verification algorithms,
    adversarial prompts, and safety guardrails
    in algorithmic isolation.
    These studies largely abstract away the
    concrete cloud systems constraints
    under which safety mechanisms must execute,
    such as interactive token latency budgets,
    dynamic continuous batching overheads,
    \ac{KV} cache co-location hazards,
    and hardware-enforced \acp{TEE}.
    As a result, the direct trade-offs
    between security enforcement,
    fault tolerance, and serving
    throughput remain largely unaddressed.

    \item \textbf{Rapid evolution and temporal obsolescence.}
    Given the extraordinary velocity of \ac{MLSys} innovation,
    surveys whose scopes concluded in 2023 or mid-2024 \citep[e.g.,][]{p106}
    necessarily miss defining advancements
    that have emerged between 2024 and 2026.
    Crucial recent breakthroughs:
    such as prefill-decoding disaggregation \citep{p38},
    \ac{CXL}-based unified memory pooling \citep{p79},
    zero-overhead serverless \ac{LLM} cold-start reduction \citep{p3,p59},
    proactive \ac{LoRA} adapter consolidation \citep{p22,p61},
    and confidential accelerated computing:
    remain absent from prior literature reviews.
\end{enumerate}

Closing these interconnected gaps
requires an architectural survey
that synthesizes the entire stack,
unifying hardware virtualization,
cluster scheduling,
runtime inference mechanics,
cloud-native resilience,
and security into a coherent systems paradigm.

\subsection{Contributions of this Survey}\label{subsec:contributions-of-this-survey}

To address the limitations of prior reviews and establish a comprehensive reference
for cloud-based \ac{LLM} infrastructure,
this survey presents an in-depth,
cross-layer synthesis of architecture,
orchestration,
and inference.
The primary contributions of this work are summarized as follows:

\begin{enumerate}[(1)]
    \item \textbf{Rigorous \acs*{PRISMA}-Guided Methodology:}
    We implement a systematic review protocol
    governed by the updated \ac{PRISMA} 2020 framework \citep{prisma-method-og-paper}.
    Our protocol establishes formal research questions,
    a structured multi-database search across
    top-tier systems and \ac{AI} venues,
    and transparent inclusion, exclusion,
    and quality assessment criteria to ensure
    reproducible literature curation.

    \item \textbf{Holistic Cross-Layer Taxonomy:}
    We formulate a unified architectural taxonomy
    that integrates physical hardware virtualization
    (e.g., \ac{MIG} and \ac{MPS}),
    cluster-wide scheduling, runtime-serving engines,
    and resilient cloud-native deployment patterns
    into an interconnected systems hierarchy.

    \item \textbf{Comprehensive Analysis of Cloud Inference and Orchestration:}
    We deliver an exhaustive technical analysis of runtime inference optimizations,
    dissecting iteration-level continuous batching,
    prefill-decoding disaggregation,
    semantic and chunk-based \ac{KV} cache reuse,
    and hierarchical memory offloading under strict
    multi-tenant \ac{TTFT} and \ac{TBT} \acp{SLO}.

    \item \textbf{Examination of Resilient, Serverless, and Confidential Architectures:}
    We provide a critical synthesis of
    next-generation cloud architectures,
    evaluating scale-to-zero serverless serving mechanisms,
    \ac{ABFT} and low-overhead checkpointing for \ac{SDC} mitigation,
    distributed agentic workflow orchestration,
    intelligent \ac{AI} gateways,
    and hardware-enforced \acp{TEE} for multi-tenant isolation.

    \item \textbf{Synthesis of Cross-Cutting Trade-Offs and Research Roadmap:}
    We systematically map the complex Pareto
    frontier governing serving throughput, token latency,
    monetary expenditure, and privacy guarantees.
    Building upon this analysis,
    we articulate a forward-looking research roadmap
    spanning \ac{CXL}-enabled disaggregated memory pools,
    in-memory \ac{HDC} \citep{hdc-og-study, hdc-latest-review},
    and autonomous decentralized multi-agent control planes.
\end{enumerate}

\subsection{Outline of The Study}\label{subsec:organization-of-the-study}

The remainder of the study is organized as follows.
\zcref{sec:related-works} reviews existing literature, related surveys,
and delineates the comparative gaps motivating this review.
\zcref{sec:method} details our \ac{PRISMA}-guided systematic review methodology,
search strategy, eligibility criteria, and study selection protocol.
\zcref{sec:orchestration} examines cloud orchestration
and resource scheduling, covering accelerator virtualization,
multi-tenant scheduling, and cluster coordination.
\zcref{sec:inference-in-cloud} analyzes runtime inference
optimization techniques, including request scaling, dynamic batching,
\ac{KV} cache reuse, and multi-tier cost management.
\zcref{sec:architecture} investigates cloud-based service architectures,
encompassing serverless paradigms, fault-tolerant service mechanisms,
agentic workflows, intelligent \ac{AI} gateways, and confidential computing.
\zcref{sec:discussion} synthesizes cross-cutting architectural trade-offs,
delineates open research horizons, and discusses
the methodological limitations of this survey.
Finally, \zcref{sec:conclusion} concludes the manuscript
by summarizing our core findings and strategic insights.

For clarity and ease of reference,
a comprehensive summary of the terminology and abbreviations employed in this study
is provided in the preliminary acronym list preceding the main text.

    \section{Related Works}
    \label{sec:related-works}
    The rapid growth of \ac{LLM} deployment on cloud
infrastructure has motivated a substantial
body of survey and review literature.
Unlike the individual systems papers discussed elsewhere in
this review, the works surveyed in this section themselves attempt to
synthesize and organize large bodies of prior research.
We group the nine reviewed survey papers into four thematic clusters that
mirror the pillars of our own review \emph{architecture},
\emph{orchestration}, and \emph{inference} and add a fourth cluster on
\emph{reliability and trustworthiness}, since several existing surveys
treat this as an inseparable concern of deploying \acp{LLM} at a cloud scale.
For each cluster, we summarize the scope,
contributions, and self-reported limitations of
the constituent surveys, before presenting
a comparative overview in \zcref{tab:related-works-summary} and closing
with an explicit discussion of the gaps that motivate our own review.

\subsection{Hardware Architecture and Acceleration for LLMs}
\label{subsec:rw-hardware}

Two surveys in our review focus primarily on the hardware and low-level
systems layer that underpins \ac{LLM} training and inference.
\citet{p27} provides a survey of hardware accelerators for \acp{LLM},
covering \acp{GPU}, \acp{FPGA}, and custom-designed \ac{ASIC} architectures.
The author argues that as \acp{LLM} workloads grow,
energy consumption in data centers becomes a
first-class design concern alongside raw
throughput and shows that specialized accelerators can reduce
energy requirements by several orders of magnitude relative to
general-purpose hardware.
The survey also highlights emerging
directions such as in-memory computing
and neuromorphic processors,
positioning them as long-term
alternatives to \ac{GPU}-centric design.
However, the survey remains largely
hardware-centric and does not
connect these accelerator-level
gains to higher-level cloud
orchestration or serving-system
concerns such as scheduling or  autoscaling.

\citet{p106} take a broader, lifecycle-oriented view.
Following a systematic literature review process based on the
\ac{PRISMA} statement, the authors analyze 65 publications
(out of 983 initially retrieved) spanning 2017 to December 2023,
and organize \ac{LLM} optimization and
acceleration techniques into three
classes: training, inference, and system serving.
Two case studies are used to demonstrate
practical trade-offs between training efficiency
and inference efficiency under limited resources.
While the taxonomy is comprehensive,
the review's own stated scope ends in late 2023,
meaning that many recent
inference-serving and cloud-orchestration
innovations discussed in newer literature,
including our own review,
are necessarily absent.

\subsection{Efficient LLM Serving and Inference Systems}
\label{subsec:efficient-llm-serving-and-inference-systems}

A second cluster of surveys concentrates specifically on the
algorithmic and systems techniques used
to serve \acp{LLM} efficiently once they are deployed.
\citet{p32} present a survey from an \ac{MLSys} perspective,
covering algorithmic modifications
(e.g., decoding strategies, model compression) as well as
systems-level changes (e.g., batching,
memory management) that jointly
determine serving latency and throughput.
The survey is positioned as a bridge between
the machine-learning and systems communities,
but by its own framing it emphasizes breadth
over a fine-grained treatment of
cloud-specific concerns such
as multi-tenant resource sharing or
cost-aware scheduling.

\citet{p48} offer a more hierarchical treatment of the same problem
space, organizing techniques into
\emph{instance-level} approaches (model
placement, request scheduling,
decoding-length prediction,
storage management, and prefill/decode
disaggregation) and
\emph{cluster-level} strategies,
including \ac{GPU} cluster deployment,
multi-instance load balancing,
and cloud service integration.
The authors explicitly acknowledge
two limitations relevant to our
work: first, that their review
does not conduct detailed comparative
experiments across the surveyed methods,
relying instead on
qualitative synthesis;
and second, that their coverage is time-bound,
concluding at the end of April 2025
and therefore excluding the most
recent serverless and \ac{KV}-cache-management literature.
Both surveys converge on the observation
that serving efficiency cannot be
optimized at the model or instance level alone,
foreshadowing the need
for cluster- and cloud-level orchestration,
which our review addresses more directly.

\subsection{Cloud Orchestration and Operational Automation}
\label{subsec:cloud-orchestration-and-operational-automation}

A third cluster addresses the orchestration
layer that governs how
cloud resources are allocated, scaled,
and operated around \ac{LLM} workloads.
\citet{p29} present a comprehensive survey of autoscaling
techniques in cloud-native computing.
The authors classify autoscaling
algorithms and mechanisms into three categories,
ranging from threshold-based reactive policies
to \ac{AI}-based proactive policies, and
further examine the experimental environments,
datasets, and metrics used to evaluate these
techniques across the application lifecycle.
Notably, the survey also reviews
security concerns specific to
autoscaling, such as cyberattacks
that exploit elastic scaling behavior.
While thorough, this survey is not written with \ac{LLM}
workloads in mind;
its taxonomy is general to cloud-native applications
and does not account for the distinctive memory footprint,
stateful \ac{KV} cache behavior,
or cold-start characteristics that make
autoscaling \acp{LLM} qualitatively
different from scaling conventional microservices.

\citet{p108} instead examine orchestration from an
operational-automation angle, surveying 183 research articles
published between January 2020 and December 2024
on the use of \acp{LLM} for \acp{AIOps}.
Structured around four research questions, the survey covers the
failure-data sources used for \ac{AIOps},
the evolving taxonomy of \ac{AIOps}
tasks, the \ac{LLM}-based methods applied to them,
and the evaluation methodologies adopted in this space.
The authors identify open
challenges specific to \ac{LLM}-driven \ac{AIOps},
including the time and cost
efficiency of \ac{LLM}-based solutions and
their integration with existing toolchains,
but they explicitly exclude general \ac{LLM} limitations such
as hallucination from their scope,
treating them as outside the boundary of the survey.
Read together, \citet{p29} and \citet{p108}
show that orchestration research has so far developed along two
largely separate tracks, namely classical
autoscaling theory and \ac{LLM}-assisted
operations, without a unified treatment
of how \acp{LLM} are themselves
autoscaled and orchestrated in the cloud.

\subsection{Reliability, Safety, and Trustworthiness}
\label{subsec:rw-reliability}

The final and largest cluster of surveys
addresses the reliability and
trustworthiness of \acp{LLM},
a concern that becomes especially important
once models are exposed
as always-on cloud services.
\citet{p103} conduct a wide-ranging survey,
drawing on more than 370
references, that reviews
known vulnerabilities and limitations of \acp{LLM}
and organizes them into inherent issues,
attacks, and unintended bugs.
The central contribution of this survey is its proposal to adapt
\ac{V-and-V} techniques,
originally developed
for traditional software and deep learning models,
to the \ac{LLM} setting
through falsification, formal verification,
runtime monitoring, and
regulatory or ethical review.
The authors note that \acp{LLM} exhibit
nondeterministic behavior and are trained
on uncurated internet-scale
data, which limits the direct
applicability of conventional \ac{V-and-V}
methods and calls for
further methodological innovation.

\citet{p104} narrow the focus to two specific
failure modes (i.e., bias and hallucination),
and present what they describe
as the first review to jointly synthesize both problems.
The survey builds parallel
taxonomies for evaluation benchmarks
and mitigation strategies for
each issue, and provides a
comparative analysis of their sources,
evaluation methods,
and mitigation techniques side by side.
\citet{p111} instead survey
the broader mechanism of
\emph{guardrails},
examining how major \ac{LLM} providers and the
open-source community implement
safeguarding mechanisms, how these
guardrails can be evaluated and strengthened,
and how they can be
circumvented through adversarial attacks.
The survey concludes that
guardrail design remains an open,
multi-disciplinary problem requiring
neural-symbolic methods and systems-level engineering,
rather than a solved component of \ac{LLM} deployment.

Taken together, these three surveys establish reliability and
trustworthiness as essential,
cross-cutting properties of deployed \acp{LLM}.
However, none of them situates these properties within the
cloud-systems context that determines how a model is actually
served; for example,
how guardrail enforcement interacts with request
batching, or how runtime monitoring
can be implemented without
violating serving-latency budgets.

\subsection{Comparative Summary}
\label{subsec:comparative-summary}

\zcref{tab:related-works-summary} summarizes the scope, methodology,
and self-reported limitations of the nine surveyed papers.
Several patterns emerge from this body of literature.
First, existing surveys are largely \emph{siloed by layer}:
hardware-focused surveys~\citep{p27,p106} rarely engage with
orchestration or reliability concerns; serving-efficiency
surveys~\citep{p32,p48} treat scheduling and resource management
primarily as algorithmic problems rather than as cloud-operational
ones; and orchestration surveys~\citep{p29,p108} are either agnostic
to \ac{LLM}-specific workload characteristics or narrowly scoped to
operational automation rather than inference serving itself.
Second, the reliability-focused surveys~\citep{p103,p104,p111} are
comprehensive within their own scope but are written largely
independently of the systems literature, leaving open the question of
how safety and trustworthiness mechanisms can be integrated into
cloud-serving pipelines without degrading latency or cost efficiency.
Third, several surveys explicitly acknowledge temporal or
methodological boundaries that limit their currency.
For instance, \citet{p106} concludes coverage in December 2023,
and \citet{p48} in April 2025 which underscores
how quickly this field moves and how narrow
the effective shelf life of any single survey can be.
\zcref{tab:comparison-existing-surveys} compares our
review against the nine surveyed papers across these dimensions,
highlighting the broader scope of our work.

\begingroup
\footnotesize
\setlength{\tabcolsep}{3pt}
\renewcommand{\arraystretch}{1.15}
\begin{longtable}{@{}
        >{\raggedright\arraybackslash}p{0.08\linewidth}
        >{\raggedright\arraybackslash}p{0.18\linewidth}
        >{\raggedright\arraybackslash}p{0.34\linewidth}
        >{\raggedright\arraybackslash}p{0.32\linewidth}
    @{}}
    \caption{Comparative summary of surveyed review papers related to LLM architecture, orchestration, and inference in the cloud.}\label{tab:related-works-summary}\\
    \toprule
    \textbf{Ref.} & \textbf{Focus Area}            & \textbf{Core Contribution}                                                                           & \textbf{Reported Limitation}                                                 \\
    \midrule
    \endfirsthead

    \multicolumn{4}{c}{\small\itshape Table \thetable{} (continued from previous page)} \\
    \addlinespace
    \toprule
    \textbf{Ref.} & \textbf{Focus Area}            & \textbf{Core Contribution}                                                                           & \textbf{Reported Limitation}                                                 \\
    \midrule
    \endhead

    \midrule
    \multicolumn{4}{r}{\itshape Continued on next page} \\
    \endfoot

    \bottomrule
    \endlastfoot

    \citet{p27}   & Hardware Architecture          & Survey of \ac{GPU}/\ac{FPGA}/\ac{ASIC} accelerators for \acp{LLM}; energy-efficiency analysis & Hardware-centric; no link to cloud orchestration or serving systems \\
    \citet{p106}  & Hardware \& System Lifecycle   & \ac{PRISMA}-based systematic review of 65 papers; taxonomy across training, inference, and system serving & Coverage ends December 2023; excludes recent serving/orchestration work \\
    \midrule
    \citet{p32}   & Inference Serving              & \ac{MLSys}-perspective survey of algorithmic and systems techniques for efficient serving & Emphasizes breadth; limited depth on cloud-specific resource sharing \\
    \citet{p48}   & Inference Serving              & Hierarchical review: instance-level and cluster-level serving techniques & No comparative experiments; coverage ends April 2025 \\
    \midrule
    \citet{p29}   & Cloud Orchestration            & Taxonomy of reactive/proactive autoscaling policies; lifecycle and security review & General cloud-native scope; not \ac{LLM}-workload aware \\
    \citet{p108}  & Operational Orchestration      & Survey of 183 papers on \ac{LLM}-driven \ac{AIOps} across four research questions & Explicitly excludes general \ac{LLM} limitations (e.g., hallucination) \\
    \midrule
    \citet{p103}  & Reliability \& Safety          & 370+ reference survey adapting verification and validation techniques to \ac{LLM} vulnerabilities & Conventional \ac{V-and-V} methods only partially applicable to nondeterministic \acp{LLM} \\
    \citet{p104}  & Reliability \& Trustworthiness & Joint review of bias and hallucination benchmarks and mitigation taxonomies & Focused solely on two failure modes; no systems-deployment context \\
    \citet{p111}  & Reliability \& Safety          & Survey of \acp{LLM} guardrail mechanisms, attacks, and defenses & Guardrail design identified as open, unresolved problem \\
\end{longtable}
\endgroup

\begingroup
\setlength{\tabcolsep}{3pt}
\renewcommand{\arraystretch}{1.1}
\begin{table}[htbp]
    \centering
    \caption{
        Comparison with existing review and survey papers
        on LLM architecture, orchestration,
        and inference in the cloud.
    }
    \label{tab:comparison-existing-surveys}
    \footnotesize
    \begin{tabular}{@{}
            >{\raggedright\arraybackslash}p{0.15\linewidth}
            >{\centering\arraybackslash}p{0.05\linewidth}
            >{\centering\arraybackslash}p{0.10\linewidth}
            >{\centering\arraybackslash}p{0.13\linewidth}
            >{\centering\arraybackslash}p{0.13\linewidth}
            >{\centering\arraybackslash}p{0.15\linewidth}
            >{\centering\arraybackslash}p{0.13\linewidth}
            >{\centering\arraybackslash}p{0.08\linewidth}
            >{\centering\arraybackslash}p{0.09\linewidth}
        @{}}
        \toprule
        \textbf{Survey}                         & \textbf{Year} & \textbf{Hardware Accel.} & \textbf{Inference Serving} & \textbf{Autoscaling} & \textbf{Orchestration} & \textbf{Reliability} & \textbf{Cloud Native} & \textbf{Future Trends} \\
        \midrule

        \citeauthor{p27}                        & 2025          & \Checkmark               & partial                    & \XSolid              & \XSolid                & \XSolid              & partial               & \Checkmark             \\
        \midrule

        \citeauthor{p106}                       & 2024          & partial                  & \Checkmark                 & partial              & partial                & partial              & \XSolid               & \Checkmark             \\
        \midrule

        \citeauthor{p32}                        & 2025          & partial                  & \Checkmark                 & \XSolid              & partial                & \XSolid              & partial               & \Checkmark             \\
        \midrule

        \citeauthor{p48}                        & 2025          & \XSolid                  & \Checkmark                 & partial              & \Checkmark             & \XSolid              & \Checkmark            & \Checkmark             \\
        \midrule

        \citeauthor{p29}                        & 2025          & \XSolid                  & \XSolid                    & \Checkmark           & \Checkmark             & partial              & \Checkmark            & \Checkmark             \\
        \midrule

        \citeauthor{p108}                       & 2025          & \XSolid                  & \XSolid                    & partial              & \Checkmark             & \XSolid              & partial               & \Checkmark             \\
        \midrule

        \citeauthor{p103}                       & 2024          & \XSolid                  & \XSolid                    & \XSolid              & \XSolid                & \Checkmark           & \XSolid               & \Checkmark             \\
        \midrule

        \citeauthor{p104}                       & 2024          & \XSolid                  & \XSolid                    & \XSolid              & \XSolid                & \Checkmark (focus)   & \XSolid               & partial                \\
        \midrule

        \citeauthor{p111}                       & 2025          & \XSolid                  & \XSolid                    & \XSolid              & \XSolid                & \Checkmark           & \XSolid               & \Checkmark             \\

        \midrule[\heavyrulewidth]
        \textbf{This survey}                    & \textbf{2026} & \Checkmark               & \Checkmark                 & \Checkmark           & \Checkmark             & \Checkmark           & \Checkmark            & \Checkmark             \\
        \bottomrule
    \end{tabular}
\end{table}
\endgroup

    \section{Methodology}
    \label{sec:method}
    To establish a rigorous, reproducible,
and transparent foundation for analyzing cloud architectures,
resource orchestration frameworks,
and inference serving systems for \acp{LLM},
this survey adopts the \ac{PRISMA}
2020 guidelines \citep{prisma-method-og-paper}.
While \ac{PRISMA} was originally established
for clinical and healthcare syntheses,
its structured principles have been widely adapted
to software engineering and computer systems research
to systematically map complex,
rapidly expanding technological paradigms \citep{p106}.
In this section, we outline the research
questions that guide our investigation,
delineate the multi-database search strategy,
define explicit inclusion and exclusion criteria,
detail the end-to-end study selection process,
present the data extraction protocol,
and analyze potential threats to validity.

\subsection{Research Questions}
\label{subsec:method-rq}
Deploying and serving \acp{LLM} at datacenter scale
introduces multidimensional systems challenges,
ranging from accelerator
virtualization and dynamic scheduling
to memory-bandwidth bottlenecks,
serverless elasticity,
and hardware fault tolerance.
To systematically categorize
and evaluate these dimensions,
we formulate four overarching \acp{RQ}
that align directly with
the primary thematic sections of this survey:

\begin{itemize}
    \item \textbf{RQ1 (Orchestration and Hardware Virtualization):}
    How do modern cloud platforms and cluster schedulers
    virtualize, partition, and coordinate heterogeneous hardware accelerators
    to maximize multi-tenant resource utilization
    while enforcing fair-share scheduling,
    workload isolation,
    and throughput guarantees for \ac{LLM} workloads?
    (Investigated in \zcref{sec:orchestration}).

    \item \textbf{RQ2 (Inference Serving Efficiency and Latency):}
    What algorithmic, memory management,
    and batching mechanisms effectively optimize cloud-based
    \ac{LLM} inference serving pipelines
    to minimize \ac{TTFT} and \ac{TBT}
    under stringent multi-tenant \ac{SLO} constraints?
    (Investigated in \zcref{sec:inference-in-cloud}).

    \item \textbf{RQ3 (Serverless and Cloud-Native Architectures):}
    How do serverless computing paradigms (\ac{FaaS}),
    container orchestration frameworks such as \ac{K8s},
    and intelligent \ac{AI} gateways mitigate cold start latency,
    optimize stateful traffic routing,
    and enable elastic horizontal scaling for distributed \ac{LLM} deployments?
    (Investigated in \zcref{sec:architecture},
    specifically \zcref{subsec:serverless-architecture}
    and \zcref{subsec:ai-gateways-and-traffic-routing}).

    \item \textbf{RQ4 (Reliability, Fault Tolerance, and Privacy):}
    What architectural primitives,
    checkpointing strategies,
    and confidential computing mechanisms ensure high availability,
    resilience against hardware faults or silent data corruption,
    and secure multi-tenant execution in cloud \ac{LLM} ecosystems?
    (Investigated in \zcref{sec:architecture},
    specifically \zcref{subsec:fault_tolerance}
    and \zcref{subsec:confidential-computing-and-privacy-preserving-ai}).
\end{itemize}

\subsection{Search Strategy and Information Sources}
\label{subsec:method-search}
To assemble an exhaustive and
peer-reviewed corpus of systems research,
we executed comprehensive automated searches
across four premier computer science
and engineering digital libraries:
IEEE Xplore, ACM Digital Library,
SpringerLink, and ScienceDirect (Elsevier).
To preserve strict peer-reviewed academic rigor
and avoid unvalidated empirical claims,
Google Scholar, arXiv preprints,
and non-archival trade publications
were explicitly excluded from the
primary systematic database queries.
This deliberate boundary guarantees
that every primary study in our core corpus
has undergone formal peer review
and rigorous scholarly evaluation.

The search window encompasses publications
from January 2020 through February 2026.
This multi-year timeframe captures
early foundation model deployment efforts
following the mainstream adoption of large Transformer architectures,
and extends through contemporary cloud-native innovations
in disaggregated serving,
serverless inference,
and hardware-assisted security.

The search strategy employed parameterized boolean queries
combining controlled vocabulary and free-text terms across three core facets:
(i) the \ac{LLM} facet, (ii) the cloud and distributed infrastructure facet,
and (iii) the systems and execution efficiency facet.
The overarching boolean query structure
was specified as follows:

\begin{quote}
    \begingroup
    \small\raggedright
    \texttt{(``large language model*'' OR ``transformer*'' OR ``foundation model*'' OR ``generative AI'' OR ``LLM*'')}\par
    \vspace{2pt}
    \textbf{AND}\par
    \vspace{2pt}
    \texttt{(``cloud computing'' OR ``cloud-native'' OR ``datacenter*'' OR ``cluster*'' OR ``multi-tenant*'' OR ``serverless'' OR ``FaaS'' OR ``Kubernetes'' OR ``microservice*'')}\par
    \vspace{2pt}
    \textbf{AND}\par
    \vspace{2pt}
    \texttt{(``orchestration'' OR ``resource scheduling'' OR ``GPU virtualization'' OR ``inference serving'' OR ``batching'' OR ``KV cache'' OR ``fault tolerance'' OR ``confidential computing'')}\par
    \endgroup
\end{quote}

The exact search syntax was systematically customized
for each digital library to accommodate
platform-specific field tags, wildcard operators,
and boolean nesting rules.
Queries were executed against document titles,
abstracts, and author-supplied keywords.

\subsection{Eligibility Criteria}
\label{subsec:method-eligibility}
To guarantee that the retrieved literature
directly addresses systems-level challenges
in cloud \ac{LLM} environments,
we established four Inclusion Criteria (IC1 to IC4)
and three Exclusion Criteria (EC1 to EC3).
Each retrieved publication was required to satisfy all inclusion criteria
and satisfy none of the exclusion criteria.

\subsubsection{Inclusion Criteria}
\begin{itemize}
    \item \textbf{IC1 (Temporal and Language Scope):}
    The article must be written in English
    and published in a peer-reviewed venue
    between January 2020 and February 2026.

    \item \textbf{IC2 (Venue Archival Status):}
    The study must appear in an archival,
    peer-reviewed international conference proceedings
    or a scholarly journal indexed within
    the four target digital libraries.

    \item \textbf{IC3 (Thematic Systems Focus):}
    The study must explicitly investigate
    systems-level cloud architecture,
    hardware accelerator virtualization,
    cluster orchestration, runtime inference serving optimizations,
    cloud-native autoscaling, fault tolerance,
    or confidential computing for \acp{LLM}.

    \item \textbf{IC4 (Empirical and Methodological Rigor):}
    The publication must propose a concrete systems architecture,
    scheduling algorithm, runtime framework,
    or empirical benchmark with
    reproducible experimental evaluation
    measuring systems-level metrics
    (such as latency percentiles,
    throughput, \ac{GPU} utilization,
    memory consumption, or monetary serving cost).
\end{itemize}

\subsubsection{Exclusion Criteria}
\begin{itemize}
    \item \textbf{EC1 (Purely Algorithmic or Linguistic Focus):}
    Studies that focus exclusively on natural language processing techniques,
    linguistic evaluations, prompt engineering heuristics,
    semantic datasets, or parameter-level fine-tuning algorithms
    without systems implementation,
    hardware profiling, or cloud infrastructure evaluation are excluded.

    \item \textbf{EC2 (Gray Literature and Short Contributions):}
    Non-peer-reviewed preprints (e.g., arXiv, SSRN, etc.),
    commercial marketing white papers,
    patents, keynote abstracts,
    doctoral symposium summaries,
    and extended abstracts or short posters (under 4 pages)
    lacking comprehensive empirical validation are excluded.

    \item \textbf{EC3 (Edge or Standalone Device Exclusivity):}
    Studies focusing solely on localized execution
    on microcontrollers, embedded sensors,
    or standalone consumer mobile devices
    without any multi-node, multi-accelerator,
    or cloud-scale orchestration context are excluded.
\end{itemize}

\subsection{Study Selection Process and PRISMA Flow}
\label{subsec:method-selection}
\begin{figure}
    \centering
    \includegraphics[scale=0.78]{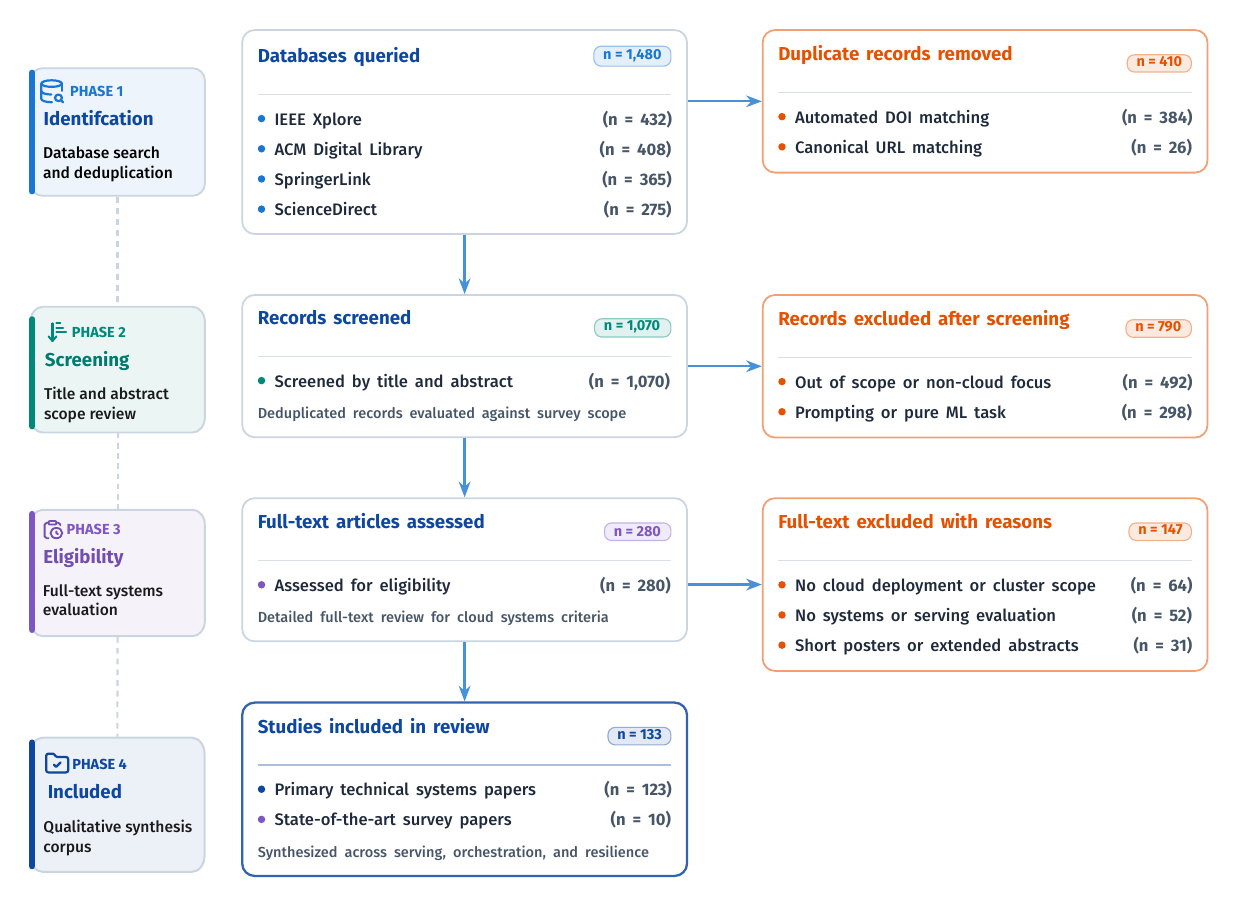}
    \caption{
        PRISMA 2020 flow diagram illustrating the study selection process,
        including identification across four digital libraries,
        automated deduplication, title and abstract screening,
        full-text eligibility assessment with explicit exclusion reasons,
        and final corpus inclusion.
    }
    \label{fig:prisma-flow}
\end{figure}

The study selection workflow was executed
in four distinct phases conforming to the \ac{PRISMA} 2020 statement:
Identification, Screening, Eligibility, and Inclusion.
The quantitative progression of literature
through each phase is illustrated in \zcref{fig:prisma-flow}.

In the Identification phase,
the automated database search queries
retrieved a total of 1,480 records
across the four target databases:
432 from IEEE Xplore, 408 from ACM Digital Library,
365 from SpringerLink, and 275 from ScienceDirect.
All retrieved bibliographic records,
including titles, author metadata,
abstracts, and unique identifiers,
were compiled into a centralized
reference management database.

Prior to screening,
automated deduplication was performed.
Records were primarily matched
using standardized \acp{DOI}.
For articles that lacked registered \acp{DOI}
(such as specific ACM conference proceedings and early access records),
canonical \ac{URL} resolution alongside exact string matching
on normalized titles and author lists was utilized.
This automated procedure successfully identified
and removed 410 duplicate records (384 via \ac{DOI} matching
and 26 via canonical persistent \ac{URL} matching),
yielding 1,070 unique articles for subsequent analysis.

During the Screening phase,
the 1,070 unique publications were evaluated
based on their titles and abstracts
against the predefined eligibility criteria.
Articles clearly outside the scope of cloud computing
or focusing strictly on linguistic modeling were eliminated.
This screening step resulted in the exclusion of 790 records:
492 studies out of scope or lacking cloud focus,
and 298 purely algorithmic, prompt engineering,
or machine learning studies lacking systems evaluation.

In the Eligibility phase,
the remaining 280 candidate publications
were retrieved in full text
and subjected to rigorous,
in-depth assessment against the complete
inclusion and exclusion criteria.
Each paper was examined to confirm the
presence of operational testbeds,
cloud scheduling mechanisms,
or concrete serving implementations.
Following full-text evaluation,
147 articles were excluded with documented reasons:
(i) 64 publications lacked cloud, datacenter,
or distributed deployment context;
(ii) 52 publications focused purely on model weights,
prompt optimization, or hyperparameter tuning
without systems or serving performance metrics;
and (iii) 31 publications were extended abstracts,
posters, or workshop summaries lacking
sufficient empirical depth.

Consequently, 133 publications met
all eligibility criteria
and were included in the final
qualitative synthesis.
This included review corpus consists
of 123 primary technical systems papers
and 10 foundational survey and taxonomy papers
(distributed and cited across \zcref{sec:related-works,
    sec:orchestration,sec:inference-in-cloud,sec:architecture}).
We explicitly note that foundational background citations,
hardware vendor datasheets (e.g., NVIDIA Hopper and AMD Instinct specifications),
and industry roadmaps cited in the Introduction,
Discussion, and Conclusion provide broad contextual orientation
and remain distinct from the systematically
indexed review corpus of 133 studies.

\subsection{Data Extraction and Synthesis Protocol}
\label{subsec:method-extraction}
To extract actionable systems insights
and enable structured qualitative synthesis across the 133 included studies,
we designed a standardized data extraction matrix.
For each included publication,
the following key architectural and operational facets were systematically extracted:

\begin{enumerate}[I.]
    \item \textbf{Cloud Abstraction Layer:}
    The primary execution environment and deployment paradigm,
    categorized into bare-metal accelerator clusters,
    virtual machine instances,
    containerized microservices managed by \ac{K8s},
    \ac{FaaS} runtimes, or multi-agent orchestration frameworks.

    \item \textbf{Hardware Virtualization and Partitioning:}
    The underlying mechanism used to share and isolate accelerators,
    such as NVIDIA Multi-Instance \ac{GPU} (\ac{MIG}), \ac{MPS},
    kernel-level interception, spatial multiplexing, or dynamic time-slicing.

    \item \textbf{Serving Optimization and Scheduling Technique:}
    The primary inference acceleration mechanisms,
    including continuous batching, chunked prefill,
    prefill-decode disaggregation, paged \ac{KV} cache management,
    prefix caching, or speculative decoding.

    \item \textbf{Reliability, Fault Tolerance, and Security:}
    The resilience mechanisms implemented,
    including checkpoint-restart overhead mitigation,
    silent data corruption detection,
    in-flight request failover,
    and \ac{TEE} enclave isolation for confidential inference.

    \item \textbf{Evaluation Environment and Workload Characteristics:}
    The experimental testbed (public cloud hyperscalers,
    on-premises multi-node clusters, or hardware emulators),
    the model families evaluated (e.g., LLaMA, OPT, Falcon, GPT),
    accelerator hardware types (e.g., NVIDIA A100, H100, AMD MI250/MI300),
    and target performance metrics (\ac{TTFT},
    \ac{TBT}, throughput, \ac{SLA} violations, and cost).
\end{enumerate}

The extracted data were synthesized qualitatively
to map the overarching system design space,
cross-compare architectural trade-offs,
and establish the taxonomies detailed throughout
\zcref{sec:orchestration,sec:inference-in-cloud,sec:architecture}.

\subsection{Threats to Validity}
\label{subsec:method-threats}
To ensure sound interpretation of our findings,
we identify and discuss three primary threats to validity,
along with the mitigation strategies applied throughout our methodology.

\subsubsection{Internal Validity}
Internal validity concerns
potential systematic errors
or subjective biases in study
selection and data extraction.
To mitigate selection bias,
we formulated precise, objective
inclusion and exclusion criteria
and enforced a standardized search
syntax across all four databases.
Deduplication errors were mitigated
through automated \ac{DOI} matching
supplemented by canonical \ac{URL} resolution.
Screening and full-text eligibility assessments were conducted
under a multi-pass protocol where ambiguous edge cases
(e.g., papers combining lightweight algorithmic adjustments
with cloud deployment benchmarks)
were cross-evaluated against IC3 and
IC4 to ensure consistent inclusion decisions.
Furthermore, data extraction followed a pre-structured matrix
to eliminate subjective interpretation of systems mechanisms.

\subsubsection{External Validity}
External validity refers to the generalizability
of the synthesized architectural patterns and trade-offs.
The cloud \ac{LLM} landscape evolves rapidly,
with frequent releases of new accelerator architectures
(e.g., NVIDIA Blackwell, AMD MI300X, Google \ac{TPU} v5)
and changing hyperscaler cloud configurations.
Our review mitigates this temporal threat
by focusing on fundamental systems-level abstractions
(such as disaggregated prefill-decode serving,
memory-hierarchy caching,
and elastic scheduling) rather than transient,
hardware-specific micro-optimizations.
These foundational architectural
principles remain widely applicable
across successive hardware generations
and heterogeneous cloud infrastructures.

\subsubsection{Construct Validity}
Construct validity addresses whether the analyzed metrics
and evaluation criteria accurately represent real-world
cloud serving performance and operational objectives.
In cloud \ac{LLM} deployments,
relying solely on aggregate throughput
can obscure severe tail-latency violations.
To mitigate this construct mismatch,
our review explicitly prioritizes studies
that characterize granular,
user-centric \ac{QoS} metrics,
specifically \ac{TTFT}, \ac{TBT},
and tail latency percentiles
under strict \ac{SLO} constraints.
By prioritizing systems evaluated under
realistic trace-driven multi-tenant workloads
and \ac{SLA}-bound operational conditions,
our synthesis preserves high fidelity
to actual cloud production environments.

    \section{Orchestration and Resource Scheduling}
    \label{sec:orchestration}
    The deployment of \acp{LLM} in public and private cloud data centers
introduces unique structural challenges \citep{private-and-public-cloud}
that diverge fundamentally
from traditional \ac{DL} inference and standard cloud microservices.
Serving modern \acp{LLM} is characterized by multi-billion parameter
model sizes, stateful \ac{KV} cache memory footprints,
and a distinct two-phase execution profile comprising a
compute-bound prefill phase (see \citet{p2})
and a memory-bandwidth-bound decode phase.
Consequently, conventional cloud resource management
frameworks that treat \acp{GPU} as monolithic \citep{gpu-resources-monolith},
stateless compute instances yield severe resource underutilization,
prohibitive tail latencies, and high operational costs.
To overcome these bottlenecks, cloud platforms rely on specialized
orchestration and resource scheduling mechanisms designed
to maximize token throughput \citep{p4},
enforce stringent \acp{SLA} \citep{review-on-server-level-agreement},
and minimize hardware idleness.

\begin{figure}
    \centering
    \includegraphics[scale=0.78]{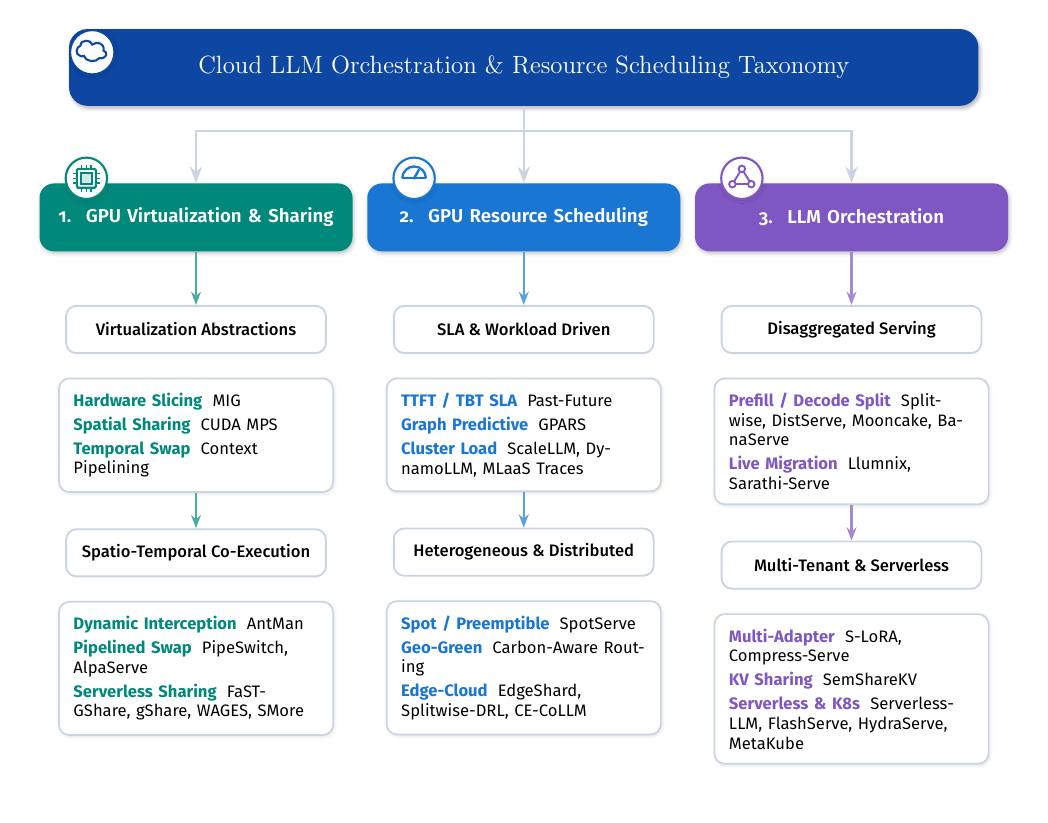}
    \caption{
        Architectural taxonomy of cloud \ac{LLM} orchestration
        and resource scheduling, illustrating the structural
        hierarchy spanning \ac{GPU} virtualization abstractions,
        cluster-level resource scheduling strategies,
        and high-level \ac{LLM} serving orchestration paradigms.
    }
    \label{fig:orchestration_taxonomy}
\end{figure}

\zcref{fig:orchestration_taxonomy} presents an architectural taxonomy
of cloud \ac{LLM} orchestration and resource scheduling,
categorizing the domain into three
interdependent architectural pillars: (1) \ac{GPU} Virtualization and Sharing,
(2) \ac{GPU} Resource Scheduling, and (3) \ac{LLM} Orchestration.
This section provides an in-depth survey of the state-of-the-art
across these three core dimensions.

\subsection{GPU Virtualization and Sharing}
\label{subsec:gpu_virtualization}
Efficient resource utilization in multi-tenant cloud
clusters requires fine-grained sharing of physical \ac{GPU} hardware.
Because transformer models exhibit dynamic resource
requirements across execution phases \citep{gpu-required-for-each-phase},
allocating dedicated physical \acp{GPU} to individual
requests leads to severe memory and compute fragmentation.
Cloud providers leverage hardware-level and
software-level \ac{GPU} virtualization abstractions
\citep{hardware-and-software-gpu-virtualization}
to enable spatial and temporal multiplexing.

\subsubsection{Hardware-Level Virtualization vs. Software Spatial Multiplexing}
Hardware-level partitioning, exemplified by NVIDIA \ac{MIG},
physically divides a single \ac{GPU} into isolated hardware
instances with dedicated \ac{HBM} slices,
crossbar paths, and compute engines.
\Citet{p118} evaluated spatial co-execution technologies,
highlighting that while \ac{MIG} guarantees strict hardware
isolation and predictable \ac{QoS},
its rigid, static slice reconfiguration
overheads hinder dynamic cloud workloads.
To address this limitation,
recent learning-augmented approaches apply \ac{DRL}
to dynamically reconfigure \ac{MIG} slice configurations
based on incoming job arrival patterns \citep{p116}.
Complementarily, ElasticBatch \citep{p15} introduces elastic
batch scheduling on \ac{MIG} instances,
adaptively tuning batch sizes to fit dynamic slice capacities.

In contrast to rigid hardware slicing,
software-level spatial multiplexing via NVIDIA \ac{MPS} allows
multiple \ac{CUDA} contexts to concurrently
submit kernel grids to a single \ac{GPU} command queue.
While \ac{MPS} achieves higher compute
density and eliminates \ac{MIG} reconfiguration overheads,
concurrent kernels contend for shared L2 caches,
memory controllers, and \acp{SM}.
This contention frequently causes
unpredictable tail latencies,
making naive \ac{MPS} sharing unsuitable
for latency-sensitive \ac{LLM} workloads
unless paired with intelligent kernel regulation.

\subsubsection{Spatio-Temporal Co-Execution and Memory Isolation}
To mitigate hardware contention in software spatial sharing,
advanced systems implement spatio-temporal co-execution frameworks.
AntMan \citep{p113} pioneered dynamic scaling on \ac{GPU} clusters
by introducing a memory manager and kernel scheduler that intercepts
\ac{CUDA} \ac{API} calls at runtime.
AntMan dynamically throttles memory allocations
and compute kernel launching for lower-priority background jobs,
guaranteeing hardware resource availability
for high-priority latency-critical tasks.
Additionally, SAGESERVE \citep{p6} introduced a
forecast-based approach to autoscale the huge demand
for cloud-based \ac{LLM} data centers.

For workloads requiring rapid switching
between distinct model architectures,
temporal context switching presents a critical
bottleneck due to the latency of copying
multi-gigabyte weight tensors across \ac{PCIe} buses.
PipeSwitch \citep{p115} resolves this challenge
by establishing pipelined context switching,
overlapping memory transfers over \ac{PCIe}
with \ac{GPU} compute execution,
thereby reducing model switching latencies
from seconds to milliseconds.
Building on tensor parallelism, AlpaServe \citep{p114}
explores statistical multiplexing in model-parallel serving,
demonstrating that dynamically placing model-parallel
slices across shared \ac{GPU} groups maximizes
aggregate cluster throughput
while satisfying latency budgets.

\subsubsection{Serverless GPU Sharing Frameworks}
\label{subsubsec:serverless-gpu-sharing-frameworks}

The rise of serverless \ac{FaaS} for \ac{DL} and \ac{LLM}
workloads has spurred dedicated
spatio-temporal \ac{GPU} sharing architectures.
In serverless environments, request burstiness and
short execution durations amplify the
overhead of dedicated \ac{GPU} allocation.
FaST-GShare \citep{p57} introduces spatial
and temporal \ac{GPU} sharing tailored for
serverless \ac{DL} inference, utilizing spatial
partition management to co-locate lightweight inference tasks.
Similarly, gShare \citep{p12} implements aggressive
scheduling on multi-tenant \ac{FaaS} platforms,
dynamically balancing \ac{GPU} memory allocations to
prevent \ac{OOM} failures under volatile loads.

To optimize energy efficiency during
multi-tenant \ac{GPU} sharing,
WAGES \citep{p11} introduces a
workload-aware \ac{GPU} sharing system
for serverless \ac{LLM} serving.
WAGES dynamically tracks \ac{GPU} frequency scaling
and memory footprint expansion,
dynamically grouping compatible \ac{LLM} inference tasks
onto shared \acp{GPU} to minimize active power consumption.
Furthermore, SMore \citep{p16} enhances \ac{GPU} utilization in production
deep learning clusters by co-locating serverless inference workloads
alongside long-running training jobs,
utilizing opportunistic
resource harvesting without violating
primary job completion deadlines.

\subsection{GPU Resource Scheduling}
\label{subsec:gpu_scheduling}
While \ac{GPU} virtualization establishes hardware-level
execution boundaries, \ac{GPU} resource scheduling
governs the global allocation, dispatching,
and load balancing of incoming \ac{LLM} inference
workloads across heterogeneous cloud clusters.

\subsubsection{Workload Characterization and Request Volatility in MLaaS}
\label{subsubsec:workload-characterization-and-request-volatility-in-mlaas}
Designing effective cluster schedulers requires
understanding real-world \ac{LLM} request dynamics.
Analysis of large-scale \ac{MLaaS} platforms,
such as Alibaba cloud traces \citep{p119},
reveals extreme request volatility,
non-stationary arrival rates,
and heavy-tailed prompt and output sequence lengths.
Traditional \ac{DL} job schedulers assume predictable
execution durations based on fixed
batch sizes and input shapes \citep{p120}.
However, autoregressive \ac{LLM} inference
violates these assumptions because token generation
length is inherently non-deterministic,
creating severe load imbalances across cluster nodes.

\subsubsection{SLA-Aware and Multi-Metric Cluster Schedulers}
In \ac{LLM} serving, user-perceived performance depends
on two distinct latency metrics: \ac{TTFT},
which measures the delay before prompt processing completes,
and \ac{TBT} or \ac{ITL},
which governs the generation rate
of subsequent autoregressive tokens.
Formally, for a request with prompt length $L_{p}$ and output length
$L_{o}$, total end-to-end latency $T_{\text{end}}(r)$ is expressed in
\zcref{eq:llm_latency}:

\begin{equation}
    T_{\text{end}}(r) = \text{\ac{TTFT}}(L_{p}) + \sum_{i=1}^{L_{o}-1} \text{\ac{TBT}}_{i}
    \label{eq:llm_latency}
\end{equation}

To enforce strict \ac{SLA} guarantees
across both \ac{TTFT} and \ac{TBT},
the Past-Future Scheduler \citep{p5} tracks
historical execution profiling alongside
predictive future context lengths,
dynamically adjusting request priority queues.
GPARS \citep{p117} employs a graph predictive
algorithm for resource scheduling in heterogeneous
\ac{GPU} clusters, capturing complex node topology
and interconnect bandwidths to
optimize job-to-\ac{GPU} mappings.

To balance user fairness alongside throughput,
\citet{p8} introduced fairness-aware
metrics for \ac{LLM} serving,
preventing long-prompt requests from starving
short-sequence interactive tasks.
ScaleLLM \citep{p81} optimizes end-to-end resource
frugality by dynamically tuning request batching and
execution allocations based on real-time memory pressure.
DynamoLLM \citep{p9} focuses on energy efficiency,
co-designing job scheduling with \ac{DVFS} to reduce
power consumption during off-peak demand periods.
Additionally, performance-aware load balancers \citep{p80}
and multi-objective evolutionary reinforcement
learning schedulers \citep{p85,p86} continuously
balance node-level cluster loads,
minimizing idle \ac{GPU} cycles in
financial and enterprise cloud environments.

\subsubsection{Heterogeneous, Preemptible Spot Instance, and Geo-Distributed Scheduling}
\label{subsubsec:heterogeneous-preemptible-spot-instance-and-geo-distributed-scheduling}
Cloud infrastructure often consists of heterogeneous
hardware generations (e.g., NVIDIA V100, A100, and H100 \acp{GPU};
see \citet{comparison-between-gpus} for more).
SLoB \citep{p17} addresses suboptimal load balancing
in local heterogeneous \ac{GPU} clusters by profiling compute
capabilities and assigning high-density tensor
operations to higher-tier accelerators.
In addition to that, \citet{gpu-perf-forecast}
developed a new framework called ``NeuSight''
which demonstrated the capability to forecast
typical \ac{DL} model performance on unseen \acp{GPU}.

To reduce cloud infrastructure expenditure,
providers leverage preemptible or spot \ac{VM} instances,
which offer substantial cost savings
but risk sudden instance revocation.
SpotServe \citep{p47} provides a resilient \ac{LLM} serving
framework on spot instances, utilizing parallelized
state migration and light-weight \ac{KV} cache checkpointing
to recover serving states within seconds upon instance preemption.

At a global cloud scale, green scheduling initiatives
optimize energy usage and carbon footprints.
Workload schedulers across geo-distributed
data centers \citep{p18,p84} route incoming
\ac{LLM} requests to regions powered by renewable energy,
leveraging cross-data-center model state
and prompt cache reuse to offset data transfer latencies.

\subsubsection{Cloud-Edge Collaborative Resource Allocation}
Deploying \acp{LLM} purely in centralized cloud data centers
incurs significant network transfer latencies and privacy concerns.
Consequently, cloud-edge collaborative resource
allocation has emerged as a promising paradigm.
Active inference resource allocation \citep{p25}
and EdgeShard \citep{p83} partition \ac{LLM}
execution between resource-constrained
edge devices and centralized cloud nodes.

Splitwise \ac{DRL} \citep{p98} utilizes
Lyapunov-assisted \ac{DRL} to dynamically
determine offloading boundaries between edge
devices and cloud servers based on
network bandwidth and energy constraints.
Similarly, CE-CoLLM \citep{p96} and
hybrid \ac{SLM} to \ac{LLM} frameworks \citep{p95}
employ edge-located \acp{SLM} for rapid initial
token generation while dynamically querying
cloud-hosted \acp{LLM} for complex reasoning tasks.
Cloud-edge collaborative serving frameworks
also incorporate strict data security protocols
\citep{p94} and parameter sharing mechanisms
\citep{p82} to prevent proprietary
prompt leakage during cloud offloading.

\subsection{LLM Orchestration}
\label{subsec:llm_orchestration}
While resource scheduling manages cluster-level
hardware allocation, \ac{LLM} orchestration coordinates
the high-level serving pipeline,
dynamic memory state transfers,
adapter switching,
and cloud-native auto-scaling.

\begin{figure}
    \centering
    \includegraphics[scale=0.78]{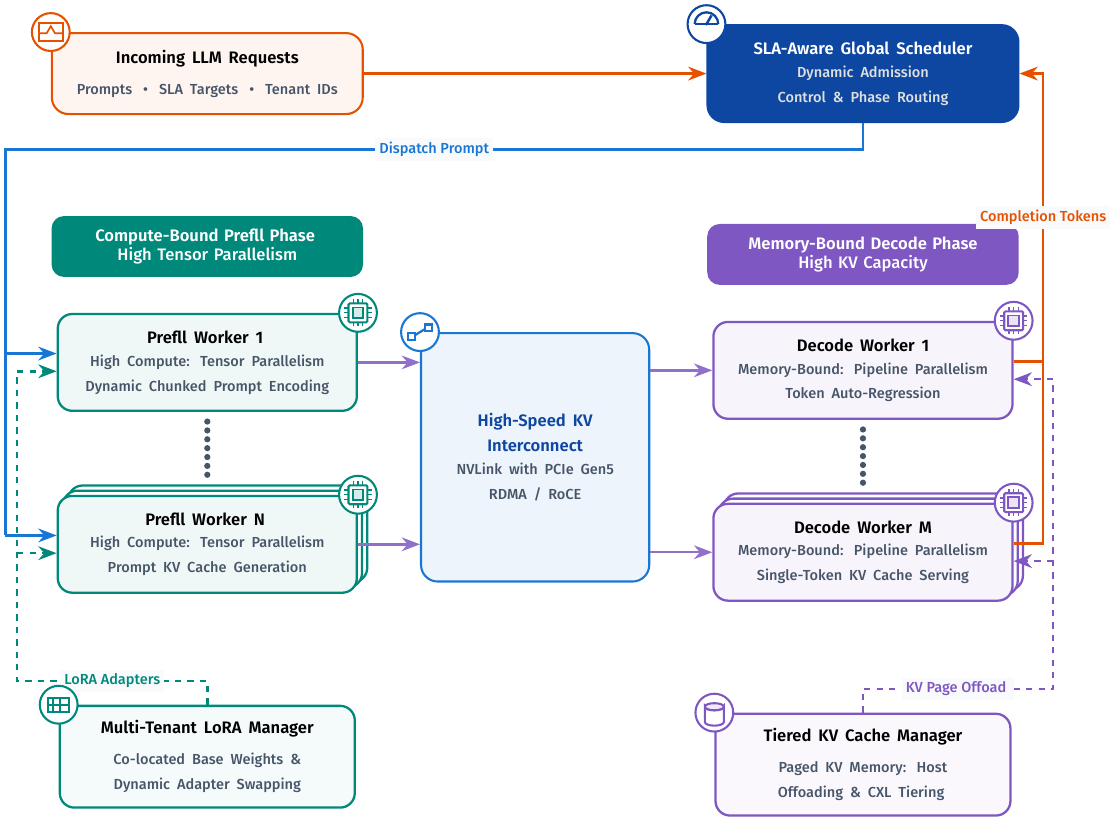}
    \caption{
        Architectural workflow of disaggregated \ac{LLM}
        serving and multi-tenant orchestration,
        illustrating the separation of compute-bound
        prefill nodes and memory-bound decode nodes,
        linked by high-speed \ac{KV} cache interconnects
        and governed by \ac{SLA} schedulers,
        multi-tenant \ac{LoRA} managers,
        and tiered memory managers.
    }
    \label{fig:disaggregated_orchestration}
\end{figure}

\subsubsection{Disaggregated Prefill-Decode Serving Architectures}
\label{subsubsec:disaggregated-prefill-decode-serving-architectures}
A major architectural breakthrough in \ac{LLM}
orchestration is the disaggregation
of the prefill and decoding phases.
In traditional monolithic serving,
prefill operations
(which compute initial prompt \ac{KV} caches in parallel)
and decode operations (which generate tokens sequentially)
run on the same physical \acp{GPU}.
This co-location causes severe mutual
interference: compute-heavy prefill operations saturate \ac{SM} execution units,
causing latency spikes in memory-bound decode steps.

As depicted in \zcref{fig:disaggregated_orchestration},
disaggregated serving disjoins these execution
phases across specialized \ac{GPU} pools:

\begin{enumerate}
    \item \textbf{Prefill Nodes:} Compute-optimized \acp{GPU}
    configured with high \ac{TP} process incoming prompt
    sequences at maximum compute throughput.
    \item \textbf{High-Speed KV Interconnect:} Once prompt encoding is complete,
    intermediate \ac{KV} cache tensors are transmitted across high-speed
    interconnects such as \ac{NVLink} (see \citet{nv-link}),
    \ac{PCIe} Gen5, or \ac{RoCE}.
    \item \textbf{Decode Nodes:} Memory-optimized \acp{GPU} configured
    with \ac{PP} receive the transferred \ac{KV} cache
    and perform single-token autoregressive generation,
    maximizing \ac{HBM} bandwidth utilization.
\end{enumerate}

Splitwise \citep{p37} and DistServe \citep{p38} demonstrated
that disaggregated prefill-decode
serving eliminates head-of-line blocking,
yielding up to 10$\times$ improvements in
goodput while adhering to strict \ac{SLA} thresholds.
Mooncake \citep{p41} extends this architectural
principle by establishing a \ac{KV} cache-centric
disaggregated architecture,
trading host memory and \ac{SSD} storage
capacity to cache reusable prompt contexts.
BanaServe \citep{p26} introduces unified
\ac{KV} cache management paired with dynamic module migration,
dynamically rebalancing prefill and
decode \ac{GPU} node pools in response to live traffic shifts.
Additionally, Llumnix \citep{p10} enables dynamic request
migration and runtime re-scheduling across serving
nodes to eliminate hotspot bottlenecks,
while Sarathi-Serve \citep{p2} integrates chunked-prefill
piggybacking to homogenize execution batch
sizes across prefill and decode iterations.

\subsubsection{Multi-Tenant LoRA Adapter and KV-Cache Orchestration}
\label{subsubsec:multi-tenant-lora-adapter-and-kv-cache-orchestration}
In enterprise cloud deployments,
thousands of domain-specific models are deployed
as lightweight \ac{LoRA} fine-tuned adapters layered
on top of a shared foundation base model.
Orchestrating multi-tenant \ac{LoRA} serving
requires dynamic loading and co-execution
of heterogeneous adapters on shared \ac{GPU} clusters.

Compress-then-Serve \citep{p22} optimizes multi-adapter
serving by compressing thousands of \ac{LoRA} adapters
into quantized memory structures,
permitting simultaneous co-location on \ac{CPU} host
memory with near-zero loading overhead.
Predictive-\ac{LoRA} \citep{p61} introduces
a proactive, fragmentation-aware serverless
inference system that anticipates adapter
access patterns and pre-fetches required
adapter weights into \ac{GPU} \ac{HBM} ahead of execution.

Concurrently, multi-tenant orchestration leverages
semantic prompt sharing to
optimize \ac{KV} cache memory footprints.
SemShareKV \citep{p23} utilizes
token-level \ac{LSH} matching to identify
semantically similar prompt
prefixes across distinct users,
sharing cached \ac{KV} representations
to conserve \ac{GPU} memory.
Cache-Craft \citep{p69} manages chunk-caches
tailored for \ac{RAG} pipelines.
However, multi-tenant \ac{KV} cache
sharing introduces critical privacy
and security vulnerabilities.
Recent investigations reveal that shared \ac{KV}
caches are susceptible to prompt leakage
attacks \citep{p76} and hardware cache
side-channel exploits \citep{p123},
necessitating secure isolation
protocols in cloud orchestration layers.

\subsubsection{Serverless and Cloud-Native Kubernetes Orchestration}
Serverless \ac{LLM} serving represents
the frontier of cloud-native orchestration,
enabling auto-scaling from zero to
thousands of instances while
maintaining cost efficiency.
The primary bottleneck in serverless
\ac{LLM} deployment is cold-start latency,
driven by loading tens of gigabytes of model
weights from remote storage into \ac{GPU}
memory during scale-from-zero events.

ServerlessLLM \citep{p3} mitigates cold-start
latency by exploiting multi-tier
host memory hierarchies and network bandwidth,
achieving low-latency checkpoint loading.
PipeCo \citep{p56} pipelines model weight
loading with container initialization,
while BlitzScale \citep{p68} achieves fast
live autoscaling via $\mathcal{O}(1)$ host caching.
HydraServe \citep{p59} and FaaScale \citep{p67} further minimize
cold-start delays in public cloud infrastructure
through proactive instance warm-up and fast tensor streaming.

Beyond cold-start acceleration,
serverless orchestration engines focus on tiered
memory management and predictive auto-scaling.
FlashServe \citep{p55} implements cost-efficient
serverless inference scheduling via tiered memory management,
dynamically offloading idle \ac{KV} caches to \ac{CPU}
host \ac{RAM} and predictive autoscaling to adjust \ac{GPU} pool sizes.
DeepServe \citep{p58} and SLINFER \citep{p65} provide
end-to-end serverless serving architectures designed
for large-scale operation, addressing hidden performance
pitfalls inherent to serverless
\ac{AI} platforms \citep{p60,p64,p66}.

At the cloud platform management layer,
\ac{K8s} serves as the industry standard for container orchestration.
Recent advances integrate \acp{LLM} directly
into cloud-native control loops.
MetaKube \citep{p49} applies an experience-aware
\ac{LLM} framework for \ac{K8s} failure diagnosis,
automatically analyzing cluster event
logs to resolve container anomalies.
Prompt-driven \ac{K8s} orchestration \citep{p51} enables natural
language system management, while pattern-aware predictive
autoscaling \citep{p28} integrates
\acp{LLM} to forecast workload traffic spikes.
Furthermore, specialized tools such as KIS-S
\citep{p35} provide \ac{GPU}-aware \ac{K8s} inference
simulation environments for training
reinforcement learning autoscalers,
complemented by empirical performance
evaluations of containerized \ac{LLM}
deployments across public
cloud platforms \citep{p30, p52, p53}.

\subsection{Comparative Summary of Orchestration and Scheduling Frameworks}
\label{subsec:comparative_summary}
To provide a consolidated reference for researchers
and system architects,
\zcref{tab:orchestration_summary}
presents a detailed comparative analysis
of important state-of-the-art cloud \ac{LLM} orchestration
and resource scheduling frameworks
surveyed in this section.
The comparison evaluates each solution across
primary technical domain, key architectural innovation,
target optimization metrics, underlying
interconnect/transfer media,
and deployment scenario.

\begingroup
\footnotesize
\setlength{\tabcolsep}{3pt}
\renewcommand{\arraystretch}{1.15}

\begin{ThreePartTable}
    \begin{TableNotes}
        \item[a] ``V'' stands for \textit{Virtualization}.
        \item[b] ``S'' stands for \textit{Scheduling}.
        \item[c] ``LO'' stands for \ac{LLM} Orchestration.
        \item[d] ``MT'' stands for \textit{Multi-Tenant}.
        \item[e] ``KO'' stands for \ac{K8s} \textit{Orchestration}.
    \end{TableNotes}

    \begin{longtable}{@{}
            >{\raggedright\arraybackslash}p{0.15\linewidth}
            >{\raggedright\arraybackslash}p{0.12\linewidth}
            >{\raggedright\arraybackslash}p{0.25\linewidth}
            >{\raggedright\arraybackslash}p{0.14\linewidth}
            >{\raggedright\arraybackslash}p{0.14\linewidth}
            >{\raggedright\arraybackslash}p{0.16\linewidth}
        @{}}
        \caption{Comparative summary of representative Cloud LLM Orchestration and Resource Scheduling frameworks.}\label{tab:orchestration_summary}\\
        \toprule
        \textbf{Framework}         & \textbf{Primary Domain} & \textbf{Key Architectural Innovation} & \textbf{Primary Optimization Metric} & \textbf{Interconnect / Transfer} & \textbf{Target Deployment} \\
        \midrule
        \endfirsthead

        \multicolumn{6}{c}{\small\itshape Table \thetable{} (continued from previous page)} \\
        \addlinespace
        \toprule
        \textbf{Framework}         & \textbf{Primary Domain} & \textbf{Key Architectural Innovation} & \textbf{Primary Optimization Metric} & \textbf{Interconnect / Transfer} & \textbf{Target Deployment} \\
        \midrule
        \endhead

        \midrule
        \multicolumn{6}{r}{\itshape Continued on next page} \\
        \endfoot

        \bottomrule
        \insertTableNotes
        \endlastfoot

        AntMan \citep{p113}        & V\tnote{a}          & Dynamic memory manager \& \ac{CUDA} \ac{API} kernel interception              & \ac{GPU} Utilization \& \ac{QoS}            & \ac{PCIe} / Host \ac{RAM}           & Co-located Cluster          \\
        PipeSwitch \citep{p115}    & V\tnote{a}          & Pipelined context switching overlapping weight copy \& compute                & Model Switching Latency                     & \ac{PCIe} Gen3/4                    & Multi-Model Serving         \\
        AlpaServe \citep{p114}     & V\tnote{a}          & Statistical multiplexing with dynamic model parallelism                       & Aggregate Throughput                        & \ac{NVLink} / \ac{PCIe}             & Model-Parallel Cloud        \\
        FaST-GShare \citep{p57}    & V\tnote{a}          & Spatio-temporal \ac{GPU} sharing for serverless inference                     & Function Startup Latency                    & Shared \ac{GPU} Memory              & Serverless \ac{FaaS}        \\
        gShare \citep{p12}         & V\tnote{a}          & Aggressive memory allocation for multi-tenant \ac{FaaS}                       & \ac{OOM} Prevention \& Latency              & Host-to-\ac{GPU} Memory             & Multi-tenant Cloud          \\
        WAGES \citep{p11}          & V\tnote{a}          & Workload-aware \ac{GPU} sharing and \ac{DVFS} tuning                          & Active Energy Consumption                   & Shared \ac{GPU} Compute             & Serverless \ac{LLM}         \\
        \midrule
        Past-Future \citep{p5}     & S\tnote{b}          & Dual-horizon profile tracking for \ac{SLA} queue management                   & \ac{TTFT} \& \ac{TBT} \ac{SLA} Goodput      & Intra-node Bus                      & Multi-tenant \ac{LLM}       \\
        GPARS \citep{p117}         & S\tnote{b}          & Graph predictive algorithm for heterogeneous topology                         & Tail Latency \& Throughput                  & Heterogeneous Network               & Public \ac{GPU} Cloud       \\
        SpotServe \citep{p47}      & S\tnote{b}          & Parallelized state migration \& \ac{KV} checkpointing                         & Cost Efficiency \& Availability             & Cloud \ac{VM} Network               & Preemptible Spot \acp{VM}   \\
        Green-Geo \citep{p18}      & S\tnote{b}          & Carbon-aware routing with model state reuse                                   & Carbon Footprint \& Cost                    & Geo-WAN / Data Center               & Geo-Distributed Cloud       \\
        EdgeShard \citep{p83}      & S\tnote{b}          & Collaborative edge-cloud model sharding \& offloading                         & Edge Latency \& Bandwidth                   & Wireless / Cloud WAN                & Cloud-Edge Continuum        \\
        Splitwise-\ac{DRL} \citep{p98}  & S\tnote{b}     & Lyapunov-assisted \ac{DRL} for edge-cloud split decision                      & Energy Consumption \& Delay                 & 5G / Edge-Cloud WAN                 & Mobile Edge Computing       \\
        \midrule
        Splitwise \citep{p37}      & LO\tnote{c}         & Phase splitting into dedicated prefill \& decode nodes                        & Goodput per Dollar                          & High-Speed \ac{RDMA}                & Enterprise \ac{LLM} Serving \\
        DistServe \citep{p38}      & LO\tnote{c}         & Disaggregated prefill/decode with \ac{SLA}-driven allocation                  & Goodput \& \ac{TTFT}/\ac{TBT} Violations    & \ac{NVLink} / \ac{RoCE}             & High-Concurrency Cloud      \\
        Mooncake \citep{p41}       & LO\tnote{c}         & \ac{KV} cache-centric architecture trading storage for compute                & Cache Hit Rate \& Throughput                & \ac{RDMA} / \ac{PCIe} \ac{SSD}      & \ac{LLM} Chatbot Service    \\
        Llumnix \citep{p10}        & LO\tnote{c}         & Dynamic runtime request migration \& live load balancing                      & Tail Latency \& Hotspots                    & Intra/Inter-Node                    & Dynamic \ac{LLM} Cloud      \\
        ServerlessLLM \citep{p3}   & LO\tnote{c}         & Fast multi-tier checkpoint loading \& tiering                                 & Cold Start Latency                          & High-Bandwidth \ac{RAM}             & Serverless \ac{FaaS}        \\
        FlashServe \citep{p55}     & LO\tnote{c}         & Tiered memory management \& predictive autoscaling                            & Serverless Cost \& Latency                  & Host \ac{RAM} / \ac{GPU} \ac{HBM}   & Serverless \ac{LLM} Cloud   \\
        Compress-Serve \citep{p22} & MT\tnote{d}         & Quantized \ac{LoRA} adapter co-location in host memory                        & Adapter Memory Overhead                     & Host-to-\ac{GPU} \ac{PCIe}          & Multi-Adapter Cloud         \\
        SemShareKV \citep{p23}     & MT\tnote{d}         & Token-level \ac{LSH} matching for prefix \ac{KV} cache sharing                & \ac{GPU} Memory Footprint                   & Shared \ac{GPU} \ac{HBM}            & Multi-User \ac{LLM}         \\
        MetaKube \citep{p49}       & KO\tnote{e}         & Experience-aware \ac{LLM} control loop for failure diagnosis                  & Diagnosis Time \& \ac{MTTR}                 & \ac{K8s} Control Plane              & Cloud-Native \ac{K8s}       \\
    \end{longtable}
\end{ThreePartTable}
\endgroup

    \section{Inference in Cloud}
    \label{sec:inference-in-cloud}
\begin{figure}
    \centering
    \includegraphics[scale=0.66]{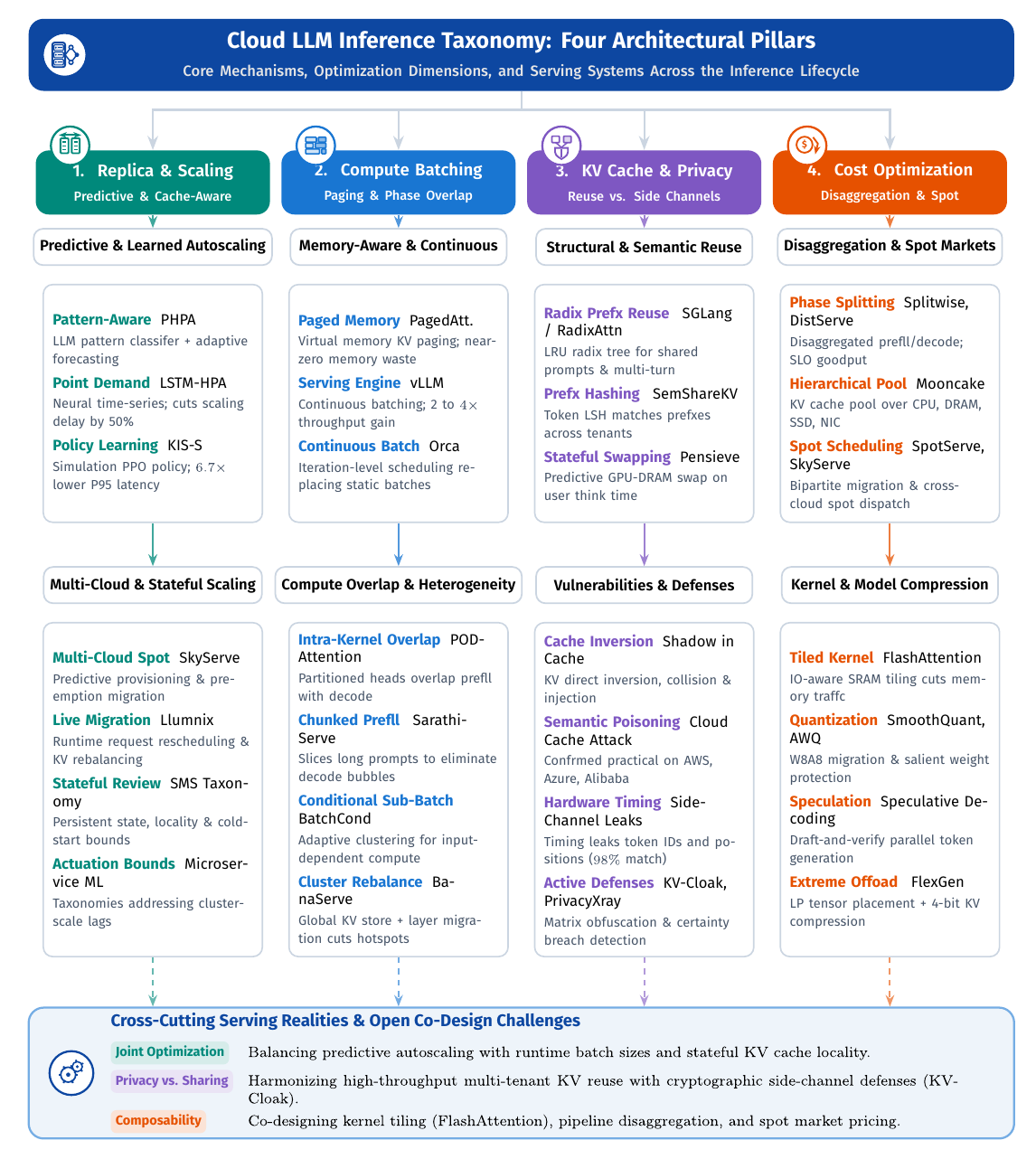}
    \caption{
        Taxonomy of cloud \ac{LLM} inference optimization across four architectural pillars:
        (1) replica and stateful scaling,
        (2) memory and compute batching,
        (3) \ac{KV} cache reuse and side-channel privacy, and
        (4) multi-tier cost optimization.
        The taxonomy highlights representative serving systems,
        underlying execution mechanisms, and cross-cutting co-design challenges.
    }
    \label{fig:inference-taxonomy}
\end{figure}

    Once a model has been placed on a \ac{GPU}
pool and its serving pipeline has
been orchestrated (see \zcref{sec:orchestration}),
the second layer of concern is what happens to an
individual request as it moves through that pipeline.
Four interlocking problems recur
across the literature on cloud-hosted \ac{LLM} inference.
First, \emph{scaling} determines how many
serving replicas should be running at any given moment,
and whether that decision is made reactively,
after the load has already changed,
or proactively, in anticipation of it \citep{sky-serve}.
Second, \emph{batching} determines
how concurrent requests are grouped for
execution so that \ac{GPU} compute and
memory bandwidth are not
left idle between arrivals \citep{pod-attention}.
Third, \emph{KV cache reuse} determines
whether the intermediate attention state computed
for one request or one turn of a conversation
can be safely and efficiently
shared with another \citep{stateful-llm-serving},
which raises both a performance
opportunity and a privacy liability.
Fourth, \emph{cost optimization} cuts
across all three, since scaling, batching,
and caching decisions are, in practice,
never made purely for throughput but
always under an implicit or
explicit dollar budget \citep{p47}.
\zcref{fig:inference-taxonomy}
summarizes these four pillars and the
representative systems discussed throughout this section.

\subsection{Scaling}
\label{subsec:scaling}
\subsubsection{Autoscaling Taxonomies and Surveys}
\label{subsubsec:scaling-surveys}

Autoscaling \citep{auto-scaling-method}
decides how many model replicas,
or how much \ac{GPU} capacity,
a serving system should hold at a given moment,
and the reviewed literature
splits sharply between reactive
and predictive approaches.
Two broad surveys frame the design space.
\citet{p29} classify
autoscaling algorithms and mechanisms
into three families, spanning
threshold-based reactive policies
to \ac{AI}-based proactive ones,
and additionally review the security posture of autoscaling itself,
including attacks that deliberately
trigger scale-out to exhaust a tenant's budget;
however, as already noted in
\zcref{subsec:cloud-orchestration-and-operational-automation},
this taxonomy is written for
cloud-native applications in
general and does not account for the
stateful \ac{KV}-cache footprint
that makes \ac{LLM} autoscaling behave differently
from scaling a stateless microservice.
\citet{p30} narrow the focus specifically
to machine-learning-based autoscaling,
surveying 60 primary studies
from 2015--2025 and organizing them along a
five-dimensional taxonomy of goal,
decision logic, scaling mode,
control scope, and deployment;
the survey's own synthesis identifies actuation
delay and telemetry lag as the
dominant unresolved bottlenecks,
since a scaling decision computed
from stale metrics arrives after the workload
shape it was meant to
respond to has already changed.

\subsubsection{Predictive and Proactive Scaling Systems}
\label{subsubsec:predictive-and-proactive-scaling-systems}

Within this design space, Kubernetes'
default \ac{HPA} is the reactive baseline
against which nearly every
proactive system is benchmarked,
and each of the reviewed
proactive systems targets a
different point in the forecasting pipeline.
\citet{p28} propose a \ac{PHPA} methodology that
combines time-series forecasting,
hyperparameter optimization, and an
\ac{LLM}-based pattern classifier;
workloads are first assigned to a taxonomy of patterns,
after which a pattern-specific
forecasting model is selected automatically.
Using Gemini~2.5~Pro \citep{gemini-2.5-pro-paper}
for pattern classification,
the system reports a 37.4\% improvement
in prediction accuracy over single-model
forecasting and 96.7\% pattern-classification
accuracy on NASA HTTP logs \citep{nasa-http-logs-1995}
and Alibaba Cluster Traces (see \citet{alibaba-clusterdata-repo,
    alibaba-cluster-traces-dataset-paper} for datasets),
with gradient-boosting models achieving
sub-2\% mean absolute percentage error on production workloads.
\citet{p34} instead
forecast demand directly with an \ac{LSTM}
network layered on top of Kubernetes
\ac{HPA}, reporting a 28\% improvement in response time,
a 50\% reduction in scaling delay,
and a 15\% reduction in energy consumption relative to
reactive \ac{HPA};
because the evaluation is confined
to \ac{LSTM} point forecasts of
\ac{CPU}-style metrics,
it is unclear how the
approach would behave under the
heavy-tailed,
non-stationary request-length
distributions that \citet{p119} report for real
\ac{MLaaS} traffic
(see \zcref{subsubsec:workload-characterization-and-request-volatility-in-mlaas}).
\citeauthor{p35}'s KIS-S framework \citep{p35}
sidesteps the forecasting problem
altogether and instead learns a
scaling \emph{policy} end-to-end: a
\ac{GPU}-aware simulator (\ac{KISim}) provides high-fidelity,
Prometheus-integrated scheduling emulation,
on top of which a \ac{PPO}-based
autoscaler (i.e., KIScaler) is trained purely in simulation.
Evaluated across ramp, periodic, random,
and spike traffic patterns,
KIScaler's moving-average
reward improves by 75.2\% over 100 training
episodes and generalizes across
all four patterns without retraining,
reducing P95 latency by up to
$6.7\times$ relative to \ac{CPU}-only baselines;
the authors are explicit,
however, that training relies on synthetic
feedback because of single-\ac{GPU}
hardware constraints,
leaving open whether the learned policy transfers to a
real multi-\ac{GPU} cluster with contention
effects the simulator does not model.
Beyond single-cluster policies, \citeauthor{sky-serve}'s
SkyServe \citep{sky-serve} tackles proactive replica scaling
across geo-distributed and multi-cloud spot markets:
by combining predictive provisioning with eviction-aware migration,
SkyServe anticipates spot-instance preemption and regional workload shifts
to preemptively launch and retire model replicas across cloud providers,
mitigating tail-latency spikes while leveraging low-cost transient capacity.

\subsubsection{Stateful and Disaggregation-Aware Scaling}
\label{subsubsec:scaling-stateful}

A second cluster of work argues that
container-level replica scaling is,
on its own, the wrong level of
abstraction for stateful \ac{LLM} services.
\citet{p33} review scaling techniques
for \ac{SMS} broadly,
arguing that persistent state, session continuity,
and data locality make reactive,
threshold-based scaling
unsuitable, and that any autoscaler for
\ac{SMS} must additionally account for
cold-start latency and state-migration
overhead during a scaling event.
Applied to \ac{LLM} serving specifically,
this concern is precisely what
motivates runtime request rescheduling
and live migration (e.g., Llumnix \citep{p10})
as well as the disaggregated,
\ac{KV}-cache-aware orchestration mechanisms
discussed in \zcref{subsubsec:disaggregated-prefill-decode-serving-architectures}
(Splitwise~\citep{p37}, DistServe~\citep{p38},
Mooncake~\citep{p41}, BanaServe~\citep{p26}):
none of these treat scaling as
adding or removing a stateless replica,
but instead migrate or rebalance
\ac{KV}-cache-bearing state across
instances and prefill/decode pools.
\citet{p32} situate the entire
scaling problem within a broader \ac{MLSys}
survey of efficient generative \ac{LLM} serving,
covering algorithmic and
systems-level techniques jointly;
as already noted in
\zcref{subsec:efficient-llm-serving-and-inference-systems},
the survey's own stated limitation is
that it emphasizes breadth over a
fine-grained treatment of
cloud-specific concerns such as multi-tenant
resource sharing, so it is best read as context for,
rather than a substitute for,
the \ac{LLM}-specific autoscaling
work surveyed here.

\subsubsection{Gaps in Predictive Scaling}
\label{subsubsec:gaps-in-predictive-scaling}

\zcref{tab:scaling-summary} synthesizes the reviewed scaling systems.
Two patterns are worth naming explicitly.
First, every predictive system
reviewed here reports substantial
gains over reactive \ac{HPA},
yet each is evaluated against a
different metric (prediction accuracy, response time,
P95 latency, reward) and a
different workload trace,
so the reported improvements are
not directly comparable to one another;
a standardized scaling benchmark,
in the spirit of the evaluation gap \citet{p30} identify,
does not yet exist for \ac{LLM}-specific autoscaling.
Second, none of the proactive systems
reviewed jointly optimizes the
scaling decision with the batching
and \ac{KV}-cache-placement decisions
discussed in \zcref{subsec:batching,subsec:kv-cache-reuse},
even though a scale-out event that
ignores in-flight \ac{KV}-cache state
is exactly the scenario the disaggregated-serving literature
(\zcref{subsubsec:disaggregated-prefill-decode-serving-architectures})
was designed to avoid.

\begin{table}
    \centering
    \caption{Comparative summary of representative cloud LLM autoscaling systems.}
    \label{tab:scaling-summary}
    \footnotesize
    \begin{tabular}{p{2.1cm}p{1.8cm}p{4.0cm}p{3.8cm}}
        \toprule
        \textbf{System}                & \textbf{Approach} & \textbf{Key Mechanism}                                                      & \textbf{Reported Gain}                                                    \\
        \midrule
        \ac{PHPA} \citep{p28}          & Predictive        & \ac{LLM}-based pattern taxonomy and adaptive forecast selection             & 37.4\% higher prediction accuracy; 96.7\% pattern-classification accuracy \\
        \cmidrule(lr){1-4}
        \ac{LSTM}-\ac{HPA} \citep{p34} & Predictive        & \ac{LSTM} demand forecasting layered on K8s \ac{HPA}                        & 28\% faster response; 50\% less scaling delay; 15\% less energy           \\
        \cmidrule(lr){1-4}
        KIS-S \citep{p35}              & Learned policy    & \ac{GPU}-aware simulator (\ac{KISim}) + \ac{PPO} autoscaler                 & 75.2\% reward increase; up to $6.7\times$ lower P95 latency               \\
        \cmidrule(lr){1-4}
        SkyServe \citep{sky-serve}     & Proactive / Spot  & Multi-cloud proactive replica provisioning and preemption-aware migration   & Mitigates spot-preemption tail latency; cuts multi-cloud deployment cost  \\
        \cmidrule(lr){1-4}
        \ac{SMS} Review \citep{p33}    & Conceptual        & State-aware reactive/proactive/hybrid taxonomy for stateful microservices   & N/A (qualitative architectural taxonomy)                                  \\
        \bottomrule
    \end{tabular}
\end{table}

\subsection{Batching}
\label{subsec:batching}
\subsubsection{Memory-Aware Batching via KV-Cache Paging}
\label{subsubsec:batching-memory}

Batching is the mechanism that
converts autoscaled \ac{GPU} capacity into
throughput: because autoregressive
generation is memory-bandwidth bound,
grouping multiple requests so they
share a single forward pass is what
allows a \ac{GPU} to approach its peak-achievable
floating point operations per second.
The central obstacle to
batching \ac{LLM} requests,
as opposed to fixed-shape
vision or classification workloads,
is that each request carries
a \ac{KV} cache whose memory footprint
expands dynamically with
every generated token until request completion.
\citet{p21} address this directly
with PagedAttention, an attention
algorithm that borrows virtual-memory
paging from operating systems to
manage the \ac{KV} cache as fixed-size,
non-contiguous blocks; built on top of this,
the \ac{vLLM} serving system achieves
near-zero \ac{KV}-cache memory waste and
enables cache sharing within and across requests,
improving throughput by $2$--$4\times$ over
FasterTransformer and Orca \citep{p36} at matched latency,
where Orca had earlier pioneered
iteration-level continuous batching
but remained constrained by memory
fragmentation under static allocation.
Because PagedAttention decouples batch size from
worst-case \ac{KV}-cache allocation,
it is the substrate that later
cache-sharing and cache-migration
systems (see \zcref{subsec:kv-cache-reuse})
build on rather than a
competing alternative to them.

\subsubsection{Compute-Aware and Heterogeneous Batching}
\label{subsubsec:batching-compute}

A second obstacle is that even within a batch,
computation is not uniform across requests.
Autoregressive decoding is strictly memory-bandwidth bound,
whereas prompt prefill is heavily compute bound;
when prefill and decode phases are batched naively,
bursty or long prefill prompts monopolize execution units,
inducing severe inter-token latency spikes (i.e., decode stalls).
Sarathi-Serve \citep{p2} mitigates this with chunked-prefill batching,
slicing long prefill prompts into uniform token chunks
that can be co-batched with decode iterations to create predictable execution times.
Extending this co-execution paradigm, \citeauthor{pod-attention}'s
POD-Attention \citep{pod-attention} unlocks full prefill-decode overlap
within the attention kernel itself:
by dynamically partitioning attention
heads and overlapping memory-bound decode
steps with compute-bound prefill
chunks on the same \ac{GPU},
it eliminates decode bubbles without
requiring physical hardware disaggregation.

Complementing phase-level heterogeneity,
\cite{p24} generalize the batching problem
to conditional neural networks,
in which computation varies with the input,
as is the case for early-exit transformers and,
more subtly, for autoregressive \acp{LLM}
whose effective sequence length differs per request.
Their BatchCond framework combines
similarity-driven batch construction,
which groups inputs likely to share a computational pattern
using a lightweight predictor,
with adaptive batch reorganization,
which dynamically splits a batch
into computationally homogeneous
sub-batches at runtime;
across a suite of conditional networks including
early-exit networks,
dynamic slimmable networks, and autoregressive transformers,
this yields up to $6.6\times$ higher
throughput (mean $2.5\times$).
Where PagedAttention solves the
\emph{memory} irregularity of batching,
BatchCond and POD-Attention solve
the \textit{compute} irregularity,
and these concerns are largely orthogonal,
suggesting they are complementary rather
than competing levers,
though few systems evaluate them
jointly in unified production runtimes.

\subsubsection{Cross-Boundary and Cluster-Level Batching}
\label{subsubsec:batching-cluster}

Batching decisions also extend past
the boundary of a single accelerator.
\cite{p25} study \ac{LLM} inference
offloading and resource
allocation in cloud-edge computing,
replacing conventional deep
reinforcement learning offloading policies,
which the authors show suffer
from data inefficiency and
are insensitive to latency requirements,
with an active-inference formulation
that adapts more readily to a fluctuating task load;
simulation results show consistent gains in data efficiency and
adaptability over mainstream \ac{DRL} baselines,
though, as with the scaling
literature above, this is validated
primarily in simulation rather than on
a production cloud-edge deployment.
At the cluster level, \citeauthor{p26}'s
BanaServe~\citep{p26} treats prefill
and decode batching as a load
that must be rebalanced continuously
rather than fixed at deployment time:
layer-level weight migration
redistributes computation coarsely,
attention-level \ac{KV}-cache migration rebalances memory finely,
and a global \ac{KV}-cache store with
layer-wise overlapped transmission decouples cache
placement from routing decisions,
which the authors argue is what
eliminates the prefix-cache-induced
routing hotspots that plague static
disaggregated deployments (see
\zcref{subsubsec:disaggregated-prefill-decode-serving-architectures}).
Under long-context, bursty,
and mixed prompt-generation workloads,
BanaServe improves throughput by
$1.2$--$3.9\times$ over \ac{vLLM}
and by $1.1$--$2.8\times$ over DistServe \citep{p38},
while cutting latency by up
to 70.1\% relative to DistServe.
Finally, \citet{p27} surveys the hardware substrate that
ultimately bounds what any
batching policy can achieve, covering \ac{GPU},
\ac{FPGA}, and custom \ac{ASIC} accelerators
for \acp{LLM} and arguing that specialized
accelerators can reduce energy consumption
by orders of magnitude relative
to general-purpose hardware;
as already noted in \zcref{subsec:rw-hardware},
this survey remains hardware-centric and
does not connect its accelerator-level
findings back to serving-level batching or scheduling policy,
which is precisely the gap that PagedAttention,
BatchCond, POD-Attention, and BanaServe fill from the systems side.

\subsubsection{Toward Jointly-Aware Batching}
\label{subsubsec:batching-synthesis}

Read together, these works
trace an evolution from static,
fixed-shape batching toward scheduling policies
that are jointly aware of memory
layout \citep{p21}, iteration-level continuous scheduling \citep{p36},
fine-grained prefill-decode overlap \citep{p2,pod-attention},
computational heterogeneity \citep{p24},
placement across the cloud-edge continuum \citep{p25},
and live cluster-wide rebalancing \citep{p26}.
What remains open is a batching runtime
that unifies all of these dimensions simultaneously;
most current serving frameworks optimize
one or two while treating the others as static constraints.

\subsection{KV Cache Reuse}
\label{subsec:kv-cache-reuse}
\subsubsection{Cache Sharing Mechanisms}
\label{subsubsec:cache-sharing-mechanisms}

Reusing the Key-Value cache,
whether across turns of the same
conversation, across users who submit semantically similar prompts,
or across \ac{GPU} and host memory tiers,
is one of the most effective levers for
reducing redundant computation in \ac{LLM} serving.
\citeauthor{p23}'s SemShareKV \citep{p23}
operationalizes prompt-level reuse
directly: token-level \ac{LSH} identifies
semantically similar prompt prefixes across distinct users,
allowing their cached \ac{KV} representations to be shared
and \ac{GPU} memory footprint to be reduced accordingly.
Beyond semantic similarity hashing, modern serving engines exploit
exact structural prefix sharing:
\citeauthor{sglang-radix}'s SGLang framework \citep{sglang-radix} introduces
RadixAttention, which maintains \ac{KV}
caches across requests as a radix tree,
enabling automatic, zero-overhead reuse
of common prompt prefixes
(such as system instructions, few-shot demonstrations,
and multi-turn dialogue histories)
governed by a least-recently-used cache eviction policy.
Similarly addressing multi-turn workloads,
\citeauthor{stateful-llm-serving}'s
Pensieve \citep{stateful-llm-serving}
establishes a stateful serving engine:
rather than continuously pinning conversation
\ac{KV} caches in scarce \ac{GPU} memory
or discarding them between user interactions,
Pensieve swaps cached conversation
states across the host \ac{DRAM}
and \ac{GPU} memory hierarchy based
on user think-time predictions,
drastically cutting \ac{TTFT} in
subsequent turns while maximizing serving capacity.
This is the same mechanism referenced in the main
orchestration discussion
(see \zcref{subsubsec:multi-tenant-lora-adapter-and-kv-cache-orchestration}),
and it is precisely this class of
cross-tenant and cross-turn sharing that the reviewed security
literature shows to be a double-edged optimization.

\subsubsection{Privacy and Security Risks of Shared KV Caches}
\label{subsubsec:kv-security}

That security literature converges
on a single finding, stated
independently across five separate venues:
the \ac{KV} cache is not just a
performance artifact but a side
channel that leaks the content it was
built to avoid recomputing.
\citet{p121} provide the first
comprehensive analysis of \ac{KV}-cache privacy risk,
demonstrating three distinct attack vectors (a
direct inversion attack, a broader collision
attack, and a semantic injection attack), each
capable of reconstructing
sensitive user input directly
from cached Key and Value tensors;
their proposed defense, \ac{KV}-Cloak,
applies a reversible matrix-based
obfuscation (see \citet{obfuscation})
combined with operator fusion,
and the authors report that it reduces
reconstruction quality to random
noise with negligible accuracy loss and minimal overhead.
\citet{p122} attack the semantic-caching layer
itself rather than the raw \ac{KV} tensors,
showing that semantic caches, of
the kind already deployed by Azure, \ac{AWS},
and Alibaba, are vulnerable to
cache poisoning: an attacker injects
crafted entries so that other users'
semantically similar queries are
served the attacker's chosen response.
The authors confirm the attack's
practicality across all three major
public clouds and find that existing
adversarial-prompting defenses do not
mitigate it, though their own
proposed countermeasure only partially closes
the gap, leaving complete mitigation,
in the authors' own words, an open problem.

A second sub-cluster shows that
the leakage is not confined to the
software caching layer but extends
down to the hardware itself.
\citet{p123} uncover hardware cache
side-channels in local \ac{LLM}
inference, showing that an
adversary can infer token values from the
cache-access pattern of the embedding
operation and token positions from
the timing of autoregressive decoding,
without any privileged access to
the victim's system; across open-source
and proprietary deployments
\citep{llm-model-deployment-and-fine-tuning},
the reconstructed input and output text reach
average cosine similarity scores of 98.7\% and 98.0\% respectively,
which is strikingly high fidelity for
a side-channel attack that never touches
the model's weights or activations directly.
\citet{p125} push the inversion attack
further up the stack, challenging the standing
assumption that a model's internal states,
exposed by collaborative-inference
and safety-auditing techniques that outsource
early layers, are irreversible;
their two-phase, white-box
optimization attack avoids the
local-minima convergence that limited prior
inversion attempts, and a
black-box variant exploits transferability
between a source model and derived fine-tunes.
Notably, a 4,112-token
medical-consulting prompt is nearly
perfectly inverted, at 86.88 F1
token-matching, from a middle layer of Llama-3,
which suggests that
depth alone, contrary to the intuition
motivating collaborative inference
designs, is not a reliable privacy boundary.
\citeauthor{p126}'s PrivacyXray \citep{p126}
inverts the framing from attack to detection,
observing that \acp{LLM} exhibit measurably
higher semantic coherence and
probabilistic certainty when they
emit correct private information, and
using four metrics, intra- and interlayer
semantic similarity together
with token- and sentence-level
probability distributions,
to flag privacy breaches without needing
an external validation dataset;
evaluated across
five \acp{LLM}, the method reports 92.69\%
average accuracy, a 20.06\%
improvement over prior detection approaches.
Finally, \citet{p127} show that the vulnerability
generalizes beyond the \ac{KV}
cache to split-learning
deployments more broadly: their \ac{BiSR}
attack exploits the fact that
fine-tuning stays close to a model's pre-trained weights (the
``not-too-far'' property) together with
the autoregressive nature of
\acp{LLM} to reconstruct private
fine-tuning data from intermediate
activations exchanged in split learning,
and the authors report that \ac{BiSR}
remains effective even against three
representative defense mechanisms.

\subsubsection{Reuse vs. Leakage}
\label{subsubsec:kv-synthesis}

\zcref{tab:kv-security-summary}
summarizes the dual landscape of \ac{KV}-cache reuse mechanisms
and the security vulnerabilities exposed by recent studies.
Taken as a whole, this cluster of work
complicates the throughput and
batching operations
(see \zcref{subsubsec:multi-tenant-lora-adapter-and-kv-cache-orchestration}
and \zcref{subsec:batching}):
SemShareKV's cross-tenant sharing,
radix-tree prefix reuse \citep{sglang-radix},
and BanaServe's global \ac{KV}-cache store are exactly the
architectural patterns that \citet{p121},
\citet{p122}, and \citet{p123} show to
be exploitable, yet none of the batching
or cache-reuse systems reviewed in
\zcref{subsubsec:multi-tenant-lora-adapter-and-kv-cache-orchestration,
    subsec:batching} evaluate against these attacks,
and none of the attack papers evaluate their proposed mitigations,
\ac{KV}-Cloak excepted, against a
production-scale multi-tenant serving stack.
Efficient \ac{KV}-cache reuse and
secure \ac{KV}-cache isolation are,
at present, developed as
largely separate research threads
rather than as a single co-designed problem.

\begin{table}
    \centering
    \caption{Comparative summary of KV-cache reuse and KV-cache security research.}
    \label{tab:kv-security-summary}
    \footnotesize
    \begin{tabular}{p{2.3cm}p{1.7cm}p{4.2cm}p{3.1cm}}
        \toprule
        \textbf{System}                       & \textbf{Type}       & \textbf{Key Contribution}                                                      & \textbf{Headline Result}                                     \\
        \midrule
        SemShareKV~\citep{p23}                & Reuse               & Token-level \ac{LSH} matching for cross-user prefix sharing                     & Reduced \ac{GPU} memory footprint                           \\
        SGLang~\citep{sglang-radix}           & Reuse               & Radix-tree prefix cache management with LRU eviction                            & Fast multi-turn and prefix \ac{KV} reuse; low \ac{TTFT}     \\
        Pensieve~\citep{stateful-llm-serving} & Reuse               & Stateful \ac{DRAM} and \ac{GPU} memory hierarchy swapping for multi-turn chats  & Lower \ac{TTFT} across turns without pinned \ac{GPU} memory \\
        \cmidrule(lr){1-4}
        Shadow in the Cache~\citep{p121}      & Attack with Defense & Inversion, collision, injection attacks; \ac{KV}-Cloak defense                  & Reconstruction reduced to random noise                      \\
        Semantic Cache Poisoning~\citep{p122} & Attack              & Poisoning of production semantic caches (Azure, \ac{AWS}, and Alibaba)          & Confirmed practical across 3 clouds                         \\
        Hardware Side-Channels~\citep{p123}   & Attack              & Cache-access timing leaks token value and position                              & 98.7\% and 98.0\% cosine similarity                         \\
        Internal-State Inversion~\citep{p125} & Attack              & White- and black-box inversion of internal states                               & 86.88 F1 on 4,112-token prompt                              \\
        PrivacyXray~\citep{p126}              & Detection           & Semantic-consistency and probability-certainty metrics                          & 92.69\% average accuracy                                    \\
        \ac{BiSR}~\citep{p127}                & Attack              & Bidirectional reconstruction in split-learning fine-tuning                      & Effective vs.\ 3 defenses                                   \\
        \bottomrule
    \end{tabular}
\end{table}

\subsection{Cost Optimization}
\label{subsec:cost-optimization}
Cost optimization is not a separate
technique so much as the objective
that scaling, batching, and cache-reuse
decisions are ultimately made
under, and the reviewed literature attacks
it from three distinct angles:
restructuring the serving pipeline itself,
compressing the model and its
cache, and exploiting cheaper,
less reliable hardware markets.

\subsubsection{Pipeline and Storage Restructuring}
\label{subsubsec:cost-pipeline}

The first angle restructures \emph{where} computation happens.
\citeauthor{p37}'s Splitwise~\citep{p37}
shows that the compute-bound prefill
phase and the memory-bound decode phase
of \ac{LLM} inference have distinct
latency, throughput, memory, and power profiles,
and that state-of-the-art
batching leaves the decode phase's
compute underutilized; splitting the
two phases onto separate, phase-appropriate machines,
connected over the
\ac{GPU} cluster's fast back-plane interconnect,
allows homogeneous and
heterogeneous cluster designs that
achieve up to $1.4\times$ higher
throughput at 20\% lower cost,
or $2.35\times$ more throughput at matched power and cost.
\citeauthor{p38}'s DistServe~\citep{p38} arrives at a similar
disaggregation but optimizes it explicitly for goodput under separate
\ac{TTFT} and \ac{TBT} (see \zcref{eq:llm_latency}) \acp{SLO},
and \citeauthor{p41}'s Mooncake~\citep{p41}
pushes the same principle further into the storage hierarchy,
trading underexploited \ac{CPU}, \ac{DRAM}, \ac{SSD},
and \ac{NIC} resources for a \ac{KV}-cache-centric
global cache that increases effective request capacity by
59\%--498\% over baselines on real traces while meeting \acp{SLO}.
Additionally, the fact that Mooncake processes over 100 billion tokens per
day in production for the Kimi chatbot service is a tangible reminder that
these architectural gains are not purely a research artifact.

\subsubsection{Model and Kernel Compression}
\label{subsubsec:cost-compression}

The second angle reduces the cost per token
through numerical and algorithmic compression.
\citeauthor{p42}'s FlashAttention~\citep{p42} makes the
attention kernel itself \ac{IO}-aware,
tiling computation to reduce the number
of reads and writes between \ac{GPU}
high-bandwidth memory and on-chip static \ac{RAM};
because this is a lower-level primitive
than the serving-system papers
above, it composes with essentially all of them;
the reported gains, a 15\% end-to-end speedup
on BERT-large relative to the MLPerf~1.1 training
record and up to $3\times$ on GPT-2,
come without any change to model outputs.
\citeauthor{p43}'s SmoothQuant~\citep{p43} and \citeauthor{p44}'s
AWQ~\citep{p44} reduce the cost
of every forward pass through
quantization: SmoothQuant enables
training-free 8-bit integer weight-and-activation
quantization by migrating the
quantization difficulty from
harder-to-quantize activations to
easier-to-quantize weights, reporting up
to $1.56\times$ speedup and $2\times$ memory
reduction with negligible
accuracy loss; meanwhile, AWQ \citep{awq-detailed-config,p44}
observes that only a small fraction of
weights are salient and protects
them by calibrating scales from
activation statistics rather than weights,
reporting over $3\times$
speedup relative to an \ac{FP16} baseline and,
notably, the first practical on-device deployment
of a 70-billion-parameter LLaMA-2 model on a mobile \ac{GPU}.
Speculative decoding \citep{p45} tackles
the same cost problem from the
decoding-algorithm side,
using a cheaper draft
model to propose candidate tokens
which the target model verifies in
parallel, yielding a $2$--$3\times$ acceleration
on T5-XXL with outputs
identical to standard autoregressive decoding.
Moreover, \citeauthor{p46}'s FlexGen~\citep{p46}
targets the extreme end of resource
constraint, running a 175B-parameter model
on a single 16 GB \ac{GPU} by
solving a linear program over how tensors
should be distributed across
\ac{GPU}, \ac{CPU}, and disk,
and additionally compressing weights and the
attention cache to 4 bits;
the system is the first to report 1
token/second throughput under this hardware budget,
at an effective batch size of 144.

\subsubsection{Market-Aware and Preemptible Scheduling}

The third angle treats cost optimization
as a market and scheduling
problem rather than a compression problem.
\citeauthor{p47}'s SpotServe~\citep{p47},
discussed already in
\zcref{subsubsec:heterogeneous-preemptible-spot-instance-and-geo-distributed-scheduling},
is the first distributed \ac{LLM} serving system designed
explicitly for preemptible cloud instances:
it dynamically reparallelizes
the model in response to fluctuating
spot-instance availability and
workload, formulates the migration
required by reparallelization as a
bipartite matching problem solved
with the Kuhn-Munkres algorithm \citep{kuhn-munkres-algorithm}
to minimize communication cost,
and introduces stateful inference recovery so
that a preempted instance can resume
from a fine-grained checkpoint rather
than restarting a request from scratch;
on real spot-instance preemption
traces, SpotServe reduces P99 tail latency
by $2.4$--$9.1\times$ relative
to the best existing serving systems
while saving 54\% of monetary cost
relative to on-demand-only deployment.
Broadening the market scope
across multiple cloud providers,
SkyServe \citep{sky-serve} exploits
cross-cloud and geo-distributed spot instances,
demonstrating that combining proactive
replica management with inter-region traffic dispatching
can achieve up to 58\% lower serving costs
compared to single-region on-demand deployments
while meeting latency \acp{SLO}.
Finally, KIS-S~\citep{p35},
introduced in \zcref{subsec:scaling}
as a scaling system, is equally a
cost-optimization system in this framing,
since its \ac{PPO}-trained policy is
explicitly rewarded for latency-aware,
resource-efficient scaling rather than throughput alone.

\subsubsection{Composability of Cost Levers}
\label{subsubsec:cost-synthesis}

\zcref{tab:cost-summary} consolidates these
systems by the layer of
the stack at which they intervene.
The pattern that emerges is that no
single technique dominates:
pipeline restructuring (Splitwise, DistServe,
Mooncake), kernel and weight-level
compression (FlashAttention,
SmoothQuant, AWQ, speculative decoding, FlexGen),
and market-aware scheduling (SpotServe, SkyServe)
address cost at different, largely composable
layers of the stack,
and the strongest production deployments reviewed
here, Mooncake in particular,
combine several of these layers rather than
relying on any one of them in isolation.
What the literature does not yet
offer is a unified accounting of
how these layers trade off against one
another; a quantization scheme that
shrinks the \ac{KV} cache, for instance,
changes the storage-versus-compute tradeoff
that Mooncake's own placement
decisions are built around,
but no work reviewed in this section evaluates
that interaction directly.

\begin{table}
    \centering
    \caption{
        Comparative summary of cloud LLM cost-optimization techniques by
        intervention layer.
    }
    \label{tab:cost-summary}
    \footnotesize
    \begin{tabular}{p{2.3cm}p{2.1cm}p{4.2cm}p{3.0cm}}
        \toprule
        \textbf{System}                  & \textbf{Layer}      & \textbf{Key Mechanism}                                                          & \textbf{Reported Gain}                                  \\
        \midrule
        Splitwise \citep{p37}            & Pipeline            & Prefill/decode phase splitting across machines                                  & Up to $1.4\times$ throughput at 20\% lower cost         \\
        DistServe \citep{p38}            & Pipeline            & \ac{SLO}-driven prefill/decode disaggregation                                   & Higher goodput under \ac{TTFT}/\ac{TBT} \acp{SLO}       \\
        Mooncake \citep{p41}             & Storage tiering     & \ac{KV}-cache-centric global cache across \ac{CPU}/\ac{DRAM}/\ac{SSD}/\ac{NIC}  & 59\%--498\% higher effective capacity                   \\
        FlashAttention \citep{p42}       & Kernel              & \ac{IO}-aware tiled attention                                                   & Up to $3\times$ speedup, exact outputs                  \\
        SmoothQuant \citep{p43}          & Quantization        & Activation-to-weight quantization migration                                     & Up to $1.56\times$ speedup, $2\times$ memory reduction  \\
        AWQ \citep{p44}                  & Quantization        & Activation-aware salient-weight protection                                      & $>3\times$ speedup vs.\ \ac{FP16}                       \\
        Speculative decoding \citep{p45} & Decoding algorithm  & Draft-and-verify parallel token generation                                      & $2$--$3\times$ acceleration, identical outputs          \\
        FlexGen \citep{p46}              & Offloading          & LP-optimized \ac{GPU}/\ac{CPU}/disk tensor placement and 4-bit compression      & 1 tok/s on 175B model, single 16GB \ac{GPU}             \\
        SpotServe \citep{p47}            & Scheduling / market & Dynamic reparallelization with checkpointing on spot instances                  & 54\% cost savings; $2.4$--$9.1\times$ lower P99 latency \\
        SkyServe \citep{sky-serve}       & Multi-cloud market  & Cross-region proactive replica migration on spot instances                      & Up to 58\% cost reduction vs.\ on-demand                \\
        \bottomrule
    \end{tabular}
\end{table}

    \section{Cloud-based Architecture}
    \label{sec:architecture}
    \subsection{Serverless Architecture}
\label{subsec:serverless-architecture}
Serverless computing,
or \ac{FaaS} \citep{p67, faas-related-recent-paper},
has emerged as an attractive deployment paradigm
for cloud-hosted \ac{LLM} inference because of its elastic,
pay-per-use resource model,
which in principle eliminates the need for
practitioners to provision and manage \ac{GPU} clusters directly.
However, the mismatch between the stateless,
ephemeral execution model of traditional \ac{FaaS}
platforms and the stateful,
multi-gigabyte-parameter nature of modern \acp{LLM}
introduces a distinct set of systems challenges
that do not arise in conventional serverless workloads.
The reviewed literature converges on four recurring
problems: (i) prohibitively long \emph{cold-start}
latency when loading model weights into \ac{GPU} memory,
(ii) inefficient and coarse-grained \ac{GPU}
sharing across concurrent functions,
(iii) the difficulty of achieving fast,
live autoscaling for large models under bursty demand,
and (iv) the general absence of standardized platforms
and evaluation methodologies for \ac{LLM}-specific serverless systems.
We organize the following review of fourteen papers
around these themes before presenting
a comparative summary in \zcref{tab:serverless-summary}.

\subsubsection{Cold-Start Latency Mitigation}

Cold-start latency, driven by the need to load large model
checkpoints into accelerator memory on every fresh container invocation,
is the most extensively studied problem in serverless \ac{LLM} serving.
\citet{p55} propose FlashServe,
which combines tiered memory snapshotting (host \ac{DRAM} staging with \ac{PCIe}
\ac{DMA} transfer to \ac{GPU} memory),
a hybrid Prophet-\ac{LSTM} predictor for proactive pod pre-warming,
and \ac{LoRA} adapter multiplexing to
support multiple fine-tuned models on shared \ac{GPU} resources.
On the Azure Functions trace,
FlashServe reduces cold-start latency by up to $49\times$ relative
to S3-based checkpoint loading and by $3.3\times$ relative to ServerlessLLM,
while cutting \ac{GPU} idle cost by 32\% and
sustaining sub-second \ac{TTFT} for 95\% of requests.
The authors note, however,
that these gains depend on sufficient host \ac{DRAM} (below 256\,GB per
server the caching benefit erodes),
are fundamentally bounded by \ac{PCIe} Gen4 bandwidth (25\,GB/s),
and that the pre-warmed container pool still
incurs idle \ac{GPU}-context overhead.

\citet{p59} address a similar problem from
a network-placement perspective with HydraServe,
which proactively distributes model
replicas across servers to parallelize fetching,
overlaps the distinct stages of cold start
(fetching, deserialization, runtime preparation),
and strategically places workers to avoid network contention;
a pipeline-consolidation mechanism subsequently
merges the resulting workers back
into single serving endpoints once warm.
HydraServe reports a 1.7--4.7 reduction in cold-start latency and a 1.43--1.74
improvement in \ac{SLO} attainment over baseline systems,
with particular benefit for long-tail (infrequently invoked) models.

Rather than optimizing the data
path for model weights specifically,
\citet{p56} target the general class of
\acp{DLIS} with PipeCo, which vertically partitions
a service into slices that are pre-warmed in a sequential,
overlapping fashion, guided by an attention-based predictor
of request periodicity and a similarity-based
matcher for reusing idle containers.
Implemented on OpenFaaS and evaluated
on three real-world \ac{DLIS} repositories,
PipeCo reduces end-to-end latency by up to 62.67\%
on \ac{CPU} and 58.81\% on \ac{GPU} clusters,
and cuts overall resource usage by 65.31\%
relative to five state-of-the-art baselines.

At a finer granularity, \citet{p63} isolate
\emph{application-level} bootstrap-framework loading,
model initialization, download, deserialization,
and memory copy-as the dominant contributor to
cold-start time as model size grows, and propose \ac{PISeL},
a layer-grouping pipelining
mechanism that overlaps model download,
deserialization/copy, and request execution via
lightweight hook points.
\ac{PISeL} is framework-agnostic
and yields cold-start speedups of
37\%/63\% (PyTorch, \ac{CPU}/\ac{GPU})
and 29\%/33\% (TensorFlow, \ac{CPU}/\ac{GPU}),
together with peak memory reductions
of up to 59\% and 30\%, respectively.

\subsubsection{GPU Sharing and Resource Multiplexing}

A second cluster of work targets the coarse-grained,
exclusive \ac{GPU} allocation typical of \ac{FaaS} platforms,
which leads to severe under-utilization.
\citet{p57} propose FaST-GShare,
a spatio-temporal \ac{GPU}-sharing architecture comprising
a FaST-Manager for resource isolation,
a FaST-Profiler that characterizes function
throughput under varying resource allocations,
and a FaST-Scheduler that performs
heuristic auto-scaling and \ac{SLO}-aware placement,
further supplemented by model sharing to reduce memory contention.
On an OpenFaaS prototype benchmarked with MLPerf workloads,
FaST-GShare improves throughput by $3.15\times$,
\ac{GPU} utilization by $1.34\times$, and \ac{SM}
occupancy by 3.13 relative to naive time-sharing.

\citet{p65} instead focus on small-to
mid-sized \acp{LLM} under private,
low-frequency serverless deployment conditions,
observing that both \acp{CPU} (with modern built-in accelerators)
and \acp{GPU} are capable of hosting multiple concurrent \acp{LLM}
but remain underutilized in existing designs.
Their system, SLINFER,
performs token-level fine-grained compute allocation,
forward-looking memory scaling to preempt
out-of-memory hazards,
and a dual-consolidation strategy combining
proactive preemption with
reactive bin-packing.
Evaluated across \ac{CPU} and A100 \ac{GPU} testbeds,
SLINFER improves serving capacity by 47\%--62\%
through sharing alone, rising to 86\%--154\%
when heterogeneous \ac{CPU}
resources are also exploited.

\subsubsection{End-to-End Serverless LLM Platforms and Live Autoscaling}

A third group of papers proposes complete,
production-oriented serverless \ac{LLM} serving platforms.
\citet{p58} present DEEPSERVE,
an industrial serverless \ac{AI} platform built around
a request-job-task abstraction that unifies
post-training and model-serving workloads,
an in-house serving engine (FLOWSERVE) with \ac{NPU} centric,
\ac{SPMD} based execution,
and scheduling policies supporting
both PD-disaggregated and PD-colocated deployment configurations.
Combined with pre-warmed pods, \ac{DRAM} pre-loading,
and an \ac{NPU}-fork mechanism,
DEEPSERVE scales to 64 instances within
seconds and has operated in production on a
large Ascend \ac{NPU} cluster for over a year,
though the reported evaluation is
largely operational rather than
benchmarked against external baselines.

Autoscaling under bursty demand is addressed directly
by \citet{p67} and \citet{p68}.
FaaScale~\citep{p67} identifies data-transfer
cost as the principal bottleneck to on-demand \ac{LLM}
scaling and introduces \emph{Pipelined Multicast Inference} \acs{PipeCast},
which adaptively multicasts model blocks
over \ac{RDMA} while dynamically forming cross-node
inference pipelines during the transfer itself;
combined with \ac{GPU}/host memory management,
this yields up to $5\times$ lower tail \ac{TTFT} and a 31.3\%
cost reduction on real-world traces.
BlitzScale~\citep{p68} makes a related but
architecturally distinct observation: that the compute
(\ac{GPU}-to-\ac{GPU}) network is underutilized and
can deliver host-cache-comparable bandwidth for
parameter loading with no or O(1) caching,
and further shows that autoscaling can
be made \emph{live} by decomposing scaling
from the instance level to the layer level,
offloading layer computation to a scaling
instance before its parameters have fully loaded.
Relative to ServerlessLLM,
BlitzScale achieves up to 94\% lower tail latency,
and relative to autoscaling-unaware serving systems
such as DistServe and \ac{vLLM} it reduces
\ac{GPU}-time consumption by 49\% under equivalent \acp{SLA};
the authors note that scaling-policy design
and intra-instance (e.g., MoE expert-level)
scaling remain open for future work.

\subsubsection{Adapter-Level and Multi-Tenant Serving}

\citet{p61} address the specific problem of serving many
\ac{LoRA}-fine-tuned variants of a shared base
model in a serverless setting,
where reactive adapter loading causes
cold-start latency and frequent swapping
fragments \ac{GPU} memory.
Their system, \ac{P-LoRA},
couples a lightweight \ac{LSTM} traffic predictor that
proactively prefetches ``hot'' adapters from host
to \ac{GPU} memory with an operating-system-inspired,
page-based adapter memory manager.
On Azure-Functions-derived workloads, \ac{P-LoRA}
reduces cold-start latency by up to 68\%,
sustains \ac{GPU} memory utilization above 87\%
under heterogeneous adapter ranks,
and achieves $1.52\times$ higher throughput
and 35\% lower average \ac{TTFT}
than S-\ac{LoRA} (scalable \ac{LoRA}; see \citet{p61})
under high concurrency;
the authors note that prediction accuracy
degrades under highly irregular
or adversarial traffic patterns.

\subsubsection{General-Purpose Serverless DNN Inference Systems}

Two papers target deep learning inference more broadly
rather than \ac{LLM} specifically,
providing useful architectural context.
\citet{p66} propose INFless,
the first ML-domain-specific serverless platform,
which exposes a unified \ac{CPU}-accelerator
resource abstraction and combines built-in
batching with non-uniform scaling and coordinated management
of queuing time, execution time,
and cold-start rate; INFless outperforms
state-of-the-art general-purpose serverless
systems by 2--5 in throughput while meeting latency \acp{SLO},
evaluated on both a local testbed and production-scale
traces from a life-service website.
\citet{p62} instead tackle the configuration-search
problem with \ac{AYCI},
a fully managed service that helps developers navigate
the large configuration space
(hardware choice, batch size, model compilation)
for optimal \ac{DNN} inference on public serverless platforms;
the prototype, however, is \ac{AWS} only,
remains in an early stage,
and reports limited gains from \ac{TVM} \citep{apache-tvm}
compilation due to restricted hardware visibility.

\subsubsection{Surveys and Conceptual Frameworks}

Finally, two contributions provide synthesis rather than new systems.
\citet{p60} present, to the authors' knowledge,
the first survey to situate \ac{LLM} inference explicitly
within the serverless computing literature,
reviewing system-level advances in cold-start mitigation,
resource utilization, cost optimization,
and scalability since 2019 and identifying adaptive batching,
hybrid scheduling, model partitioning, and \ac{GPU}
sharing as the dominant technique families.
\citet{p64} instead offer a forward-looking,
largely conceptual architecture
for serverless \ac{LLM} systems,
proposing a three-layer design
(declarative interface,
adaptive orchestration engine,
specialized runtime)
intended to jointly address cold-start mitigation,
state management,
resource sharing with isolation,
and pipeline optimization;
the paper explicitly does not
provide standardized
evaluation metrics or empirical validation,
positioning it as a roadmap
rather than a benchmarked system.

\subsubsection{Synthesis}

\zcref{tab:serverless-summary} summarizes the reviewed systems.
Two patterns are evident.
First, the field has converged on
a small set of complementary levers for
cold-start reduction-tiered/host-memory staging~\citep{p55},
network-aware and multicast-based model
distribution~\citep{p59,p67,p68},
pipelining of bootstrap stages~\citep{p56,p63},
and demand prediction for pre-warming~\citep{p55,p56,p61},
which are frequently combined
rather than mutually exclusive.
Second, a growing subset of work moves
beyond static container-level scaling
toward \emph{fine-grained} (layer- or token-level) elasticity~\citep{p68,p65},
suggesting a broader shift from coarse instance-based
\ac{FaaS} abstractions toward
abstractions native to model internals.
Nonetheless, several gaps persist across
the reviewed literature: evaluation is
almost universally confined to
text-generation workloads on
a single cloud provider's
trace data (most commonly Azure Functions),
with limited attention to multimodal or
retrieval-augmented pipelines~\citep{p55};
prediction-based prewarming approaches generally
assume regular or quasi-periodic traffic and
are acknowledged to degrade under adversarial
or highly bursty patterns~\citep{p55,p61};
and conceptual frameworks that identify
the full scope of the problem~\citep{p64}
remain empirically unvalidated relative to the narrower,
benchmarked systems they motivate.
This gap between vision papers and deployed systems,
together with the absence of
standardized cross-platform
benchmarks noted by \citet{p60},
represents a significant open direction
for future serverless \ac{LLM} research.

\begingroup
\footnotesize
\setlength{\tabcolsep}{3pt}
\renewcommand{\arraystretch}{1.15}
\begin{longtable}{@{}
        >{\raggedright\arraybackslash}p{0.16\linewidth}
        >{\raggedright\arraybackslash}p{0.14\linewidth}
        >{\raggedright\arraybackslash}p{0.27\linewidth}
        >{\raggedright\arraybackslash}p{0.30\linewidth}
        >{\raggedright\arraybackslash}p{0.13\linewidth}
    @{}}
    \caption{Comparative summary of representative serverless architecture frameworks for \ac{LLM} and \ac{DNN} inference.}\label{tab:serverless-summary}\\
    \toprule
    \textbf{Framework}               & \textbf{Category}                   & \textbf{Key Architectural Innovation}                                                             & \textbf{Primary Optimization Metric} & \textbf{Evaluation Platform} \\
    \midrule
    \endfirsthead

    \multicolumn{5}{c}{\small\itshape Table \thetable{} (continued from previous page)} \\
    \addlinespace
    \toprule
    \textbf{Framework}               & \textbf{Category}                   & \textbf{Key Architectural Innovation}                                                             & \textbf{Primary Optimization Metric} & \textbf{Evaluation Platform} \\
    \midrule
    \endhead

    \midrule
    \multicolumn{5}{r}{\itshape Continued on next page} \\
    \endfoot

    \bottomrule
    \endlastfoot

    FlashServe \citep{p55}           & Cold-Start Mitigation               & Tiered \ac{DRAM} and \ac{PCIe} snapshotting, Prophet-\ac{LSTM} prewarming, \ac{LoRA} multiplexing  & 49 cold-start reduction over S3; 3.3 over ServerlessLLM; 32\% lower \ac{GPU} idle cost     & Azure Functions trace                    \\
    PipeCo \citep{p56}               & Cold-Start Mitigation               & Vertical slice partitioning, attention-based prewarm scheduling, container matching                & 62.67\%/58.81\% E2E latency reduction (\ac{CPU}/\ac{GPU}); 65.31\% resource reduction      & OpenFaaS                                 \\
    HydraServe \citep{p59}           & Cold-Start Mitigation               & Proactive multi-server model distribution, overlapped cold-start stages, pipeline consolidation    & 1.7-4.7 cold-start reduction; 1.43-1.74 \ac{SLO} attainment                                & Public cloud                             \\
    \ac{PISeL} \citep{p63}           & Cold-Start Mitigation               & Layer-grouping pipeline of download, deserialize, and execute                                      & 37-63\% cold-start speedup; up to 59\% peak-memory reduction                               & PyTorch / TensorFlow (\ac{CPU}/\ac{GPU}) \\
    \midrule
    FaST-GShare \citep{p57}          & \ac{GPU} Sharing \& Multiplexing    & \ac{FaaS}-oriented spatio-temporal \ac{GPU} sharing (Manager, Profiler, and Scheduler)             & 3.15 throughput; 1.34 \ac{GPU} util.; 3.13 \ac{SM} occupancy                               & OpenFaaS, MLPerf                         \\
    SLINFER \citep{p65}              & \ac{GPU} Sharing \& Multiplexing    & Token-level allocation, forward-looking memory scaling, preemption and bin-packing                 & 47-62\% capacity gain (sharing); 86-154\% with \ac{CPU} use                                & \ac{CPU} and A100 \ac{GPU} testbed       \\
    \midrule
    DEEPSERVE \citep{p58}            & End-to-End Platforms \& Autoscaling & Request-job-task abstraction; FLOWSERVE \ac{NPU}-centric \ac{SPMD} engine; pre-warmed pods        & Scales to 64 instances in seconds; production-deployed                                      & Ascend \ac{NPU} cluster                  \\
    FaaScale \citep{p67}             & End-to-End Platforms \& Autoscaling & \ac{PipeCast} over \ac{RDMA}                                                                      & Up to 5 lower tail \ac{TTFT}; 31.3\% cost reduction                                         & Real-world \ac{LLM} traces               \\
    BlitzScale \citep{p68}           & End-to-End Platforms \& Autoscaling & O(1)-caching compute-network parameter loading; layer-level live autoscaling                      & Up to 94\% lower tail latency over ServerlessLLM; 49\% \ac{GPU}-time reduction              & Real-world workloads                     \\
    \midrule
    Predictive-\ac{LoRA} \citep{p61} & Adapter-Level \& Multi-Tenant       & \ac{LSTM} adapter-demand prediction, page-based adapter memory management                         & 68\% cold-start reduction; 1.52 throughput over S-\ac{LoRA}; 35\% lower \ac{TTFT}           & Azure Functions trace                    \\
    \midrule
    INFless \citep{p66}              & General-Purpose \ac{DNN} Serving    & Unified \ac{CPU} and accelerator abstraction, built-in batching, non-uniform scaling              & 2-5 throughput over SOTA general-purpose serverless                                         & Local testbed and production trace       \\
    \ac{AYCI} \citep{p62}            & General-Purpose \ac{DNN} Serving    & Automated search over hardware, batch, and compilation configurations                             & Prototype demonstrates configuration-space difficulty                                       & \ac{AWS}                                 \\
    \midrule
    Survey \citep{p60}               & Survey / Conceptual                 & Systematic review of serverless \ac{AI} and \ac{LLM} inference literature (post-2019)             & First survey covering \ac{LLM} inference under serverless computing                         & N/A                                      \\
    Vision Study \citep{p64}         & Survey / Conceptual                 & Three-layer architecture: declarative interface, orchestration, runtime                           & Conceptual roadmap; no empirical validation                                                 & N/A                                      \\
\end{longtable}
\endgroup

\subsection{Fault Tolerant Service Architectures}
\label{subsec:fault_tolerance}
As \acp{LLM} transition from experimental
research prototypes into
mission-critical cloud backbones,
ensuring continuous availability,
high operational reliability,
and end-to-end correctness becomes paramount
\citep{llm-ops-cloud-edge-fog}.
In conventional cloud computing,
microservices achieve fault tolerance primarily
through stateless instance replication,
active load balancing,
and rapid container failover.
However, cloud-hosted \ac{LLM} services present
a fundamentally different operating environment
governed by three distinct structural properties:

\begin{enumerate}[I.]
    \item \textbf{Heavy Statefulness and Prohibitive Failover Costs}:
    Unlike stateless web endpoints,
    generative inference accumulates multi-gigabyte
    \ac{KV} caches in accelerator memory across prolonged conversational sessions.
    Migrating these ephemeral state tensors across nodes
    or recomputing them from scratch imposes severe \ac{TTFT} penalties,
    rendering conventional stateless container restarts impractical for real-time applications.

    \item \textbf{Accelerator Transient Faults and Silent Data Corruption}:
    Modern cloud serving infrastructures aggregate dozens of
    tensor-parallel and pipeline-parallel \acp{GPU}.
    At this operational scale,
    physical accelerators are increasingly susceptible to hardware transient faults,
    such as single-event upsets caused by thermal fluctuations and voltage drops,
    which manifest as \acp{SDC} during autoregressive decoding.

    \item \textbf{Non-Deterministic and Semantic Failure Modes}:
    Unlike binary node crashes or timeout exceptions,
    \ac{LLM} failures frequently manifest as subtle semantic confabulations,
    hallucinated factual assertions,
    or abrupt quality degradation under adversarial input perturbations,
    all while execution completes with successful HTTP 200 status codes.
\end{enumerate}

\begin{landscape}
    \begin{figure}
        \centering
        \includegraphics[scale=0.8]{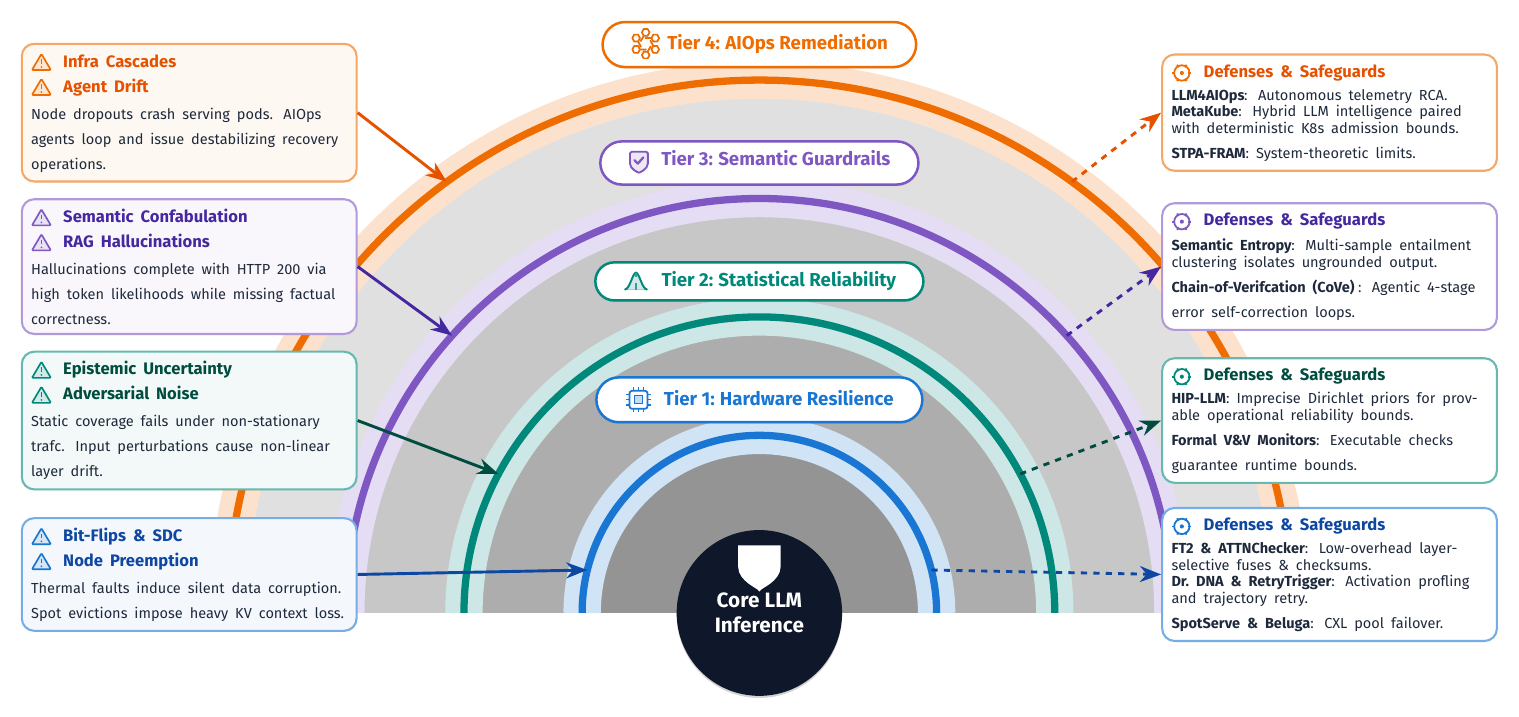}
        \caption{
            Multi-tier defense-in-depth architecture
            for cloud \ac{LLM} dependability.
            The defense pipeline is organized across
            four interrelated strata
            (hardware resilience, statistical reliability,
            semantic guardrails, and \ac{AIOps} remediation),
            progressively intercepting corresponding vulnerability vectors.
            Representative systems providing algorithmic
            safeguards for each layer are summarized on the right.
        }
        \label{fig:fault-tolerance-taxonomy}
    \end{figure}
\end{landscape}

To address these vulnerabilities,
cloud-native \ac{LLM} architectures deploy a multi-layered
defense model spanning four interrelated tiers:
hardware transient fault mitigation,
statistical reliability engineering
and formal verification \citep{p101},
semantic fault detection and
self-correcting guardrails \citep{semantic-fault-detection},
and automated cluster incident
remediation via \ac{AIOps} \citep{ai-ops-log}.
These four tiers do not constitute disjoint design alternatives;
rather, they form a defense-in-depth hierarchy.
A single serving pipeline can fail at the silicon level,
the distributional level,
or the semantic reasoning level.
Consequently, a system that protects only
a single layer cannot guarantee operational dependability.
\zcref{fig:fault-tolerance-taxonomy}
illustrates this four-tier defense hierarchy
alongside representative systems.

\subsubsection{Hardware Transient Faults and Silent Data Corruption Mitigation}

At the physical accelerator layer,
transient hardware errors induce \acp{SDC},
where bit-flips corrupt internal matrix multiplications
without triggering hardware
exceptions or operating-system kernel faults.
In generative \acp{LLM},
an unmitigated \ac{SDC} propagates across
successive autoregressive transformer layers,
producing corrupted token logits,
$\text{NaN}$ activations,
or degenerated output text.
Historically, hardware resilience
relied on \ac{ABFT} or full compute redundancy.
However, classical \ac{ABFT} requires
intrusive modifications to vendor
\textit{general matrix multiply} libraries,
while dual-modular replication doubles
operational accelerator expenses.

To overcome the high overhead of full algorithmic protection,
\citet{p_ft2} develop FT2,
an online fault tolerance framework designed
specifically for generative \acp{LLM}.
FT2 observes that error propagation resilience
is highly uneven across transformer architectures:
early layers and specific attention
operations exhibit high vulnerability,
whereas deeper layers demonstrate natural damping characteristics.
Capitalizing on this observation,
FT2 leverages first-token attention trajectory profiles
during the prefill phase to dynamically classify
and selectively protect only the
critical layers through lightweight checksum verification.
By shielding only sensitive operational zones,
FT2 provides comprehensive error recovery against
transient faults while reducing execution
overhead to a fraction of full-model \ac{ABFT}.

Complementing layer-selective protection,
\citet{p_attnchecker} introduce ATTNChecker,
a highly optimized,
lightweight fault-tolerance mechanism focused
specifically on the multi-head attention kernel.
Because attention calculations dominate \ac{LLM}
execution time and are particularly prone
to soft errors during matrix multiplications,
ATTNChecker constructs specialized mathematical
checksum relations tailored to flash-attention
and standard attention operations.
By fusing checksum generation and boundary
checking directly into the fused attention kernels,
ATTNChecker detects and localizes transient bit-flips
with near-zero latency penalty and
minimal memory overhead during live execution.

Taking an activation-distribution perspective,
\citet{p_drdna} present Dr.~DNA,
a non-intrusive framework that combats \acp{SDC}
by monitoring neuron activation distributions.
Rather than instrumenting low-level arithmetic instructions,
Dr.~DNA profiles the statistical distributions
and valid bounding intervals of
intermediate activations during offline calibration.
During online inference, it continuously
evaluates runtime activations against
these empirical distribution profiles,
successfully catching subtle numerical
drifts and soft faults before they
manifest as corrupted user-facing outputs.

At the decoding trajectory level,
\citet{p102} propose RetryTrigger,
a hardware-free,
data-driven anomaly detector.
RetryTrigger extracts lightweight statistical
features during each autoregressive decoding step:
maximum token probabilities,
top-$k$ probability divergence gaps,
logits distribution entropy,
and step latency.
A trained \ac{LightGBM} meta-model evaluates these
runtime indicators in real time to detect
anomalous decoding trajectories
indicative of transient hardware faults.
Upon detecting a likely corruption,
the runtime triggers a localized inference retry,
suppressing up to 95.33\% of \acp{SDC}
at an average execution overhead of only 4.12\%.

Beyond intra-kernel and intra-instance soft errors,
hardware fault tolerance must
address sudden instance evictions,
such as node dropouts and spot-instance
preemptions in public cloud fleets.
\citet{p47} address spot-market
volatility with SpotServe,
which implements dynamic reparallelization
and parallelized state migration.
SpotServe utilizes lightweight \ac{KV}-cache
checkpointing to preserve generative
context across preempted instances,
migrating live conversational states to
newly allocated instances within seconds.
Decoupling model state from physical accelerators
is essential for avoiding multi-node
pipeline stalls during hardware failures \citep{p106}.
To achieve resilient state persistence without
saturating \ac{GPU} high-bandwidth memory,
\citet{p75} propose a multi-tier dynamic storage
architecture that organizes \ac{KV} caches
across a hierarchy of accelerator memory,
host \ac{DRAM}, and network storage,
ensuring rapid state recovery under
severe resource constraints.
Pushing disaggregated state persistence further,
\citet{p79} introduce Beluga,
a memory architecture leveraging
\ac{CXL} links to create a shared,
disaggregated memory pool for \ac{KV} caches.
By enabling multiple compute instances to access
shared \ac{CXL} memory without
traversing host \ac{PCIe} bottlenecks,
Beluga enables instantaneous state failover
and zero-copy context recovery when a
serving node experiences
catastrophic hardware failure.

\subsubsection{Statistical Reliability Engineering and Formal Verification}

While hardware-level mechanisms
safeguard numerical integrity,
cloud serving systems require
guarantees that models remain
statistically reliable across
millions of heterogeneous user requests.
Standard benchmark suites like MMLU \citep{mmlu-details}
and GSM8K \citep{llm-benchmark-review} provide static snapshot scores,
failing to model performance drift or quantify
confidence intervals under
non-stationary production traffic.

To establish rigorous operational guarantees,
\citet{p101} introduce \ac{HIP}-\ac{LLM},
a statistical reliability assessment framework
grounded in hierarchical imprecise probability theory.
Rather than reporting scalar accuracy metrics,
\ac{HIP}-\ac{LLM} defines operational reliability
as the probability of failure-free execution
over a sequence of $N$ future requests.
By applying imprecise Dirichlet priors to
capture epistemic uncertainty and
structuring sub-domain tasks hierarchically,
\ac{HIP}-\ac{LLM} computes robust posterior reliability
envelopes with explicit confidence bounds
across diverse enterprise application domains.

In parallel with statistical modeling,
formal verification techniques are increasingly
applied to generative \ac{AI} runtimes.
Comprehensive surveys by \citet{p103}
and \citet{p111} systematize the
application of software engineering
\ac{V-and-V} methodologies to neural architectures.
These frameworks formalize safety
and operational invariants into executable
runtime monitors that continuously
inspect input prompts and generated responses,
guaranteeing that outputs remain strictly
confined within certified operational boundaries.
While \ac{HIP}-\ac{LLM} offers
probabilistic envelopes over future workloads,
formal monitors enforce deterministic
runtime safety bounds.
However, both paradigms remain sensitive to input noise.
As demonstrated by \citet{p105},
adversarial perturbations and noisy human
inputs propagate non-linearly
across deep attention layers,
causing severe output quality degradation unless
paired with structured input sanitization
and semantic error-correction layers.

\subsubsection{Semantic Fault Detection and Self-Correcting Guardrails}

Statistical envelopes and formal monitors
characterize aggregate reliability,
yet they cannot evaluate whether an individual generated
response contains subtle factual
inaccuracies or hallucinations.
Cloud gateways must therefore deploy real-time
semantic validation guardrails
that identify and remediate reasoning
failures without violating strict latency \acp{SLO}.

To detect semantic faults without the
overhead of external reference retrieval,
\citet{p109} develop Semantic Entropy.
Semantic Entropy measures model uncertainty
at the level of semantic meaning
rather than token-level probabilities.
The system samples multiple candidate
completions for a given prompt,
clusters them into semantic equivalence
classes using bidirectional textual entailment,
and computes entropy over the resulting clusters.
A high semantic entropy score reliably flags
confabulations and ungrounded statements,
allowing cloud \ac{API} gateways to discard
erroneous outputs or dynamically escalate
queries to higher-capacity foundation models.

For mission-critical tasks requiring active remediation,
\citet{p110} introduce \ac{CoVe}.
\ac{CoVe} executes an automated four-stage self-correction cycle:
(i) drafting an initial baseline completion,
(ii) generating targeted verification questions
to challenge factual premises,
(iii) executing verification queries independently
to prevent hallucination contamination,
and (iv) synthesizing an error-corrected final response.
Although \ac{CoVe} significantly reduces factual
errors across complex reasoning tasks,
its iterative multi-query cycle incurs substantial
token and latency overheads.
In enterprise deployments,
systems such as RAGVA \citep{p112} balance latency
and correctness by applying dynamic verification guardrails
selectively to high-stakes document chunks
in retrieval-augmented workflows.

\subsubsection{AIOps and Autonomous Cloud Self-Healing Loops}

At the cluster management layer,
service availability depends on rapid anomaly detection,
fault diagnosis, and automated remediation across containers,
worker nodes, and network fabrics.
Addressing cloud operations complexity,
\citet{p108} introduce LLM4AIOps,
which integrates fine-tuned language models as
autonomous control agents within cloud monitoring pipelines.
By ingesting multi-modal telemetry metrics,
distributed tracing spans, and Kubernetes event logs,
LLM4AIOps automates \ac{RCA},
anomaly triage, and incident response
across distributed microservices.

However, deploying autonomous language models
within critical operational loops
introduces distinct safety hazards.
System-theoretic safety assessments using
\ac{STPA} and the \ac{FRAM} by \citet{p107}
reveal that autonomous \ac{LLM} agents
are prone to semantic context drift
during cascading cluster failures,
potentially issuing destabilizing remediation commands.
To mitigate this operational risk,
resilient cloud architectures combine \ac{LLM}-driven
reasoning with deterministic control frameworks.
For example, MetaKube \citep{p49} couples an
experience-aware \ac{LLM} diagnostic engine with
deterministic Kubernetes admission controllers
and explicit verification policies.
This hybrid architecture leverages language models
for rapid diagnostic synthesis and root-cause
localization while ensuring that all automated
recovery actions are validated against
strict cluster admission rules,
substantially reducing mean time to
recovery (\ac{MTTR}) while preventing
erroneous operational changes.

\zcref{tab:fault-tolerance-summary}
synthesizes representative fault-tolerance
mechanisms across the four defense layers.
The literature reveals that ensuring cloud
\ac{LLM} dependability requires an
integrated approach spanning physical hardware,
statistical distributions,
semantic reasoning, and cluster orchestration:
lightweight checksums and activation
profiling detect silicon-level transient faults;
probabilistic envelopes and runtime
monitors (\ac{HIP}-\ac{LLM}, \ac{V-and-V})
bound distributional uncertainty;
semantic clustering and self-correction
(Semantic Entropy, \ac{CoVe}) eliminate hallucinations;
and deterministic guardrails
paired with \ac{AIOps} (MetaKube)
ensure robust infrastructure self-healing.
A critical open research
challenge is cross-layer fault propagation:
investigating how hardware-level soft
errors affect semantic drift,
and developing unified fault-tolerance
frameworks capable of coordinating
recovery across all four architectural
tiers simultaneously.

\begin{table}
    \centering
    \caption{Comparative summary of key representative fault-tolerant cloud LLM serving mechanisms.}
    \label{tab:fault-tolerance-summary}
    \footnotesize
    \setlength{\tabcolsep}{2.5pt}
    \renewcommand{\arraystretch}{1.15}
    \begin{tabular}{@{}p{0.18\linewidth}p{0.17\linewidth}p{0.33\linewidth}p{0.24\linewidth}@{}}
        \toprule
        \textbf{System}                               & \textbf{Defense Layer}     & \textbf{Key Mechanism}                                                                       & \textbf{Reported Outcome}                                                    \\
        \midrule
        FT2 \citep{p_ft2}                             & Hardware (\ac{SDC})        & First-token trajectory profiling with layer-selective checksum protection                    & Comprehensive transient fault recovery at low execution overhead             \\
        ATTNChecker \citep{p_attnchecker}             & Hardware (\ac{SDC})        & Fused mathematical checksum relations tailored for multi-head attention                      & Near-zero latency overhead fault detection in attention kernels              \\
        Dr.~DNA \citep{p_drdna}                       & Hardware (\ac{SDC})        & Neuron activation distribution profiling and runtime bound checking                          & Detects soft data corruption without low-level arithmetic hooks              \\
        RetryTrigger \citep{p102}                     & Hardware (\ac{SDC})        & \ac{LightGBM} anomaly detection on runtime decoding features and localized retry             & Up to 95.33\% fewer \acp{SDC} at 4.12\% average overhead                     \\
        SpotServe \citep{p47}                         & Hardware (Instance)        & Dynamic reparallelization and lightweight \ac{KV}-cache checkpointing                        & Recovers from cloud \ac{VM} preemption within seconds                        \\
        Beluga \citep{p79} / Wang et al.\ \citep{p75} & Hardware (Statefulness)    & \ac{CXL} shared disaggregated memory pooling and multi-tier storage                          & Sub-millisecond context failover across host boundaries                      \\
        \midrule
        \ac{HIP}-\ac{LLM} \citep{p101}                & Statistical Reliability    & Hierarchical imprecise-probability model of failure-free execution over $N$ requests         & Robust posterior reliability envelopes with confidence bounds                \\
        \ac{V-and-V} Surveys \citep{p103,p111}        & Formal Verification        & Systematized runtime-monitor techniques adapted from software and deep learning \ac{V-and-V} & Formal execution monitors bounding operational behavior                      \\
        \midrule
        Semantic Entropy \citep{p109}                 & Semantic Detection         & Bidirectional entailment clustering and entropy over semantic equivalence classes            & Identifies confabulations without requiring ground-truth verifiers           \\
        Chain-of-Verification \citep{p110}            & Semantic Self-Correction   & Four-stage draft, verification-question, independent-answer, and correction cycle            & Decreases hallucinations across closed-book and reasoning tasks              \\
        \midrule
        LLM4AIOps \citep{p108}                        & \ac{AIOps} Diagnosis       & Fine-tuned \ac{LLM} control agents over multimodal telemetry, traces, and Kubernetes logs    & Automates \ac{RCA}, anomaly triage, and predictive incident mitigation       \\
        \ac{STPA}/\ac{FRAM} \citep{p107}              & \ac{AIOps} Safety Critique & System-theoretic and functional-resonance safety assessment of \ac{LLM} agents               & Identifies context drift vulnerabilities during cascading failures           \\
        MetaKube \citep{p49}                          & Cloud-Native Remediation   & Experience-aware \ac{LLM} diagnostic engine paired with deterministic Kubernetes admission   & Decreases diagnosis latency and \ac{MTTR} safely                             \\
        \bottomrule
    \end{tabular}
\end{table}

\subsection{Agentic Workflows and Multi-Agent Systems}
\label{subsec:agentic-workflows-and-multi-agent-systems}
The rapid evolution of generative intelligence has
precipitated a fundamental paradigm shift from static,
single-turn query-response serving toward
autonomous agentic workflows and
collaborative \acp{MAS} \citep{multi-agent-systems-study}.
In conventional cloud-serving deployments,
inference requests are treated as stateless,
independent transactions routed to isolated worker
processes that terminate context upon response generation.
In sharp contrast,
an agentic workflow executes iterative
loops of autonomous reasoning,
action generation,
environment observation,
and self-reflection.
In these stateful loops,
the \ac{LLM} dynamically queries external tools,
invokes enterprise \acp{API},
accesses external knowledge stores,
and spawns dependent sub-tasks \citep{mult-agent-task-workflows}.
When multiple specialized agents collaborate
within a shared environment,
the computational complexity and execution dynamics
diverge radically from traditional microservice architectures.
Managing multi-turn state persistence,
dynamic execution graph topologies,
asynchronous agent-to-agent communication channels,
and compounded operational latencies requires fundamentally
re-architecting the underlying
cloud-serving infrastructure \citep{multi-agent-architecture}.
The following review systematizes communicative topologies,
cloud scheduling frameworks,
and operational vulnerability mitigation strategies,
before summarizing representative systems in \zcref{tab:agentic-systems-summary}.

\subsubsection{Architectural Paradigms: Communicative Topologies and Workflow Decomposition}
\label{subsubsec:agentic-paradigms-and-topologies}

The design of agentic workflows centers on structuring
how autonomous entities decompose
high-level objectives into executable operations.
Early autonomous agent frameworks relied
on monolithic prompt-chaining loops,
such as ReAct and Reflexion,
where a single model instance sequentially plans,
acts, and evaluates feedback.
However, as demonstrated by \citet{agentic-workflow-in-llms},
monolithic single-agent workflows exhibit severe
performance degradation when confronted
with multi-step reasoning horizons,
frequently suffering from hallucination spirals,
unbounded execution loops,
and cognitive overload.
To overcome these limitations, \citet{agentic-workflow-in-llms}
formalize four foundational agentic design patterns,
namely reflection, tool use, planning,
and multi-agent collaboration,
demonstrating that decomposing complex tasks
across structured agent roles systematically
enhances solution accuracy and operational robustness.

In multi-agent collaborative environments,
orchestration topologies dictate
communication dynamics and synchronization overheads across cloud instances.
A prominent architectural
paradigm is the communicative agent collective,
exemplified by ChatDev \citep{chatdev}.
ChatDev structures software engineering
workflows through the lens of
human organizational hierarchies and \acp{SOP}.
By decomposing the software development lifecycle into sequential phases,
such as designing, coding, testing, and documenting,
ChatDev establishes communicative chat chains
where pairs of specialized agent personas
(e.g., chief executive officer, programmer, and code reviewer)
engage in multi-turn dialogues to reach consensus.
Each phase utilizes localized communicative reflection,
enabling agents to iteratively verify intermediate artifacts,
detect syntax anomalies, and rectify algorithmic
logic before advancing the global workflow state.
This communicative chat-chain topology demonstrates
that role specialization and peer review mechanisms
substantially reduce task-level hallucination
rates while keeping single-turn context windows bounded.

Beyond software development,
multi-agent orchestration has expanded into complex cyber-physical,
autonomous systems,
and industrial engineering domains.
\citet{shih-sac25} propose a multi-agent framework tailored
for multi-objective \ac{DSE} in autonomous driving systems.
Their cloud-native architecture partitions
the search space among specialized agent personas,
including parameter interpreters,
system configuration generators,
simulation evaluators,
and Pareto-optimality analyzers.
By leveraging structured inter-agent dialogue
to navigate complex trade-off surfaces
(e.g., inference latency, trajectory safety, and thermal limits),
the multi-agent framework identifies Pareto-optimal
configurations significantly faster than
classical evolutionary or Bayesian search heuristics.
Similarly, \citet{huang-smc25} introduce an
\ac{LLM}-driven hierarchical multi-agent
workflow for mission control
in unmanned autonomous platforms.
Their system decomposes high-level
mission goals into granular sub-tasks,
employing a centralized mission planner
agent that dynamically routes
execution directives to specialized operational agents
(e.g., path planning, payload management,
and environmental sensing).
By incorporating feedback loops between
physical telemetry streams and reasoning modules,
the architecture achieves real-time adaptation
against unexpected environmental disturbances.
In parallel, \citet{lim-case24} explore the
integration of \acp{MAS} within smart
manufacturing and industrial automation.
Their framework maps high-level production
schedules into deterministic execution
graphs executed by distributed robot controllers,
proving that semantic reasoning combined with
strict validation protocols bridges the gap
between probabilistic generative
intelligence and safety-critical physical processes.

At cluster and network scales, \citet{lee-commmag25}
extend multi-agent architectures to decentralized network management.
By deploying distributed \ac{LLM} agents
across heterogeneous network switches,
cloud gateways, and edge controllers,
the authors demonstrate that communicative
agent teams can autonomously
optimize dynamic routing tables,
mitigate traffic congestion spikes,
and coordinate load distribution
without relying on a single,
fragile centralized orchestrator.
These diverse implementations underscore
that communicative topologies,
whether organized as hierarchical trees,
sequential chat chains,
or decentralized peer-to-peer federations,
enable complex collective problem-solving
that single-instance models cannot match.

\subsubsection{Cloud Scheduling, Resource Allocation, and Memory Fabrics}
\label{subsubsec:agentic-scheduling-and-memory}

While communicative workflows
offer superior reasoning capabilities,
their deployment across cloud infrastructure
introduces severe scheduling and
resource provisioning bottlenecks \citep{p7}.
Traditional cloud orchestrators schedule
workloads based on predictable resource footprints (e.g.,
fixed \ac{CPU}, \ac{GPU} memory, and execution deadlines).
Agentic workflows, however, generate non-deterministic
execution trees characterized by variable loop iterations,
sporadic external \ac{API} calls,
and branching sub-agent invocations.
During external tool execution or
human-in-the-loop verification \citep{human-in-the-loop},
the underlying \ac{LLM} agent enters an idle or waiting state,
leading to severe \ac{GPU} underutilization
if hardware accelerators
remain statically bound to the container.

To address these inefficiencies,
dynamic task scheduling tailored specifically
for agentic and \ac{LLM}-driven workloads is essential.
\citet{p19} investigate
\ac{LLM}-based cost-aware task scheduling within cloud computing systems.
They formulate an optimization
framework that evaluates both direct computing costs
(e.g., instance runtime leasing and token processing fees)
and communication overheads across heterogeneous cloud nodes.
By utilizing an \ac{LLM} to model the complex
interdependencies of multi-task workflows
and predict task execution runtimes,
their scheduler dynamically places dependent
sub-tasks across available cloud resources
to minimize overall financial expenditure
while satisfying strict end-to-end makespan deadlines.
Such intelligent scheduling frameworks
illustrate that agentic workflows
must be co-designed with the cloud control plane,
dynamically deallocating and re-attaching
\ac{GPU} resources across active reasoning
phases and passive input/output waiting periods.

Equally critical is the design of
persistent memory fabrics \citep{agents-memory-context}.
Unlike single-turn inference where
context can be discarded immediately,
an agent collective requires multi-tiered
memory architectures consisting
of at least the following three primary layers:

\begin{enumerate}[(i)]
    \item \textit{Working Memory:} Active context windows
    maintained in ephemeral \ac{GPU} or host memory,
    housing current dialogue history,
    active tool call schemas,
    and immediate intermediate reflections.
    \item \textit{Episodic Memory:} High-dimensional vector
    embeddings stored in scalable cloud vector databases,
    enabling agents to retrieve relevant historical interactions,
    past execution trajectories,
    and lessons learned across preceding execution
    epochs through semantic similarity search.
    \item \textit{Semantic Memory:} Structured knowledge
    graphs and relational stores encapsulating
    domain guidelines, static schemas,
    and enterprise policies that
    remain invariant across workflow runs.
\end{enumerate}

Synchronizing these memory fabrics across
distributed agent workers requires
efficient state-checkpointing mechanisms.
If an agent pod experiences a
hardware failure or
container eviction mid-execution \citep{midway-failure},
the orchestrator must restore
the agent state without
re-executing previous \ac{LLM} calls,
which would incur substantial
financial cost and latency penalties.

\subsubsection{Operational Vulnerabilities: Context Drift, Cascading Errors, and Systemic Bottlenecks}
\label{subsubsec:agentic-vulnerabilities}

Deploying agentic systems in production
cloud environments exposes several
architectural failure modes and systemic
vulnerabilities that do not exist in conventional microservice pipelines.
The most pervasive failure mode is cascading error propagation.
In multi-agent pipelines where the output of
an upstream agent serves directly as
the context and instruction for a downstream agent,
minor semantic inaccuracies or
hallucinations compound exponentially \citep{p109, p110}.
If a planning agent generates a subtly flawed sub-goal,
downstream code generation, validation,
and execution agents will exhaust
extensive compute cycles
attempting to fulfill an unviable trajectory.

A related challenge is context drift
and token explosion \citep{context-drift}.
As agents engage in multi-turn dialogues,
repeated tool feedback,
trace outputs, and inter-agent messages rapidly saturate the model
context window \citep{context-shit-and-hallucination}.
Long-context saturation degrades the
reasoning capabilities of the underlying \ac{LLM},
leading to instruction-following failure,
repetitive tool invocation loops,
and dramatic increases in latency and operational cost.
Without automated summarization,
selective context pruning \citep{token-pruning},
and deterministic loop-detection circuit breakers,
multi-agent workflows can become trapped in
circular ping-pong conversations that silently drain cloud \ac{API} quotas.

Finally, security perimeters face
unique threats in agentic clouds.
Because agents possess autonomy to invoke shell commands,
write code, and access internal database endpoints,
they present an expanded attack surface
for indirect prompt injection and tool
privilege escalation \citep{prompt-injection-attack}.
If untrusted external data retrieved
via a tool call contains malicious embedded instructions,
the agent may hijack the workflow,
exfiltrating internal credentials or
altering downstream system configurations \citep{work-flow-hijacking}.
Securing agentic clouds therefore
demands strict privilege sandboxing,
read-only permission envelopes,
deterministic policy verification guards,
and complete audit logging of every
inter-agent message and tool
invocation \citep{case-study-on-prompt-qa}.

\zcref{tab:agentic-systems-summary} provides
a comparative synthesis of key representative
agentic workflows and \acp{MAS} deployed within cloud ecosystems.
Analyzing these works reveals three central architectural patterns.
First, communicative topologies dictate both
reasoning capability and cloud communication overheads.
Sequential chat chains such as ChatDev \citep{chatdev}
and structured workflow patterns \citep{agentic-workflow-in-llms}
excel at phased, reflective problem solving,
whereas hierarchical planner-worker frameworks \citep{huang-smc25}
and decentralized collectives \citep{lim-case24,lee-commmag25}
provide the flexibility needed for dynamic physical telemetry
and distributed network control.
Second, agentic workloads fundamentally disrupt
conventional static resource provisioning.
Because agents alternate between compute-heavy reasoning
and passive waiting on external tools or environment feedback,
static accelerator binding yields severe underutilization.
Intelligent, dependency-aware schedulers \citep{p19}
paired with tiered memory fabrics (working, episodic, and semantic)
are therefore essential for balancing compute costs against makespan deadlines.
Finally, operational dependability requires systematic safeguards.
Cascading hallucination compounding across agent turns,
context drift from unbounded history growth,
and prompt injection risks during tool invocations
necessitate rigorous privilege sandboxing,
context pruning, and deterministic runtime guardrails.

\begin{table}
    \centering
    \caption{Comparative summary of key representative agentic workflows and multi-agent systems in cloud environments.}
    \label{tab:agentic-systems-summary}
    \footnotesize
    \begin{tabular}{p{2.2cm}p{2.3cm}p{3.8cm}p{3.5cm}}
        \toprule
        \textbf{Study}                   & \textbf{Topology}              & \textbf{Primary Focus}                                                 & \textbf{Key Architectural Mechanism}                                         \\
        \midrule
        \citet{chatdev}                  & Sequential chat chain          & Collaborative software engineering lifecycle                           & Role-playing communicative reflection across multi-phase \acp{SOP}           \\
        \cmidrule(lr){1-4}
        \citet{agentic-workflow-in-llms} & Modular workflow patterns      & Pattern formalization and workflow benchmarking                        & Deconstructs workflows into reflection, tool use, planning, and consensus    \\
        \cmidrule(lr){1-4}
        \citet{shih-sac25}               & Multi-role collaborative       & Multi-objective \ac{DSE} in autonomous driving                         & Partitioned agent dialogue optimizing Pareto-trade-off search spaces         \\
        \cmidrule(lr){1-4}
        \citet{huang-smc25}              & Hierarchical planner-worker    & Mission planning and telemetry adaptation for autonomous platforms     & Closed-loop feedback between physical telemetry and semantic sub-task graphs \\
        \cmidrule(lr){1-4}
        \citet{lim-case24}               & Decentralized industrial       & Smart manufacturing and robotic process control                        & Translates semantic goals into deterministic execution graphs for robots     \\
        \cmidrule(lr){1-4}
        \citet{lee-commmag25}            & Decentralized collective       & Scalable autonomous network management and traffic optimization        & Distributed peer agents coordinating real-time routing and load balancing    \\
        \cmidrule(lr){1-4}
        \citet{p19}                      & \ac{LLM}-driven task scheduler & Cost-aware task scheduling in heterogeneous clouds                     & Evaluates compute costs, communication latency, and task dependency graphs   \\
        \bottomrule
    \end{tabular}
\end{table}

\subsection{AI Gateways and Traffic Routing}
\label{subsec:ai-gateways-and-traffic-routing}
The rapid proliferation of generative foundation models
within enterprise cloud ecosystems has
elevated the \ac{AI} Gateway from a passive
network ingress into a mission-critical
control plane \citep{ai_gateway_routing_review}.
In classical microservice deployments,
Layer-4 and Layer-7 reverse proxies (e.g., NGINX, Envoy, or HAProxy)
distribute traffic using uniform connection metrics,
such as round-robin, least-connections, or static path matching.
However, these conventional routing policies
fail catastrophically when applied to generative \ac{LLM} serving.
Generative workloads exhibit extreme operational
asymmetry that invalidates traditional
assumptions of uniform request
execution \citep{p38, llm_inference_asymmetry_2024}:

\begin{enumerate}[I.]
    \item \textit{Token Length Asymmetry:} The computational
    and memory cost of an
    \ac{LLM} request is governed by the length
    of its input prompt and generated response,
    meaning two requests routed to identical hardware
    can differ in processing duration
    by multiple orders of magnitude \citep{llm_workload_heterogeneity_latest}.

    \item \textit{Phase Asymmetry:} As detailed in previous sections,
    the initial prefill phase is heavily
    compute-bound and memory-intensive,
    whereas subsequent autoregressive token decoding
    is strictly memory-bandwidth-bound and sequential \citep{p38, gateway_kv_routing_latest}.

    \item \textit{Cache Locality and Prefix Affinity:} Incoming requests
    frequently share long common prompt prefixes,
    such as system instructions,
    retrieved documents in \ac{RAG} pipelines,
    or multi-turn conversation histories.
    Routing a request to a worker node that already caches
    the corresponding \ac{KV} tensors eliminates
    the entire prefill phase,
    reducing \ac{TTFT} from seconds to milliseconds.
    Conversely, naive load balancing scatters
    identical prefixes across disparate workers,
    causing redundant compute duplication
    and thrashing \ac{GPU} memory \citep{kv_prefix_affinity}.

    \item \textit{Dual Latency \acp{SLO}:} Cloud providers must concurrently
    guarantee independent thresholds for \ac{TTFT} and \ac{ITL},
    which frequently demand contradictory scheduling decisions \citep{p2}.
\end{enumerate}

Consequently, modern cloud architectures require specialized,
context-aware \ac{AI} gateways capable of dynamic prefix tracking,
\ac{KV}-cache affinity routing, multimodel cascading,
and predictive queue management \citep{gateway_kv_routing_latest}.
We organize the following discussion around prefix-aware dispatching,
heterogeneous model cascades, and ingress security perimeters,
culminating in a comparative synthesis in \zcref{tab:ai-gateway-summary}.

\subsubsection{Prefix-Aware Dispatching and KV-Cache Centric Routing}
\label{subsubsec:prefix-aware-routing}

To maximize hardware throughput
and satisfy stringent \acp{SLO},
the frontline defense of
the \ac{AI} gateway is prefix-aware dispatching.
Under this paradigm,
the gateway inspects incoming prompt
tokens or computes cryptographic
hash digests over prompt segments
(e.g., system prompts and few-shot exemplars)
to match requests against the
cached \ac{KV} inventories of active worker instances.

A prominent breakthrough in cache-centric
architecture is Mooncake \citep{mooncake-tos25}.
Mooncake reorganizes the entire
cloud serving fabric around a distributed,
hierarchical \ac{KV} cache
store spanning \ac{GPU} high-bandwidth memory,
host \ac{CPU} memory,
and high-speed local solid-state drives.
The gateway control plane within Mooncake
maintains a global metadata registry of prefix locations.
When a request arrives,
the gateway executes cache-centric dispatching,
routing the query to the node containing
the longest common prefix match.
If the matched node lacks sufficient compute capacity for generation,
Mooncake leverages high-speed \ac{RDMA}
to stream the cached \ac{KV} tensors to an idle decoding node,
effectively decoupling cache
retention from compute execution.

To further alleviate network bottlenecks
during distributed cache migration,
CacheGen \citep{cachegen-sigcomm24}
introduces custom \ac{KV} cache compression and streaming protocols.
Because transferring raw 16-bit
floating-point \ac{KV} tensors across
cloud data center networks
can saturate cross-rack bandwidth,
CacheGen compresses \ac{KV} activations
using distributional quantization and
delta encoding tailored to the
mathematical structure of transformer attention layers.
By streaming compressed \ac{KV}
representations between gateway proxies and worker pools,
CacheGen enables gateways to dynamically
offload and rebalance requests across
distant servers without incurring prohibitive prefill latencies.

In retrieval-augmented workloads,
prompt prefixes do not follow static,
hierarchical templates but instead consist of unordered,
dynamically retrieved document chunks.
To resolve this limitation,
CacheBlend \citep{cacheblend-eurosys25}
provides an advanced knowledge fusion and routing mechanism.
When an \ac{AI} gateway routes a query
with multiple retrieved document contexts,
standard attention kernels would normally
require recomputing the full prefill
attention matrix due to cross-token positional encodings.
CacheBlend dynamically fuses pre-computed
\ac{KV} segments of disparate knowledge
chunks by selectively recomputing only
the cross-chunk attention dependencies,
allowing the gateway to achieve up to
80\% cache reuse even under
unpredictable retrieval combinations.

From a resource provisioning perspective,
\citet{p72} examine how \ac{AI} gateways can
optimize cloud \ac{VM} selection when
serving \ac{LLM} workloads with \ac{KV} cache offloading.
They observe that offloading cached
contexts from expensive \ac{GPU} memory
to cheaper host \ac{CPU} \ac{DRAM} creates
complex trade-offs between instance
leasing costs and memory bandwidth saturation.
Their framework dynamically steers incoming
requests toward optimal \ac{VM} configurations
(e.g., high-bandwidth \ac{PCIe} instances
versus memory-dense compute instances)
based on real-time prefix lengths
and expected generation durations,
drastically reducing total monetary cost
while preserving tight latency guarantees.

\subsubsection{Heterogeneous Model Cascades and QoS-Aware Dispatching}
\label{subsubsec:qos-and-cascading-routing}

Beyond routing requests for a single model family,
cloud \ac{AI} gateways increasingly
orchestrate heterogeneous model cascades.
In enterprise deployments,
queries range from simplistic factual lookups
to complex mathematical derivations.
Dispatching every transaction to a frontier,
parameter-dense model is economically unsustainable.
Consequently, the gateway acts as an
intelligent classifier and semantic router,
dispatching simple queries to compact,
low-cost \acp{SLM} and reserving massive
models for difficult prompts.

\citet{yang-tmc25} develop a comprehensive
\ac{QoS}-aware \ac{LLM} routing framework
designed for edge-cloud collaborative
computing environments featuring multiple expert models.
Their system constructs a multi-dimensional
routing policy that evaluates prompt semantic complexity,
available edge/cloud compute budgets,
and hard deadline constraints.
By formulating the dispatching challenge
as an online optimization problem,
their gateway dynamically decides whether to
execute inference locally on
resource-constrained edge servers
or forward the query to high-capacity cloud clusters,
guaranteeing end-to-end \ac{QoS}
compliance while minimizing
cloud egress and compute charges.

Within homogeneous or hybrid clusters,
handling mixed workloads composed of short
conversational exchanges and long-document
summarization presents severe head-of-line blocking.
\citet{p80} address this challenge by
introducing a performance-aware
\ac{LLM} load balancer for mixed workloads.
Their system monitors the real-time
execution states of distributed worker nodes,
tracking active prefill token volume
and autoregressive decoding throughput.
By predicting the queuing delay and
memory fragmentation induced by incoming requests,
their gateway dynamically directs
prefill-heavy requests away from
nodes actively generating long token streams,
preventing latency spikes and ensuring
consistent \ac{ITL} across all active sessions.

Complementing reactive balancing,
predictive traffic routing
anticipates future demand spikes.
\citet{p6} present SAGESERVE,
a forecast-aware auto-scaling and
traffic dispatching framework for cloud data centers.
SAGESERVE leverages predictive time-series
models to forecast incoming prompt
volumes and sequence lengths minutes in advance.
The gateway utilizes these forecasts to
proactively spin up worker instances and pre-warm model weights,
directing incoming request flows onto
pre-allocated hardware before queue
buildup can violate contractual service level agreements.

\subsubsection{Cluster Service Reliability, Ingress Protection, and Network Security}
\label{subsubsec:gateway-reliability-and-security}

As the single point of entry
into massive computing infrastructure,
the \ac{AI} gateway must enforce rigorous
reliability and defensive security controls.
The operational scale of modern \ac{LLM} clusters
creates substantial vulnerability to
hardware degradation and network partition events.
\citet{p100} conduct an extensive
empirical study on the service reliability
of intelligent computing cluster
systems serving \acp{LLM}.
Their analysis reveals that high-density
computing clusters experience frequent node dropouts,
silent network packet corruption,
and intermittent accelerator faults
during sustained high-throughput inference.
To maintain continuous availability,
the \ac{AI} gateway must
integrate real-time health checks,
rapid circuit-breaking policies,
and transparent request retry mechanisms
that divert active sessions around
failing accelerators without losing user state.

Simultaneously, the gateway functions as the
security perimeter defending serving
clusters against external exploitation.
\citet{p50} present a two-tier intrusion detection
framework tailored for containerized
environments orchestrated via \ac{SDN} and Kubernetes.
Their architecture deploys an \ac{LLM}-enhanced
inspection engine at the network
ingress to monitor traffic flow patterns,
detect anomalous packet payloads,
and identify \ac{DDoS} attempts designed to
exhaust cluster memory through prompt-bombing attacks.
By coupling \ac{SDN} flow control
with container-level telemetry,
the gateway dynamically isolates compromised
ingress pods and redirects legitimate traffic,
ensuring robust multi-tenant defense.

\zcref{tab:ai-gateway-summary} unites
representative \ac{AI} gateway and traffic routing architectures
across their routing dimensions,
target optimizations, and core dispatching mechanisms.
The literature demonstrates that the \ac{AI} gateway
has transformed from a conventional reverse proxy
into an active, stateful inference control plane.
Three dominant design principles characterize this transformation.
First, cache-centric prefix dispatching represents
the most effective architectural lever for latency reduction.
Frameworks such as Mooncake \citep{mooncake-tos25},
CacheGen \citep{cachegen-sigcomm24},
and CacheBlend \citep{cacheblend-eurosys25}
exploit prefix locality and compressed tensor streaming
to maximize \ac{KV}-cache reuse across distributed memory pools,
while InferSave \citep{p72} complements cache offloading
by optimizing underlying virtual machine selection.
Second, heterogeneous model cascades and workload-aware routing
address the economic and throughput limits of uniform serving.
By dynamically matching query complexity to model capacity \citep{yang-tmc25},
decoupling prefill from decoding streams to prevent head-of-line blocking \citep{p80},
and forecasting traffic surges to proactively warm instances \citep{p6},
modern gateways satisfy dual \ac{TTFT} and \ac{ITL} \acp{SLO}.
Finally, operational robustness requires tightly coupling traffic dispatching
with reliability and defense mechanisms.
By incorporating real-time health monitoring and rapid circuit-breaking \citep{p100},
alongside containerized intrusion detection \citep{p50},
modern gateways safeguard serving clusters against hardware faults,
performance degradation, and malicious ingress exploits.

\begin{table}
    \centering
    \caption{Comparative summary of representative AI gateway and traffic routing architectures.}
    \label{tab:ai-gateway-summary}
    \footnotesize
    \begin{tabular}{p{2.4cm}p{2.3cm}p{3.8cm}p{3.3cm}}
        \toprule
        \textbf{System / Architecture}                             & \textbf{Routing Dimension}   & \textbf{Target Optimization}                                 & \textbf{Core Dispatching Mechanism}                                                \\
        \midrule
        Mooncake \citep{mooncake-tos25}                            & Global cache affinity       & \ac{TTFT} reduction and disaggregated cluster throughput     & Routes queries to longest prefix match across multi-tier storage pools             \\
        \cmidrule(lr){1-4}
        CacheGen \citep{cachegen-sigcomm24}                        & Streaming \ac{KV} transfer  & Cross-node network bandwidth and dynamic load balancing      & Quantized and delta-encoded \ac{KV} cache streaming between workers                \\
        \cmidrule(lr){1-4}
        CacheBlend \citep{cacheblend-eurosys25}                    & Composable \ac{RAG} caching & Cross-document knowledge reuse in dynamic retrieval          & Fuses selective \ac{KV} chunks, recomputing only cross-chunk attention             \\
        \cmidrule(lr){1-4}
        InferSave \citep{p72}                                      & Hardware instance selection & Total monetary leasing cost with \ac{KV} offloading          & Evaluates \ac{PCIe} and memory bandwidth to steer requests to optimal \acp{VM}     \\
        \cmidrule(lr){1-4}
        \ac{DRL} based \ac{QoS} \ac{LLM} Router \citep{yang-tmc25} & Multi-expert edge-cloud     & \ac{QoS} compliance and latency deadline satisfaction        & Online optimization matching query complexity to edge versus cloud models          \\
        \cmidrule(lr){1-4}
        Heuristic-Guided \ac{DRL} \ac{LLM} Router \citep{p80}      & Mixed-workload balancing    & Dual-tier \ac{SLO} protection (\ac{TTFT} and \ac{ITL})       & Isolates prefill-heavy streams from active autoregressive decoding nodes           \\
        \cmidrule(lr){1-4}
        SAGESERVE \citep{p6}                                       & Predictive autoscaling      & Queue mitigation and proactive capacity reservation          & Time-series forecasting directing traffic onto pre-warmed instance pools           \\
        \cmidrule(lr){1-4}
        RDpipe \citep{p100}                                        & Cluster fault tolerance     & End-to-end service reliability under hardware faults         & Real-time health monitoring and transparent circuit-breaking failover              \\
        \cmidrule(lr){1-4}
        Two-Tier \ac{SDN} \citep{p50}                              & Ingress perimeter defense   & Containerized intrusion detection and \ac{DDoS} mitigation   & Two-tier \ac{SDN} and Kubernetes flow monitoring identifying anomaly bursts        \\
        \bottomrule
    \end{tabular}
\end{table}

\subsection{Confidential Computing and Privacy-Preserving AI}
\label{subsec:confidential-computing-and-privacy-preserving-ai}
The pervasive adoption of generative foundation models
within enterprise clouds is severely constrained by confidentiality,
compliance, and intellectual property concerns.
When organizations dispatch proprietary corporate data,
patented source code, or sensitive patient
health records to a public cloud \ac{LLM} service,
they operate within a shared multi-tenant
substrate that exposes data to several critical threat vectors:

\begin{enumerate}[a.]
    \item \textit{Untrusted Infrastructure Operators:}
    Cloud infrastructure providers,
    hypervisor software stacks,
    and rogue platform administrators with
    root privileges can inspect raw system memory,
    extracting proprietary prompt
    context or model weights.

    \item \textit{Adversarial Co-Tenants:} Malicious
    tenants sharing identical physical servers or
    accelerators can mount microarchitectural
    side-channel attacks, timing probes,
    or memory-bus sniffing to
    reconstruct private inferences.

    \item \textit{Model Weight Exfiltration:} From the
    model provider perspective,
    proprietary weights representing
    millions of dollars in pre-training
    compute are vulnerable to intellectual property
    theft when deployed on
    third-party cloud infrastructure.
\end{enumerate}

To resolve this multi-tenant privacy dilemma,
the cloud architecture community has
embraced Confidential Computing.
By establishing hardware-enforced \acp{TEE},
cloud systems can guarantee memory encryption,
integrity protection,
and cryptographic remote attestation.
However, adapting \acp{TEE} to generative
foundation models introduces
unprecedented architectural bottlenecks,
as multi-billion-parameter models easily
exceed conventional enclave memory boundaries
and suffer immense latency penalties
during encrypted data transport.
The following review examines
hardware-enforced accelerator enclaves,
collaborative fine-tuning and inference frameworks,
and emerging architectural privacy threats,
before presenting a comparative summary
in \zcref{tab:confidential-computing-summary}.

\subsubsection{Hardware-Enforced TEEs and Accelerator Enclaves}
\label{subsubsec:tee-accelerators}

The architectural landscape of confidential
machine learning is comprehensively
systematized by \citet{mo-csur24}.
Their taxonomy categorizes hardware-assisted
enclaves across classical \ac{CPU} architectures
(e.g., Intel \ac{SGX}, Intel \ac{TDX}, AMD \ac{SEV}, and Arm \ac{CCA})
and analyzes the transition
toward emerging \ac{GPU} \acp{TEE}.
While \ac{CPU} enclaves provide
mature memory isolation primitives,
their execution engines are fundamentally
unsuited for the dense matrix-multiplication
operations demanded by modern transformers,
creating a critical tension between
cryptographic security and computational throughput.

To quantify whether hybrid confidential
computing is viable for foundation models,
\citet{mohan-cloud24} conduct an empirical
evaluation of cloud-based \ac{CPU}-\ac{GPU}
confidential computing environments.
Benchmarking enterprise models deployed
within Intel \ac{TDX} host enclaves
coupled to NVIDIA Hopper H100 \ac{GPU} \acp{TEE},
the authors reveal that encrypted
\ac{PCIe} data transfers constitute
the primary performance bottleneck.
Because every activation tensor, weight block,
and \ac{KV} cache element must be encrypted
and cryptographically authenticated
before traversing the physical \ac{PCIe} bus,
standard inference pipelines experience substantial
throughput degradation compared to non-confidential baselines,
particularly at large batch
sizes where memory bus saturation occurs.

Providing a rigorous economic and performance characterization,
\citet{chrapek-iiswc25} evaluate end-to-end
confidential inference across both
\ac{CPU} and \ac{GPU} \acp{TEE} using the
LLaMA-2 model family up to 70-billion parameters.
Their experiments demonstrate that
for smaller models (e.g., 7B parameters),
modern \ac{CPU} enclaves equipped
with hardware matrix acceleration
(such as Intel \ac{AMX}) achieve viable latencies at
significantly lower financial leasing
costs than dedicated \ac{GPU} instances.
However, for large-scale 70B models,
\ac{CPU} enclaves suffer from memory-bandwidth starvation,
rendering \ac{GPU} \acp{TEE} necessary despite
higher hardware acquisition expenses
and \ac{PCIe} encryption penalties.
Their findings provide actionable sizing envelopes
for cloud architects designing
confidential \ac{LLM} infrastructure.

To overcome the severe latency degradation
caused by encrypted \ac{PCIe} data movement,
\citet{pipellm-asplos25} introduce PipeLLM.
PipeLLM observes that in autoregressive generation,
memory transfers between the host
system and the \ac{GPU} enclave
follow deterministic scheduling dependencies.
By introducing speculative pipelined encryption,
PipeLLM anticipates upcoming tensor access patterns,
pre-encrypting and staging required model layers
and \ac{KV} activations across
the \ac{PCIe} bus concurrently
with active \ac{GPU} computation.
This asynchronous cryptographic overlap
successfully masks transfer latency,
allowing PipeLLM to achieve near-native
serving throughput across large models
up to 175-billion parameters without
compromising hardware-enforced isolation guarantees.

At the client and mobile edge,
protecting proprietary models deployed in decentralized
environments presents distinct architectural constraints.
\citet{wang-tzllm-eurosys26} present TZ-\acs{LLM},
a framework designed to protect
on-device \acp{LLM} using Arm TrustZone.
Because mobile TrustZone memory partitions
are constrained to mere megabytes,
storing full foundation models inside
the secure enclave is physically impossible.
TZ-\acs{LLM} resolves this restriction through a minimal
neural processing unit co-driver split between the
untrusted rich execution environment and the secure world.
By employing pipelined parameter restoration
and on-the-fly decryption of model layers during forward passes,
TZ-\acs{LLM} prevents physical memory dump attacks on
proprietary edge weights while
introducing minimal latency overhead.

\subsubsection{Collaborative Inference, Enclave Fine-Tuning, and Prompt Sanitization}
\label{subsubsec:collaborative-inference-and-sanitization}

Beyond monolithic enclave execution,
hybrid architectures partition computational
workflows across trusted enclaves and
untrusted cloud workers to optimize performance.
For privacy-preserving parameter-efficient fine-tuning,
\citet{lin-tsc26} propose PriPTune.
Recognizing that full-parameter fine-tuning
inside \ac{CPU} \acp{TEE} is computationally intractable,
PriPTune leverages a collaborative \ac{TEE}-\ac{GPU}
architecture for \ac{LoRA} adaptation.
The framework maintains all sensitive
enterprise training samples, activations,
gradients, and low-rank adapter matrices
securely inside \ac{CPU} \ac{TEE} enclaves.
Meanwhile, computationally intensive
matrix multiplications over frozen,
non-sensitive base weights are offloaded to untrusted,
high-performance cloud \acp{GPU}
using block-mixed tensor transformations.
By integrating blockchain smart
contracts to verify execution integrity,
PriPTune enables organizations to
fine-tune proprietary adapters
on untrusted cloud infrastructure
without risking corporate data exfiltration.

In distributed enterprise environments,
orchestrating containerized secure
workflows across heterogeneous
cloud providers is critical.
\citet{p53} investigate the deployment
of secure \ac{LLM} workflows
across multi-cloud infrastructures.
Their framework standardizes
containerized deployment pipelines,
enforcing encrypted inter-pod communication,
ephemeral secret management,
and zero-trust network policies that
prevent lateral data leakage
across untrusted cloud segments.

Complementing hardware-based isolation,
algorithmic data protection at the ingress boundary
provides a defense-in-depth layer against cloud provider snooping.
\citet{p124} develop Pr$\epsilon\epsilon$mpt,
a framework that sanitizes sensitive user prompts
before they are dispatched to public foundation model endpoints.
Pr$\epsilon\epsilon$mpt employs local differential
privacy and semantic entity replacement to identify
and obscure personally identifiable information
and proprietary enterprise entities within prompt texts.
By replacing sensitive tokens with mathematically
calibrated synthetic substitutes while preserving semantic coherence,
Pr$\epsilon\epsilon$mpt allows cloud \acp{LLM} to
generate accurate completions without
ever receiving raw sensitive attributes.

Finally, cloud architects must address
covert architectural vulnerabilities
unique to optimized \ac{LLM} serving engines.
As uncovered by \citet{p121} in their analysis
of \ac{KV}-cache vulnerabilities (termed Shadow in the Cache),
popular multi-tenant optimization techniques
such as cross-request prefix caching
introduce severe privacy vulnerabilities.
When multiple tenants share an underlying
serving instance that caches common prompt prefixes,
an adversarial tenant can measure inference execution
latency to infer whether another
tenant has recently queried specific
documents or private tokens.
\citet{p121} demonstrate that timing-based
side-channel probing allows attackers
to reconstruct private conversational
context with high fidelity.
Mitigating these risks requires cloud
providers to strictly isolate \ac{KV}
caches across tenant security domains,
enforcing cryptographic access controls
and time-padding defenses at the gateway dispatcher.

\zcref{tab:confidential-computing-summary} consolidates
representative confidential computing systems
and privacy-preserving architectures
across hardware trust bases,
workload targets, and core security mechanisms.
Three architectural insights emerge
from the surveyed literature.
First, hardware-enforced accelerator enclaves
face severe bandwidth and capacity constraints.
While \ac{CPU} enclaves offer cost-effective execution
for smaller models \citep{chrapek-iiswc25},
scaling to foundation models demands \ac{GPU} \acp{TEE}
where encrypted \ac{PCIe} bus transfers create
significant overheads \citep{mo-csur24,mohan-cloud24}.
Pipelined speculative pre-encryption in
PipeLLM \citep{pipellm-asplos25}
and on-the-fly parameter restoration in
TZ-\acs{LLM} \citep{wang-tzllm-eurosys26}
demonstrate that overlapping cryptographic
transfers with forward execution
is essential for achieving near-native serving throughput.
Second, collaborative partitioning
and ingress sanitization
provide flexible alternatives to
monolithic enclave execution.
Frameworks such as PriPTune \citep{lin-tsc26}
isolate sensitive adapters within secure enclaves
while delegating compute-intensive tensor
operations to untrusted cloud accelerators,
and algorithmic defenses like Pr$\epsilon\epsilon$mpt \citep{p124}
sanitize prompts before cloud transmission.
Finally, hardware isolation alone does not eliminate
algorithmic and cache side channels.
The timing vulnerabilities demonstrated
by \citet{p121} in shared \ac{KV} caches
underscore that architectural
optimizations must be co-designed
with cryptographic access boundaries,
mandating tenant-level cache isolation
and zero-trust policies \citep{p53} in multi-tenant clouds.

\begin{table}
    \centering
    \caption{Comparative summary of representative confidential computing and privacy-preserving AI systems.}
    \label{tab:confidential-computing-summary}
    \footnotesize
    \begin{tabular}{p{2.3cm}p{2.2cm}p{3.8cm}p{3.4cm}}
        \toprule
        \textbf{Study}                   & \textbf{Hardware and Trust Base}         & \textbf{Target Workload}                  & \textbf{Core Mechanism and Protection}                                                 \\
        \midrule
        \citet{mo-csur24}                & Multi-\acs{TEE} taxonomy                 & Comprehensive machine learning workflows  & SoK covering \ac{CPU} enclaves, \ac{GPU} \acp{TEE}, and side-channel threats           \\
        \cmidrule(lr){1-4}
        \citet{mohan-cloud24}            & Intel \ac{TDX} and NVIDIA H100           & Enterprise \ac{LLM} cloud inference       & Quantifies \ac{PCIe} encryption latency and batching bottlenecks in \ac{GPU} \acp{TEE} \\
        \cmidrule(lr){1-4}
        \citet{chrapek-iiswc25}          & Intel \ac{TDX} and NVIDIA H100           & LLaMA-2 inference scaling up to 70B       & Architectural cost-performance analysis of \ac{CPU} \ac{AMX} and \ac{GPU} \acp{TEE}    \\
        \cmidrule(lr){1-4}
        \citet{pipellm-asplos25}         & \ac{GPU} \ac{TEE} (Hopper)               & Large-scale \ac{LLM} serving (up to 175B) & Speculative pipelined pre-encryption masking \ac{PCIe} bus data transfers              \\
        \cmidrule(lr){1-4}
        \citet{wang-tzllm-eurosys26}     & Arm TrustZone                            & On-device mobile \ac{LLM} inference       & Pipelined parameter restoration protecting weights with minimal enclave \ac{RAM}       \\
        \cmidrule(lr){1-4}
        \citet{lin-tsc26}                & \ac{CPU} \ac{TEE} and Untrusted \ac{GPU} & Collaborative \ac{LoRA} fine-tuning       & Keeps adapter states in enclave while offloading tensor operations to \acp{GPU}        \\
        \cmidrule(lr){1-4}
        \citet{p53}                      & Multi-cloud containers                   & Enterprise multi-cloud \ac{LLM} pipelines & Containerized zero-trust deployment and inter-pod cryptographic isolation              \\
        \cmidrule(lr){1-4}
        \citet{p124}                     & Algorithmic and Ingress                  & Cloud prompt dispatching                  & Differential privacy and entity substitution sanitizing sensitive prompts              \\
        \cmidrule(lr){1-4}
        Shadow in the Cache \citet{p121} & Shared \ac{KV} Cache                     & Multi-tenant serving clusters             & Characterizes cache timing side-channels and mandates tenant-level isolation           \\
        \bottomrule
    \end{tabular}
\end{table}

    \section{Discussion}
    \label{sec:discussion}
\begin{figure}
    \centering
    \includegraphics[scale=0.75]{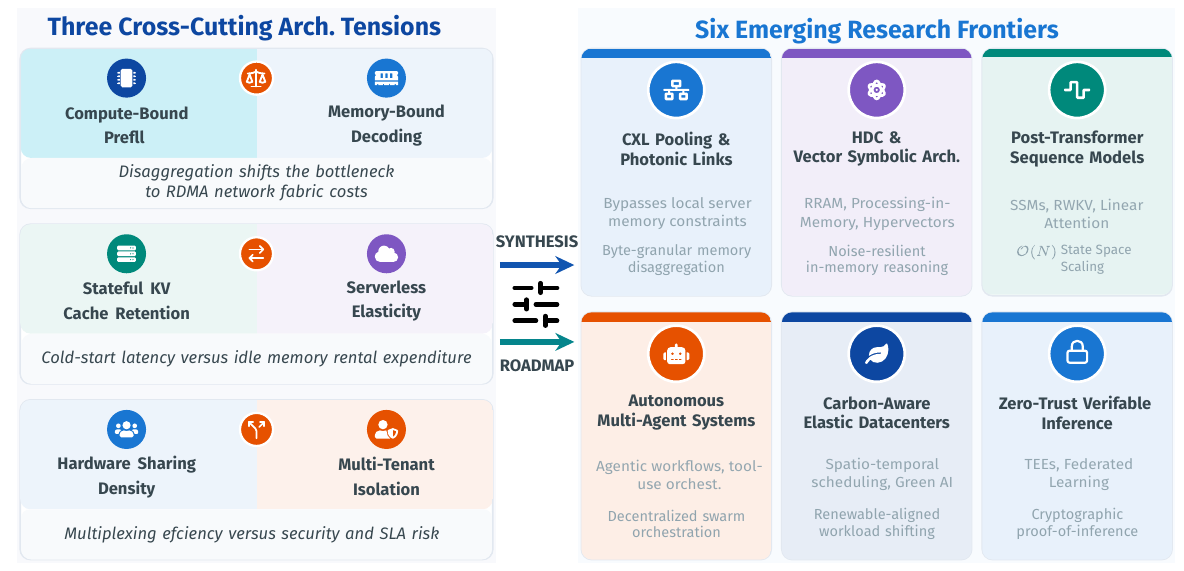}
    \caption{
        A synthesized roadmap illustrating the intersection
        of core architectural tensions in cloud foundation
        models and the emerging hardware and
        algorithmic frontiers they inform.
    }
    \label{fig:discussion_roadmap}
\end{figure}

    This section synthesizes the
core architectural paradigms,
resource scheduling frameworks,
and inference serving optimizations
analyzed across the survey.
We first provide a holistic
synthesis of cross-cutting design tensions
that govern modern cloud-hosted
foundation model ecosystems,
directly connecting our findings
to the four primary research questions
formulated in \zcref{subsec:method-rq}.
We subsequently present a forward-looking research roadmap
outlining promising hardware and algorithmic horizons,
including disaggregated memory fabrics,
\ac{HDC} \citep{hdc-latest-review,
    hdc-rram-review-cloud-with-potential},
and autonomous multi-agent orchestration.
Finally, we critically delineate the
methodological limitations of this study
and outline concrete directions
for future systematic reviews.
\zcref{fig:discussion_roadmap} visually
consolidates this synthesis,
illustrating how core architectural trade-offs intersect
with emerging cloud systems frontiers.

\subsection{Architectural Synthesis and Cross-Cutting Trade-offs}
\label{subsec:discussion-synthesis}
The operationalization of foundation
models within cloud datacenters
has triggered a fundamental shift
from conventional homogeneous compute clusters
toward specialized, disaggregated,
and memory-centric execution architectures.
Across \zcref{sec:orchestration},
\zcref{sec:inference-in-cloud},
and \zcref{sec:architecture},
our systematic analysis reveals
that optimizing cloud \ac{LLM} systems
is fundamentally governed by multi-objective design tensions
rather than isolated algorithmic improvements.
In this subsection, we synthesize
our answers to \ac{RQ}1 through \ac{RQ}4
by analyzing three primary architectural trade-offs that dictate
modern cloud serving efficiency, elasticity, and reliability.

\subsubsection{Compute Saturation versus Memory-Bandwidth Bottlenecks}
\label{subsubsec:tradeoff-compute-memory}

Addressing both \ac{RQ}1 (resource scheduling)
and \ac{RQ}2 (inference efficiency),
the reviewed literature demonstrates
that the generative execution pipeline
is partitioned by a deep operational
asymmetry between the prompt prefill phase
and the autoregressive token decoding phase.
The prefill phase processes all
prompt tokens concurrently,
forming compute-bound matrix multiplications
that fully saturate \ac{GPU} tensor cores.
Conversely, the autoregressive decoding
phase generates tokens sequentially,
retrieving substantial parameter weights
and expanding \ac{KV} cache states
at arithmetic intensities that fall
well below the operational
roofline of modern accelerators.

When these heterogeneous phases are
colocated on identical physical accelerators,
large batch prefill operations
introduce head-of-line blocking,
which severely inflates \ac{TBT} and
violates latency \acp{SLO} for
concurrent decoding queries.
To resolve this structural conflict,
pioneering frameworks such as Splitwise \citep{p37},
DistServe \citep{p38}, and Mooncake \citep{p41}
disaggregate execution onto specialized physical pools.
However, disaggregation transforms
an on-chip memory bandwidth bottleneck
into an inter-node datacenter networking bottleneck.
Transferring multi-gigabyte \ac{KV}
caches across prefill and decoding pools
requires dedicated high-bandwidth \ac{RDMA} fabrics,
such as \ac{RoCE} or InfiniBand.
As demonstrated by BanaServe \citep{p26},
under heavy multi-tenant request arrival rates,
the network transmission latency
incurred during \ac{KV} cache transfer
can approach or even exceed the execution
savings achieved by phase disaggregation.
Consequently, cloud providers face
an inescapable architectural trade-off:
they must balance compute hardware
utilization and token generation throughput
against the monetary and physical interconnect
capital required to sustain low-latency state migration.

\subsubsection{Stateful KV Cache Retention versus Serverless Elasticity}
\label{subsubsec:tradeoff-state-elasticity}

Addressing \ac{RQ}3 (cloud-native serverless systems),
the prevailing cloud execution
paradigm has long relied on stateless,
ephemeral containers managed by
frameworks such as Kubernetes (\ac{K8s})
and \ac{FaaS} platforms \citep{faas-related-recent-paper}.
However, \ac{LLM} workloads
are inherently stateful,
characterized by tens of gigabytes of model weights
and dynamically expanding,
per-request \ac{KV} cache allocations.
This creates severe friction with
serverless scale-to-zero semantics.

When instances are dynamically
de-allocated to conserve idle compute costs,
subsequent requests incur
prohibitive cold-start latencies,
often spanning tens of seconds
to retrieve, deserialize,
and stage model parameters
into accelerator \ac{HBM} \citep{p55, p56}.
While tiered caching frameworks like FlashServe \citep{p55},
multicast streaming
pipelines like FaaScale \citep{p67},
and layer-level autoscalers like BlitzScale \citep{p68}
dramatically truncate initialization latency,
retaining warm model states and persistent \ac{KV} caches
inevitably consumes expensive host \ac{DRAM} and accelerator memory.
Furthermore, multi-turn conversational agents and \ac{RAG} workflows
amplify this tension by generating prolonged session states.
Systems such as SemShareKV \citep{p23} and Cache-Craft \citep{p69}
demonstrate that sharing cached prefix states across concurrent requests
yields significant reductions in \ac{TTFT}.
Nevertheless, persistent cache retention limits
the dynamic elasticity of the cloud fabric,
forcing cloud architects to negotiate an
operational compromise between cold-start avoidance,
memory rental expenditure,
and instantaneous scale-out agility.

\subsubsection{Hardware Sharing Density versus Multi-Tenant Isolation and Security}
\label{subsubsec:tradeoff-sharing-isolation}

Addressing \ac{RQ}1 (accelerator virtualization)
and \ac{RQ}4 (reliability and privacy),
datacenter operators seek to
maximize hardware sharing density
to amortize the immense capital
expenditure of enterprise accelerator infrastructure.
Fine-grained spatial partitioning via NVIDIA \ac{MIG},
software-level spatial multiplexing via \ac{CUDA} \ac{MPS},
and dynamic co-execution schedulers such as AntMan \citep{p113},
FaST-GShare \citep{p57}, and SLINFER \citep{p65}
achieve impressive hardware saturation
by interleaving compute-heavy and
memory-heavy kernels on shared hardware.

However, this spatial and temporal colocation
severely compromises tenant isolation.
Shared memory buses, caches, and power
domains create inter-tenant performance interference,
introducing tail latency jitter that
jeopardizes strict \ac{SLA} commitments.
More critically, multi-tenant sharing
exposes severe security and reliability vulnerabilities.
Recent studies highlight that shared \ac{KV} cache storage
and collocated execution engines create side-channel vectors
that enable malicious tenants to
reconstruct private prompt contexts \citep{p76}.
Furthermore, hardware faults and
\ac{SDC} originating within a single faulty compute unit
can propagate silently across
colocated tenant workloads,
corrupting intermediate embeddings
without raising hardware exceptions \citep{p50}.
While deploying hardware \acp{TEE}
(e.g., AMD SEV-SNP, Intel TDX, and NVIDIA Confidential Computing)
enforces cryptographic isolation and
protects model intellectual property \citep{p53},
it incurs non-trivial memory encryption
overheads and complicates dynamic page sharing.
Cloud providers must therefore
navigate a fundamental trade-off between
aggressive multiplexing efficiency
and provable tenant isolation.

\subsection{Future Study Prospects for LLMs in Cloud}
\label{subsec:discussion-future}
The rapid acceleration of generative
\ac{AI} deployment reveals critical horizons
where conventional cloud architectures
must evolve to sustain future computing demands.
Based on the synthesis of our surveyed literature,
we identify six pivotal research frontiers
that will shape the next generation
of cloud-scale foundation model systems.

\subsubsection{CXL Memory Pooling and Photonic Interconnects}
\label{subsubsec:future-cxl-photonics}

As model scales expand into
hundreds of billions of parameters,
physical accelerator memory boundaries
remain the primary scaling ceiling.
While high-bandwidth memory provides
exceptional on-chip throughput,
its capacity per package remains severely
constrained by thermal and physical limits.
Emerging standards such as \ac{CXL}
offer memory disaggregation across host processors,
enabling unified, shared physical memory
pools that can be accessed at byte granularity
with near-native latency \citep{p79}.
Future cloud schedulers must develop
\ac{CXL}-native memory orchestrators
capable of dynamically mapping and
evicting \ac{KV} caches across heterogeneous tiers,
spanning on-chip \ac{HBM}, \ac{CXL}-attached
\ac{DRAM}, and non-volatile flash storage.
Concurrently, replacing electronic packet-switched
network fabrics with optical circuit switching
and co-packaged silicon photonics will
fundamentally alter distributed
prefill and decoding pipelines,
reducing inter-node serialization
latency by orders of magnitude
and unlocking near-instantaneous
cross-accelerator parameter sharing.

\subsubsection{Hyperdimensional Computing and Vector Symbolic Architectures}
\label{subsubsec:future-hdc-vsa}

An emerging post-von Neumann computing
paradigm with substantial potential
for cloud-scale foundation model acceleration is \ac{HDC}
and \ac{VSA} \citep{hdc-og-study, hdc-2022-survey}.
By projecting continuous representations into ultra-wide,
pseudo-orthogonal binary or bipolar hypervectors,
\ac{VSA} algebraic operators enable single-pass,
noise-resilient symbolic binding and bundling
\citep{hdc-2022-survey-second-portion, hdc-diff-arc-comp}.
When integrated into cloud inference serving fabrics,
in-memory \ac{HDC} processing architectures \citep{in-memory-hd}
can offload repetitive associative \ac{KV} cache similarity matching
and exact-match classification kernels
\citep{different-hdc-classification-analysis}
directly onto ultra-low-power memristive
or photonic memory arrays.
This architectural hybridization offers
an attractive research path
to bypass memory-wall bottlenecks and
radically slash datacenter energy consumption.

\subsubsection{Post-Transformer and Linear-Complexity Sequence Architectures}
\label{subsubsec:future-post-transformer}

The quadratic time and memory complexity
of standard self-attention
imposes severe operational costs
during extended-context processing.
While attention-free and sub-quadratic architectures,
such as State Space Models (e.g., Mamba),
recurrent memory transformers,
and linear attention variants,
exhibit linear $\mathcal{O}(N)$
computational scaling,
their operational footprint in multi-tenant
cloud environments remains largely uncharacterized.
Unlike transformer decoders that
rely on expanding \ac{KV} caches,
state space architectures
maintain fixed-size hidden states,
fundamentally eliminating decoding memory bloat.
However, their operational profile shifts the
primary execution bottleneck back to compute throughput,
altering existing batching heuristics
and disaggregated prefill-decode paradigms.
Investigating specialized cloud orchestrators
that can transparently host,
co-locate, and dynamically route traffic
across hybrid transformer-Mamba model ensembles
represents a vital future research trajectory.

\subsubsection{Autonomous Multi-Agent Systems and Decentralized Swarm Orchestration}
\label{subsubsec:future-agentic-orchestration}

Modern cloud infrastructure is
predominantly coordinated
by centralized schedulers,
such as Kubernetes control
planes or custom cluster brokers.
As agentic workflows proliferate,
cloud systems must handle bursty, multi-round,
and nondeterministic interaction
patterns that overwhelm centralized controllers.
Future research must explore decentralized,
swarm-based orchestration architectures
where autonomous software agents
dynamically negotiate compute leases,
exchange intermediate context
embeddings via peer-to-peer side-channels,
and collectively optimize cluster
throughput without central coordination bottlenecks.
Developing formal verification
frameworks that bound cyclic execution,
detect cascading workflow deadlocks,
and enforce strict monetary budgets
across autonomous agent colonies
will be essential for robust
production deployment.

\subsubsection{Sustainable, Carbon-Aware, and Energy-Elastic Datacenters}
\label{subsubsec:future-green-serving}

The exponential surge in \ac{LLM} inference
requests poses severe environmental
and power grid challenges.
While static power capping and
\ac{DVFS}) offer localized remedies,
future cloud architectures must embrace
carbon-budgeted and energy-elastic
scheduling paradigms \citep{p9, p18}.
This involves real-time spatial
and temporal request shifting,
where non-urgent inference jobs
(such as batch summarization or offline code analysis)
are dynamically routed across
geographically distributed datacenters
to follow peaks in local renewable
solar and wind generation.
Furthermore, integrating embodied carbon
tracking and thermal-aware accelerator placement
into cluster schedulers will
transition cloud \ac{AI} platforms
from simple cost minimization
toward sustainable, climate-aligned operations.

\subsubsection{Zero-Trust, Verifiable, and Confidential Inference Fabrics}
\label{subsubsec:future-verifiable-inference}

As proprietary foundation models increasingly
interact with sensitive enterprise data,
ensuring end-to-end data confidentiality
and inference integrity is paramount.
Future cloud architectures
must integrate hardware \acp{TEE}
with cryptographic proof systems,
such as zero-knowledge proofs of inference
and verifiable machine learning \citep{zero-knowledge-ml-review},
to provably verify that a given output
was generated by an uncorrupted,
specific model checkpoint without disclosing
proprietary model weights or client prompts.
Overcoming the prohibitive computational latency
of cryptographic verification
and designing zero-trust container runtimes
capable of managing attestation
at microsecond serving scales constitute critical
systems challenges for secure enterprise cloud \ac{AI}.

\subsection{Limitations of this Study}
\label{subsec:discussion-limitations}
While this systematic review provides a rigorous
and holistic mapping of cloud \ac{LLM} architectures,
orchestration systems, and inference optimizations,
several methodological and structural
boundaries must be explicitly acknowledged.
Identifying these limitations
clarifies the operational scope of our findings
and contextualizes the state of current systems research.

\subsubsection{PRISMA Protocol Boundaries and Rapid Literature Turnover}
\label{subsubsec:limitations-prisma-turnover}

To guarantee reproducible and transparent evidence synthesis,
this survey strictly adhered to the \ac{PRISMA}
2020 protocol \citep{prisma-method-og-paper},
focusing on peer-reviewed academic literature
published up to our search cutoff date.
However, generative \ac{AI} systems research
moves at an unprecedented evolutionary velocity.
Numerous breakthrough frameworks
and systems engineering innovations
are initially disseminated through
open-source software repositories,
technical reports, and industrial white
papers that deliberately bypass traditional peer-review channels.
While our eligibility criteria incorporated
foundational non-peer-reviewed systems
that achieved widespread community adoption and verification,
certain cutting-edge industrial optimizations deployed
in bleeding-edge production environments
may postdate our synthesis window or remain
uncaptured by formal academic indexing databases.

\subsubsection{The Proprietary Datacenter Opacity Barrier}
\label{subsubsec:limitations-proprietary-opacity}

A persistent systemic challenge
in cloud systems literature
is the significant discrepancy
between academic testbed evaluations
and large-scale commercial hyperscaler reality.
Tier-one cloud operators
(e.g., Google Cloud, Microsoft Azure, and Amazon Web Services)
deploy proprietary accelerator \acp{ASIC},
custom optical network topologies,
and bespoke telemetry platforms that
are strictly guarded as commercial secrets.
Consequently, most surveyed papers
evaluate architectures on modest academic clusters
(typically 4 to 64 accelerators)
or rely on synthetic load generators
that approximate idealized production distributions.
Real-world operational complexities,
including unpredictable multi-tenant burstiness,
cross-datacenter fiber outages,
accelerator degradation over
prolonged multi-month operations,
and complex billing contracts,
are rarely fully modeled in published literature.

\subsubsection{Scope Delimitation: Serving versus Pre-Training and Ultra-Edge Platforms}
\label{subsubsec:limitations-scope-delimitation}

To maintain rigorous technical depth,
our study explicitly scoped its inquiry
around cloud-hosted model orchestration,
inference serving efficiency,
and cloud-native service architectures.
Consequently, we deliberately excluded
foundational pre-training infrastructure,
large-scale dataset pipelining,
and post-training alignment
algorithms (such as RLHF and DPO)
except where they directly intersect
with inference serving systems.
Similarly, while we reviewed
collaborative cloud-edge offloading architectures,
we excluded ultra-low-power embedded microcontrollers,
on-device mobile operating systems,
and disconnected edge sensory nodes
that operate without cloud coordination.

\subsubsection{Standardized Multi-Tenant Dynamic Benchmarking Deficits}
\label{subsubsec:limitations-benchmarking-gaps}

Across the surveyed corpus,
a notable limitation is the marked
heterogeneity of experimental
benchmarking methodologies.
Different research groups
report performance improvements
evaluated on conflicting datasets,
varying model families
(e.g., LLaMA, Mistral, Falcon, or OPT),
distinct batch sizes,
and divergent definitions of
service-level compliance.
Furthermore, publicly available
production serving traces remain extremely sparse,
forcing most researchers to
extrapolate request arrival patterns
from generic serverless datasets
(such as older Azure Functions traces).
The absence of universally recognized,
standardized dynamic benchmark suites that
simultaneously capture multi-turn
conversational context,
agentic iterative calling,
and multimodal queries hinders
fair cross-framework quantitative comparison.

\subsection{Future Literature Review Opportunities}
\label{subsec:discussion-future-reviews}
As foundation model ecosystems continue to mature,
subsequent systematic reviews must expand upon
the architectural foundation established in this work.
We recommend two critical methodological and thematic avenues
for future literature surveys and meta-analyses.

\subsubsection{Multimodal Foundation Model Orchestration and Serving Infrastructure}
\label{subsubsec:future-reviews-multimodal}

The rapid convergence of language
models with computer vision,
audio processing, video generation,
and embodied robotics introduces
heterogeneous computational
modalities that break text-only serving assumptions.
Unlike text tokens, which exhibit
uniform dimensional footprints,
multimodal tokens (such as high-resolution
image patches or dense audio spectrograms)
impose radically asymmetric memory and compute
demands on cloud clusters.
Future literature reviews should systematically
investigate the orchestration mechanisms,
caching hierarchies, and scheduling policies
tailored specifically for
multimodal foundation model serving.
Key questions include how cloud platforms
manage cross-modal attention caching,
how they balance multi-stage vision-encoder
and language-decoder execution pipelines,
and how spatial compute graphs are dynamically
scheduled across heterogeneous accelerator fleets.

\subsubsection{Longitudinal Trace-Driven Meta-Analyses and Empirical Benchmarking Surveys}
\label{subsubsec:future-reviews-empirical}

While this survey establishes a
structured qualitative taxonomy
and comparative mapping of systems innovations,
the field urgently requires
comprehensive empirical meta-analyses.
Future survey efforts should coordinate
multi-institution benchmarking initiatives
that deploy open-source serving systems
(such as \ac{vLLM}, DistServe, SGLang, and ServerlessLLM)
on identical physical hardware fabrics
under standardized workload conditions.
Such empirical reviews should synthesize
longitudinal cluster telemetry,
evaluating Pareto-optimal frontiers across throughput,
energy expenditure, tail latency compliance, and dollar cost.
Establishing reproducible, trace-driven empirical
surveys will provide cloud architects
with definitive empirical reference
models that bridge the gap
between academic prototypes and
industrial production deployment.

    \section{Conclusion}
    \label{sec:conclusion}
    In this systematic literature review,
we provided a comprehensive,
methodologically rigorous examination
of large language model serving,
orchestration, and system architectures
across modern cloud environments.
Guided by the \ac{PRISMA} 2020 framework,
we critically synthesized contemporary
advances across four foundational research questions,
elucidating the architectural mechanisms required to deploy,
scale, and sustain generative foundation
models at an enterprise scale.

Our investigation demonstrates
that cloud \ac{LLM} serving
has transitioned from conventional
homogeneous accelerator clusters
toward deeply specialized, heterogeneous,
and disaggregated execution fabrics.
In response to \ac{RQ}1, we analyzed hardware
virtualization paradigms and cluster-level scheduling frameworks,
revealing how spatial and temporal accelerator
sharing (via \ac{MIG} and \ac{MPS})
and predictive resource consolidation mitigate
severe compute underutilization.
Addressing \ac{RQ}2, we examined inference serving efficiency,
highlighting how the operational dichotomy
between compute-bound prefill
and memory-bandwidth-bound decoding
has catalyzed phase-disaggregated architectures,
chunked prefilling,
and aggressive parameter quantization.
Addressing \ac{RQ}3, we evaluated cloud-native
and serverless paradigms,
demonstrating how tiered memory snapshotting,
layer-level streaming, and predictive pre-warming
reconcile stateless scale-to-zero semantics
with the multi-gigabyte memory footprints
of stateful model weights and \ac{KV} caches.
Finally, in response to \ac{RQ}4,
we charted the landscape of system reliability,
hardware fault tolerance, and tenant privacy,
underscoring the imperative of mitigating silent data corruption
and side-channel prompt leakage through
hardware-enforced trusted execution environments.

Looking forward, the trajectory of the cloud
foundation model serving points toward a
fundamental architectural evolution.
Overcoming the impending memory
wall and datacenter power constraints
demands the convergence of disaggregated \ac{CXL} memory pooling,
photonic interconnects,
linear-complexity sequence architectures,
and post-von Neumann paradigms
such as in-memory \ac{HDC} and \ac{VSA}.
Concurrently, the transition toward
autonomous multi-agent swarms
and carbon-budgeted geographical workload routing
will redefine cluster orchestration from reactive task scheduling
to self-healing, climate-aware decentralized control planes.
By formalizing the foundational trade-offs between compute saturation,
state retention, and tenant isolation,
this survey provides a definitive architectural blueprint
to guide system designers, cloud practitioners,
and researchers in constructing the next generation of robust,
scalable, and sustainable cloud \ac{AI} infrastructure.

    \backmatter

%
%

    \bmhead{Conflict of interest/Competing interests}
    The authors declare that they have no known competing financial
    interests or personal relationships that could have appeared to influence
    the work reported in this paper.

    \bibliography{sn-bibliography}

\end{document}